\documentclass[a4paper,12pt,times,numbered,index,custombib,twoside]{Classes/PhDThesisPSnPDF}

\input{Preamble/preamble}

\title{On the Reconstruction of Dark Energy Models}
\author{Ankan Mukherjee\\(10IP10)}

\renewcommand{\submissiontext}{\bf Supervisor: Prof. Narayan Banerjee}

\dept{Department of Physical Sciences}

\university{ Indian Institute of Science Education and Research Kolkata}

\degreetitle{\small Thesis submitted to IISER Kolkata 
\\
for the Degree of Doctor of Philosophy in Science}
\crest{\centering \includegraphics[width=0.25\textwidth]{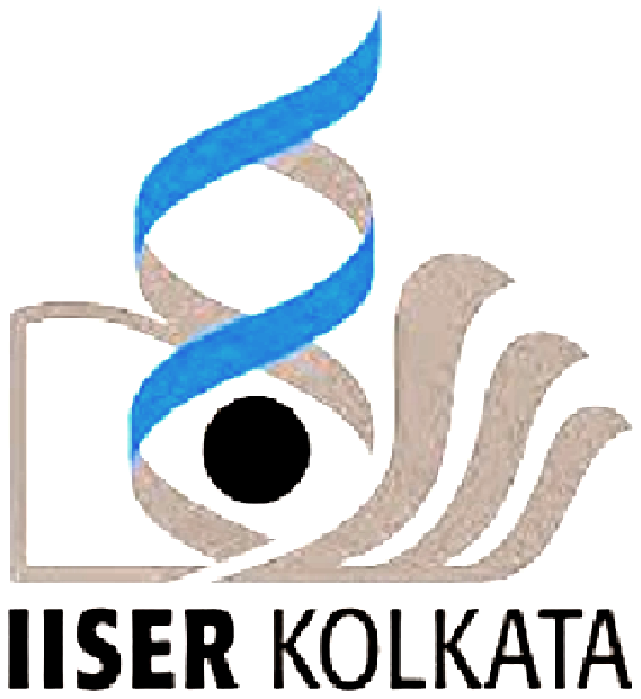}}

\subject{LaTeX} 

\ifdefineAbstract
\fi

\begin{document}

\frontmatter
\begin{titlepage}
  \maketitle
\end{titlepage}

% ******************************* Thesis Dedidcation ********************************

\begin{dedication} 

{\Large To

My loving parents 

\&

My teachers.  }

\end{dedication}

\begin{dedication}

\begin{center}

{ \Large \bf \underline{Certificate} }

\end{center}

\begin{flushleft}

{\centering This is to certify that the Ph.D. thesis entitled “ {\bf  On the Reconstruction of Dark Energy Models}” submitted by {\bf Ankan Mukherjee } is absolutely based upon his own work under the supervision of  {\bf Prof. Narayan Banerjee} at Indian Institute of Science Education and Research (IISER) Kolkata and that neither this thesis nor any part of it has been submitted for either any degree/diploma or any other academic award anywhere before.}

\end{flushleft}

~~

\begin{flushleft}
Mohanpur, India.
\end{flushleft}

\begin{flushright}
(Prof. Narayan Banerjee)
\end{flushright}

\end{dedication}

% ******************************* Thesis Declaration ***************************

\begin{declaration}

This thesis is a presentation of my original research work. Whenever contributions of others are involved, every effort is made to indicate this clearly, with due reference to the literature and acknowledgement of collaborative research and discussions. The work is original and has not been submitted earlier as a whole or in part for a degree or diploma at this or any other Institution or University. This work has been done under the guidance of Prof. Narayan Banerjee, at Indian Institute of Science Education and Research (IISER) Kolkata.

\vskip 2.0 cm

% Author and date will be inserted automatically from thesis.tex \author \degreedate

\end{declaration}

% ************************** Thesis Acknowledgements **************************

\begin{acknowledgements}      

First and foremost I would like to thank my supervisor, Prof. Narayan Banerjee for precious guidance, support and infinite patience which have made the thesis possible. Besides, his incredible teaching and inspirations are always highly motivating in the way of life. 

I am also greatful to Prof. Anjan Ananda Sen (Centre for Theoretical Physics, Jamia Millia Islamia, New Delhi), for his valuable suggestions and helps regarding the research works. 

My special thanks to Prof. Prasanta K. Panigrahi, Dr. Dibyendu Nandi, Dr. Golam Mortuza Hossain, Dr. Ritesh K. Singh for valuable discussions, teaching and suggestions.

I would also like to thank  my seniors Arghya Choudhury, Kumar Abhinav, Barun Majumder, Soumitra Hazra, Nandan Roy, Dyuti Bhattacharya, Aveek Sarkar, Subhrajit Modak. Very special thanks to my friends and co-researchers at IISER Kolkata Soumya, Debmalya, Gopal, Santanu, Diptesh,  Subhajit, Chiranjeeb, Anushree, Nivedita, Vivekananda, Santanu, Anuradha, Anirban. I would like to thank Md. Wali Hossain,  Sumit Kumar for useful discussions during my visit to Centre for Theoretical Physics, Jamia Millia Islamia.

I express my gratitude to my parents for their enormous support and inspirations.

\end{acknowledgements}

% ************************** Thesis Abstract *****************************
% Use `abstract' as an option in the document class to print only the titlepage and the abstract.
\begin{abstract}
The present work, which has been discussed in the following chapters, is devoted to the reconstruction of the dark energy models using diverse observational data sets. The parametric approach has been adopted for the reconstruction of cosmological models. The reconstruction of kinematical quantities and the possibility of interaction between dark energy and dark matter has also been emphasised. In first chapter, a brief introduction to cosmology has been presented. In the second chapter, a reconstruction of the dark energy equation of state parameter for a quintessence scalar field model has been discussed. The nature of evolution of the relevant cosmological parameters and the quintessence potential has also been studied for the reconstructed model.  In the third chapter, a parametric reconstruction of the effective or total equation of state has been discussed.  The reconstructed model mimics the $\Lambda$CDM model for a particular value of  the model parameter, thus the reconstruction indicates the consistency or deviation form $\Lambda$CDM. A comparative study of the reconstructed model and the $w$CDM model has been done by representing both the models on the same parameter space. It has been shown that the reconstructed model ensures tighter constraints on the kinematic quantities like the deceleration parameter and the jerk parameter. In the fourth chapter, a kinematic approach in the reconstruction of dark energy model through the parametrization of the cosmological jerk parameter has been discussed. Four different parametric form of the jerk parameter have been studied. The evolution of the deceleration parameter, dark energy equation of state and also the nature of associated quintessence potential have been studied for these models. The fifth chapter is also about a kinematic approach to the reconstruction of dark energy. The reconstruction has been done with an assumption that the jerk parameter is a very slowly varying function. This model invokes the possibility of interaction between dark matter and the dark energy. The interaction term has also been reconstructed and it shows that the possibility of interaction is high in the past, but it is very small at the present time. The sixth chapter is devoted to the reconstruction of the interaction rate in holographic dark energy model. The interaction rate has been reconstructed for three different parametrization of the deceleration parameter. The evolution of the interaction rate, the nature of dark energy equation of state parameter etc. have been studied. The cosmic coincidence problem has also been addressed in the context of holographic dark energy.  Finally, chapter seven contains the concluding remarks and relevant discussions regarding the overall work presented in the thesis.    

\end{abstract}

  \begin{dedication}
  \vskip -2.0 cm
  \textbf{\Large Preface}

  \begin{flushleft}
  
  The research work contained in this thesis was carried out at the Department of Physical Sciences,  Indian Institute of Science Education and Research (IISER) Kolkata, India . \\

Chapter 1 contains an introduction and the other chapters are based on the papers as follows:

  \begin{itemize}
  
  \item Chapter 2 
  
  A. Mukherjee and N. Banerjee, {\it ``A reconstruction of quintessence dark energy"},\\ Eur. Phys. J. Plus {\bf 130}, 201 (2015).
 
  \item Chapter 3  
 
  A. Mukherjee, {\it ``Acceleration of the universe: a reconstruction of the effective equation of state"}, Mon. Not. R. Astron. Soc. {\bf 460}, 273 (2016).

 \item Chapter 4

    A. Mukherjee and N. Banerjee, {\it ``Parametric reconstruction of the cosmological jerk from diverse observational data sets"},  Phys. Rev. D {\bf 93}, 043002 (2016).

  \item Chapter 5
  
    A. Mukherjee and N. Banerjee, {\it  ``In search of the dark matter dark energy interaction: a kinematic approach''},  Class. Quantum Grav. {\bf 34}, 035016 (2017).

  \item Chapter 6 
  
   A. Mukherjee, {\it ``Reconstruction of interaction rate in holographic dark energy''},\\ JCAP {\bf 11}(2016)055. 
  
\end{itemize}    

\end{flushleft}
  
  \end{dedication}

\begin{dedication}

\textbf{\Large List of Publications}

\begin{flushleft}

1.  A. Mukherjee and N. Banerjee, {\it  ``In search of the dark matter dark energy interaction: a kinematic approach''}, Class. Quantum Grav. {\bf 34}, 035016 (2017);  {\footnotesize arXiv:1610.04419 [astro-ph.CO]}. 

\vskip 0.6 cm

2.  A. Mukherjee, {\it ``Reconstruction of interaction rate in holographic dark energy''},\\ JCAP {\bf 11}(2016)055; {\footnotesize arXiv:1608.00400 [astro-ph.CO]}.\\

\vskip 0.6 cm

3.   A. Mukherjee, {\it ``Acceleration of the universe: a reconstruction of the effective equation of state"}, Mon. Not. R. Astron. Soc. {\bf 460}, 273 (2016); {\footnotesize arXiv:1605.08184 [gr-qc]}.\\

\vskip 0.6 cm

4.  A. Mukherjee and N. Banerjee, {\it ``Parametric reconstruction of the cosmological jerk from diverse observational data sets"},  Phys. Rev. D {\bf 93}, 043002 (2016);  {\footnotesize arXiv:1601.05172 [astro-ph.CO]}.\\

\vskip 0.6 cm

5.  A. Mukherjee and N. Banerjee, {\it ``A reconstruction of quintessence dark energy"},\\ Eur. Phys. J. Plus {\bf 130}, 201 (2015); {\footnotesize arXiv:1311.4024 [gr-qc]}.\\

\vskip 0.6 cm

6.  A. Mukherjee and N. Banerjee, {\it `` Acceleration of the universe in f(R) gravity models''} , Astrophys. Space Sci. {\bf 352}, 893 (2014);
    {\footnotesize arXiv:1405.6788 [gr-qc]}.\\
    {\it (This paper is not included in the thesis.)}

\end{flushleft}

\end{dedication}

% *********************** Adding TOC and List of Figures ***********************

\tableofcontents

\listoffigures

\listoftables

% \printnomencl[space] space can be set as 2em between symbol and description
%\printnomencl[3em]

\printnomencl

% ******************************** Main Matter ********************************* 
\mainmatter

%*******************************************************************************
%*********************************** First Chapter *****************************
%*******************************************************************************

\chapter{Introduction}  %Title of the First Chapter

\ifpdf
    \graphicspath{{Chapter1/Figs/Raster/}{Chapter1/Figs/PDF/}{Chapter1/Figs/}}
\else
    \graphicspath{{Chapter1/Figs/Vector/}{Chapter1/Figs/}}
\fi

%********************************** %First Section  **************************************
\section{Introduction to Cosmology}
The quest of knowing the beginning, the evolution and the ultimate fate of the Universe is the prime motivation of the subject of cosmology. 

\par Earlier, it was believed that we occupy a very special location in the Universe and it was believed to be the centre of the Universe. But in modern cosmology, this idea has been abandoned. We are at a place which is no way special than the other places in the Universe. This simple and reasonable idea is called the {\it cosmological principle}.

\par It is important to note that the viability of the cosmological principle depends upon the length scale  of interest. Even if the length scale is of the order of that of a single galaxy (10 to 100 kpc, where 1 pc = 3.0857$\times10^{16}$ meters or 3.26 light-years), this principle lacks the viability. But observations indicate quite strongly that at a length scale of 10 Mpc or more, which is still 3 order of magnitude less than the visible Universe, the cosmological principle appears to be true with a great accuracy. 

\par The standard Big Bang picture represents the Universe as an evolving entity, the present Universe has actually evolved from a condition different from what prevails now. This evolving picture of the Universe is highly supported by the cosmological observations which practically rule out stationary models of the Universe.

\par Cosmological principle implies important properties, spatial {\bf homogeneity} and {\bf isotropy} of the Universe. Homogeneity states that the Universe is similar at every spatial points and isotropy states that the Universe is similar in all spatial directions. 

\par Edwin Hubble, in 1929, observed a redshift in the spectral line of nearby galaxies and nebula and concluded that those objects are moving away from each other. Hubble expressed the velocity of recession ($\overrightarrow{v}$) to be proportional to the distance of the object ($\overrightarrow{r}$),

\begin{equation}
\overrightarrow{v}=H_0\overrightarrow{r},
\end{equation}   

where the proportionality constant $H_0$ is called the Hubble constant. This is known as the {\bf Hubble's law}. It is the pioneering observation which advocates the time evolution of the Universe.

\subsubsection{Friedmann Cosmology} 
The mathematical framework of cosmology is mainly based on the the way of defining the distance between two points in the Universe at cosmological scale. Let us consider a coordinate system where the coordinate separation between two points is $\overrightarrow{x}$, then the real distance between these two points are 

\begin{equation}   
\overrightarrow{r}=a(t)\overrightarrow{x},
\end{equation}

where the $a(t)$, called the {\it scale factor}, takes care of the time evolution of the real distance. As the Universe expands, the real distance increases, even between two comoving objects for whom the coordinate distance ($\overrightarrow{x}$) remains the same. The idea of homogeneity ensures that the scale factor $a$ is only a function of time.  

 The most general way of defining the line element, which satisfies the cosmological principle, is the Friedmann-Robertson-Walker (FRW) metric \cite{c1, c2, c3, c4}, which is written as 
 \begin{equation} \label{frw}
 ds^2 = - dt^2 + a^2(t) \Big[\frac{dr^2}{1- k r^2} + r^2(d \theta^2 + \sin^2{\theta} d \phi^2)\Big],
 \end{equation}
 where $t$ is cosmic time and ($r, \theta, \phi$) are the spatial coordinates and $k$ is the curvature parameter which determines the nature of the spatial geometry.

\par Along with the cosmological principle, two other basic assumption in the modelling of the Universe are, (i) general relativity is the correct theory of gravity, and (ii) the matter distribution of the Universe can be represented as ideal fluid. Einstein's field equations which  are the key equations of general relativity, are written as,

\begin{equation}
G_{\mu\nu} \equiv R_{\mu\nu} - \frac{1}{2}g_{\mu\nu}R = 8\pi GT_{\mu\nu},
\label{efieldeq}
\end{equation}
where $G_{\mu\nu}$ is the Einstein tensor, $R_{\mu\nu}$ is the Ricci tensor, $R$ is the Ricci scalar and $T_{\mu\nu}$ is the energy momentum tensor. Introducing the FRW metric to the Einstein's field equations (\ref{efieldeq}), one arrives at the two basic equations of cosmology,

\begin{equation}
3\frac{\dot{a}^2}{a^2}+3\frac{k}{a^2}=8\pi G\rho,
\label{friedmann1}
\end{equation}  

and

\begin{equation}
2\frac{\ddot{a}}{a}+\frac{\dot{a}^2+k}{a^2}=-8\pi Gp.
\label{friedmann2}
\end{equation}  

These are known as the Friedmann equations. Here, $\rho$ and $p$ are respectively the energy density and pressure of perfect fluid for which the energy momentum tensor is $T^{\mu}_{\nu}=Diag(-\rho,p,p,p)$.

\par The Hubble parameter, which is the fractional rate of expansion of the linear size of the Universe, is defined as,

\begin{equation}
H(t)=\frac{\dot{a}}{a}.
\end{equation}
The Friedmann equations can be written in terms of Hubble parameter and its derivatives as, 
\begin{equation}
3H^2+3\frac{k}{a^2}=8\pi G\rho,
\label{friedmannhubble}
\end{equation}

\begin{equation}
2\dot{H}+3H^2+\frac{k}{a^2}=-8\pi Gp.
\label{friedmannhubble2}
\end{equation}

From {\bf contracted Bianchi identity} ($G^{\mu\nu}_{~~~;\mu}$), the continuity equation yields,

\begin{equation}
\dot{\rho}+3H(\rho+p)=0.
\label{coneq}
\end{equation}
The energy density ($\rho$) and pressure ($p$) are related through an equation called the equation of state, $p=w\rho$, where the $w$ is called the equation of state parameter. For pressureless dust matter the equation of state parameter $w=0$ and for a distribution of photons, the value of equation of state parameter is $w=1/3$.  For a certain component, if the equation of state parameter $w$ is known, the dependence of the energy density ($\rho$) upon the scale factor ($a(t)$) can be determined. For instance, dust matter density $\rho_{matter}\propto\frac{1}{a^3(t)}$, radiation or photon energy density $\rho_{photon}\propto\frac{1}{a^4(t)}$. 

\vskip 1.0 cm

\par Now let us define a quantity called the {\it critical density} as,

\begin{equation}
\rho_c(t)=\frac{3H^2(t)}{8\pi G}.
\end{equation}

From equation (\ref{friedmannhubble}),

\begin{equation}
\Omega(t)=1+\frac{k}{3a^2H^2},
\end{equation}

where $\Omega(t)=\frac{\rho(t)}{\rho_c(t)}$. If the energy density of the Universe is exactly equal to the critical density, then the curvature parameter $k=0$, that means that spatial geometry is flat. If $\rho(t)>\rho_c(t)$, then the value of $k$ is positive and it can be scaled as $+1$. Similarly for $\rho(t)<\rho_c(t)$, the value of $k$ is negative and it can be scaled as $-1$. For $k=+1$, the spatial geometry of the Universe is closed and for $k=-1$, the spatial geometry if open.

\par Observations suggest that the geometry of the spatial part of the  Universe is very close to be flat, i.e. $\rho(t)\approx\rho_c(t)$ \cite{komatsuWMAP,planck2013,planck2015}. It is theoretically argued that the flatness of spatial geometry of the Universe is a consequence of an early inflation. The idea is that the rapid expansion during the inflation washed out the spatial curvature.

\subsubsection{Cosmological redshift, luminosity distance and angular diameter distance}

As the electromagnetic radiation or photon can travel a finite distance in the time equal to the age of the Universe, we can only see a certain part of the Universe. It is known as the {\it observable Universe}.

\par Due to the expansion of the Universe, the wavelength of radiated photon increases. The photon wavelength is proportional to the scale factor, i.e. $\lambda\propto a(t)$. Now the redshift of photon is defined as,

\begin{equation}
z=\frac{\lambda_{observed}-\lambda_{emitted}}{\lambda_{emitted}},
\end{equation}
which leads to
\begin{equation}
1+z=\frac{a_0}{a(t_e)},
\end{equation}
where $a_0$ is scale factor at present (when the photon is  observed) and $a(t_e)$ is the scale factor when the photon was emitted. It is convenient to use the redshift ($z$), instead of cosmic time $t$ in the study of the late time dynamics of the Universe as $z$ is dimensionless quantity.

\par The {\it luminosity distance} is a way of representing the observed photon flux from a distant object. It is not exactly the actual or the physical distance of the object from the observer. Let us consider an object with a total power output $4\pi L$, where $L$ be the luminosity, that is the energy emitted per unit solid angle. The radiation flux ($S$) is the amount of energy received per unit area per unit time. The distance (which is actually the luminosity distance $d_{L}$), of  the object and the observed photon flux can be connected through the following equation, 

\begin{equation}
S=\frac{4\pi L}{4\pi d^2_{L}}.
\end{equation} 
Hence the luminosity distance is defined as,

\begin{equation}
d^2_{L}=\frac{L}{S}.
\label{lumdis}
\end{equation}

In a static Universe, the luminosity distance is exactly equal to the physical distance ($d_{phys}$) of the object from the observer. As the Universe expands, the energy loss of the of the photon is proportional to $(1+z)$, and the observed photon becomes less frequent which  is also proportional to $(1+z)$. Thus the final relation between the luminosity distance ($d_{L}$) and physical distance ($d_{phys}$) yield as,

\begin{equation}
d^2_{L}=d^2_{phys}(1+z)^2,
\end{equation} 
or, 
\begin{equation}
d_{L}=d_{phys}(1+z).
\label{dlumdphys}
\end{equation} 

It is clear from the relation (equation (\ref{dlumdphys})), that for a nearby object (i.e. $z<<1$), $d_{L}\approx d_{phys}$. However, an object at a long distance seems to be farther away than it really is ($d_{L}> d_{phys}$).

\par Angular diameter distance is a distance estimated from the appearance of an object in angular extent. If an object of physical extent '$l$' be observed at an angle $d\theta$, then the angular diameter distance is written as (assuming the object lies perpendicular to the line of sight),

\begin{equation}
d_{A}\simeq\frac{l}{d\theta}.
\end{equation} 

Now the angular diameter distance can be written as, $d_{A}=r_0a(t_e)$, where $r_0$ be the coordinate distance and $a(t_e)$ be the scale factor at the time of photon emission. The physical distance is $d_{phys}=r_0a_0$. Thus the relation between $d_{A}$ and $d_{phys}$ can be written as, 

\begin{equation}
d_{A}=\frac{d_{phys}}{(1+z)}.
\end{equation}

It is interesting to note that the that for a distant object $d_{A}<d_{phys}$, that means a distant object appears to be larger at angular extent.

\par The relation between luminosity distance ($d_{L}$) and angular diameter distance ($d_{A}$) is,

\begin{equation}
d_{A}=\frac{d_{L}}{(1+z)^2}.
\end{equation} 

The luminosity distance is related to the observed photon flux and angular diameter distance is related to the appearance of an object at angular extent. For a nearby object ($z<<1$), both the luminosity distance and angular diameter distance are equal to the physical distance of the object from the observer.

\subsubsection{Success and problems of standard cosmology}
As it has already been mentioned that observations suggest that the Universe is an evolving entity rather than a stationary one. The evolution picture of the Universe is developed from the Bag Bang model based on the theory of general relativity. The success of the Big Bang model is based on three observational facts, the expansion of the Universe, the abundance of light element in the Universe and the presence of the relic blackbody radiation called  the cosmic microwave background (CMB). There  are certainly other very important successes of the model. 

\par The expansion of the Universe was first observed by Hubble (in 1929) from the redshift measurement of distance galaxies. The fact that galaxies are moving away from each other, that is the Universe is expanding, gives a big justification of the Big Bang cosmology. 

\par The abundance of light elements in the Universe can be well explained by the Big Bang cosmology. At the very beginning, the temperature of the Universe was much higher than the binding energy of atoms or nuclei. Due to the expansion of the Universe, the temperature gradually decreases and at a temperature well below the binding energy of typical nuclei, the formation of light element started. This scenario is dubbed as the {\bf Big Bang Nucleosynthesis}. For the conditions of the early Universe and the relevant nuclear scattering-crossections, the relative abundance of different element can be estimated. This matches the observed relative abundance of different elements in the Universe with great accuracy. This is another big success of the Big Bang cosmology.

\par The most astounding success of the Big Bang model is the prediction of cosmic microwave background (CMB) radiation \cite{alpherherman}. These are the relic photons that decoupled from the baryonic matter when neutral hydrogen atoms formed at the {\it era of recombination}  at redshift $z\sim1100$. This was detected in 1965 by Penzias and Wilson \cite{penziascmb}. This CMB is very highly isotropic black body radiation and acts as the most powerful probe of the Universe.

\par Though these observations have brought enormous success to the standard Big Bang cosmology, there are certain issues which can not be explained by it.  The tiny fluctuation in the observed CMB temperature and the source of initial perturbation in the matter distribution, which led to the formation of large scale structure in the Universe, can not be explained by the Big Bang model. The early inflation, which is the very rapid expansion of the Universe just after Big Bang (from $10^{-36}sec$ to $10^{-33}sec$), is essential to explain  the flatness problem and horizon problem ({\it detail discussions are in the next section}) in cosmology. The recent surprise is the alleged accelerated expansion of the Universe at present epoch. The Big Bang model does not have any viable explanation about genesis of these two phases of accelerated expansion of the Universe. The existence of {\it dark matter}, which gravitates like the ordinary baryonic matter but does not have other types of interaction, is also a problem not fully resolved as yet.

\section{Accelerated expansion of the Universe}
The observations suggest that the Universe is expanding with an accelerating rate at present. The accelerated expansion of the Universe is very strange because gravity is known to be attractive. From the Friedmann equations (equation (\ref{friedmann1}) and (\ref{friedmann2})), $\ddot{a}/a$ can be expressed as,

\begin{equation}
\frac{\ddot{a}}{a}=-\frac{4\pi G}{3}(\rho+3p).
\label{addot}
\end{equation} 

As the scale factor $a(t)$ is the scaling of distance between two object in the Universe at cosmological scale, the second order time derivative of $a(t)$ is actually the acceleration. The knowledge of standard cosmology suggests that the normal components of the energy budget of the Universe (mainly the dark matter, the ordinary luminous matter, the relativistic particles ({\it photon and neutrino})) respect the strong energy condition that is $\rho+3p>0$. Thus it shows that the rate of expansion should decrease (decelerated expansion) which is not consistent with the present observations. 

\par The cosmic acceleration is represented in a dimensionless way using the deceleration parameter $q(t)$, defined as,

\begin{equation}
q(t)=-\frac{1}{H^2(t)}\frac{\ddot{a}}{a}.
\label{decel}
\end{equation}

If the value of $q(t)$ is negative, then the Universe is accelerating and if $q(t)$ is positive, the expansion is decelerating.

\par The Universe apparently has two phases of accelerated expansion. One is the early inflation which is required to explain certain observational phenomena. The second one is the late time acceleration which is the present state of evolution of the Universe.  In between this two phases of accelerated expansion, there prevailed a phase of decelerated expansion.

\par The cosmological inflation is the very rapid expansion of the Universe just after the Big Bang. It started after $10^{-36}$ sec. of the Big Bang and ended around $10^{-33}$ sec. to $10^{-32}$ sec. Inflationary models have been introduced by Alan Guth \cite{guthinflation} for theoretical requirement to explain the flatness problem and the horizon problem. The flatness problem is related to the reason of observed flatness of the spatial geometry of the Universe and the horizon problem is related to  the isotropy of the cosmic microwave background (CMB) radiation.

\par Late time acceleration is an observed phenomenon. It was first discovered in late nineties by two supernova observing groups independently (Supernova Search Team led by Riess \cite{riess1} and Supernova Cosmology Project headed by Permutter \cite{perlmutter}).  The prime observation was of the magnitude redshift curve of type Ia supernovae and it was highly consistent with the present accelerated expansion of the Universe. Later on, many other observations have confirmed the late time cosmic acceleration \cite{sn2,sn3,snsuzuki, allenSW}. Analysis of observational data also suggest that the acceleration has started very recently, at around redshif $z\approx0.5$ \cite{riess2}. It has emerged as the most puzzling phenomenon of modern cosmology.  

\par As already mentioned, the accelerated expansion of the Universe is surprising because for all the well known components of the Universe, even for the dark matter, gravity is attractive. But the accelerated expansion invokes the possibility of repulsive gravity at cosmological scale. There are different theoretical prescriptions in literature attempting to answer this puzzle. The following section is devoted to the discussions about those theoretical attempts to explain the phenomenon of cosmic acceleration.

\section{Theoretical attempts to explain the cosmic acceleration}

As already mentioned, for the known components of the Universe it is not possible to generate the accelerated expansion. There are different theoretical prescriptions in the literature to explain the cosmic acceleration. Though, still now,  none of them has been universally accepted as the one which is flawless, has a theoretical support and a clue towards direct detection. The attempts are broadly classified into two classes. 

\par One is the dark energy model, where some exotic component, dubbed as {\it dark energy} is introduced in the matter sector which has an effective negative pressure, making $\rho+3p<0$ and thus giving rise to a negative value of the deceleration parameter ($q$) (see equations (\ref{addot}) and (\ref{decel})). Dark energy approach is based on the assumption that the GR is the appropriate theory of gravity.

\par The second way is to look for a suitable modification of GR to accommodate the cosmic acceleration without introducing any exotic component to the energy budget of the Universe. \\

Both of these approaches are applicable in the modelling of cosmic inflation and late time acceleration.  These two approaches have been discussed in the following.

\subsection{Dark energy models}

The dark energy models are the attempts to explain the cosmic acceleration assuming  GR to be the appropriate theory of gravity\cite{shstatoAA,paddyLCDM,copelandsami,ruiz,durrermaartens}. Equation (\ref{addot}) shows that an accelerated expansion can be generated if some component of the energy sector has sufficient negative pressure. The exotic component , dubbed as dark energy, generates the acceleration by its characteristic negative pressure. It is essential to mention that this pressure is not the fluid pressure which is caused by the motion of particles. Rather, it is actually the contribution of the exotic component to the energy momentum tensor. It is convenient to introduce the density of different components in a dimensionless way by scaling them with critical density ($\rho_c$) which is defined as, 

\begin{equation}
\rho_c=\frac{3H^2}{8\pi G}.
\end{equation}

The dimensionless representations of the energy densities of different components are,

\begin{equation}
\Omega_i=\frac{\rho_i}{\rho_c},
\end{equation} 

where $i=m$ denotes the dust matter, $i=r$ denotes the radiation and $i=DE$ denotes the dark energy and $\Omega_k=-\frac{k}{a^2\rho_c}$ is the contribution of  the spatial curvature. For a spatially flat model, $\Omega_k=0$.  The $\Omega_i$ are called the density parameter of the corresponding component. Now the Hubble parameter $H$ scaled by its present value $H_0$ can be expressed in terms of the density parameters as (for a spatially flat model),

\begin{equation}
h^2\equiv\frac{H^2}{H_0^2}=\Omega_m+\Omega_r+\Omega_{DE}.
\end{equation} 

It is clear from equation (\ref{addot}) that for accelerated expansion, the effective or total equation of state parameter ($w_{eff}=p/\rho$) should be less than $-\frac{1}{3}$. The matter component present in the energy budget is mainly in the form of dust which is pressureless, that is $p_m=0$. So the equation of state parameter for dust matter $w_m=0$. Now the dark energy equation of state parameter ($w_{DE}$) is defined as,

\begin{equation}
w_{DE}=\frac{p_{DE}}{\rho_{DE}}.
\end{equation}    

The limiting value of $w_{DE}$ required for cosmic acceleration can be estimated from the deceleration parameter $q$ (defined in equation (\ref{decel})). Ignoring the radiation energy density as it has only a negligible contribution compared to the other components at late time, the deceleration parameter can be expressed as,

\begin{equation}
q\simeq\frac{1}{2}(1+3w_{DE}\Omega^{-1}_{DE}).
\end{equation} 

Hence for a negative value of $q$, 

\begin{equation}
w_{DE}<-\frac{1}{3}\Omega^{-1}_{DE}.
\end{equation} 

The recent cosmological observations suggests that the present value of dark energy density parameter $\Omega_{DE}(z=0)\simeq0.7$ and thus the present value of the dark energy equation of state parameter should be $w_{DE}(z=0)<-0.5$. It is important to note that the dark energy has started  dominating the dynamics of the Universe at recent past \cite{sn2,snsuzuki,sn3,allenSW,riess2}. Earlier it was mostly dominated by dust matter and radiation. This invokes the cosmic coincidence problem, that is, why the dark energy has started dominating the evolution so recently. The other issue is the physical identity of dark energy. In the following, different dark energy models are briefly discussed.

\subsubsection{Cosmological constant ($\Lambda$)} 
The simplest model of dark energy is the cosmological constant model. Dark energy model with cosmological constant is called the $\Lambda$CDM (cosmological constant $\Lambda$ with pressureless cold dark matter). The constant energy density associated to $\Lambda$ is, 

\begin{equation}
\rho_{\Lambda}=\frac{\Lambda}{8\pi G}=-p_{\Lambda}.
\end{equation} 

The dark energy equation of state parameter for cosmological constant $w_{DE}=-1$. The cosmological constant model is preferred by most of the observations at a good level of accuracy. But there are certain issues related to cosmological constant. The constant vacuum energy density is the only possible candidate for the cosmological constant. But it suffers from the humongous discrepancy between the observationally estimated value ($\rho_{\Lambda}^{obs}$) and the theoretically calculated value ($\rho_{\Lambda}^{th}$) of the energy density. The ratio is $\rho_{\Lambda}^{obs}/\rho_{\Lambda}^{th}\sim10^{-120}$.  This is the {\it fine tuning problem} of cosmological constant model. Different aspects of the cosmological constant model have been discussed in great detail by Carroll \cite{carrollLCDM} and by Padmanabhan \cite{paddyLCDM}.

\subsubsection{Constant dark energy equation of state model ($w_{DE}\neq-1$)} 

A constant dark energy equation of state parameter, with a value other than $w_{DE}=-1$, is also a relevant option for phenomenological study of dark energy. It is the $w$CDM model where $w_{DE}=constant$, but not necessarily equal to $-1$. It is a phenomenological approach to investigate whether observational evidence of any deviation from $\Lambda$CDM model. 

\par Though this phenomenological model ($w$CDM) seems to be very similar to the $\Lambda$CDM model, there are certain differences. It allows the deviation of the value of $w_{DE}$ from $-1$.  The energy density ($\rho_{DE}$) does not remain constant if $w_{DE}\neq-1$. So, this phenomenological model allows the evolution of the dark energy density though $w_{DE}$ remains constant.

\subsubsection{Quintessence scalar field model}

Quintessence scalar field model is the most popular one among the different models of time evolving dark energy. The idea is to introduce a homogeneous time dependent scalar field  $\phi(t)$, minimally coupled to the matter field and the nature of the field is characterized by a potential $V(\phi)$ associated to the scalar field.

\par The relevant action of a scaler field is written as,

\begin{equation}
S=\int d^4x\sqrt{-g}\Big(-\frac{1}{2}g^{\mu\nu}\partial_{\mu}\phi\partial_{\nu}\phi-V(\phi)\Big).
\end{equation}

The energy momentum tensor of the scalar field is defined as, 

\begin{equation}
T_{\mu \nu} = - \frac{2}{\sqrt{- g}} \frac{\delta S}{\delta g^{\mu \nu}} 
            = \partial_{\mu} \phi \partial_{\nu} \phi - g_{\mu \nu} \Big[\frac{1}{2} g^{\alpha \beta} \partial_{\alpha} \phi \partial_{\beta} \phi  + V(\phi)\Big].
\end{equation}  

Thus for FRW space time, the component of the energy momentum tensor of the quintessence scalar field ($\phi(t)$) are obtained as,

\begin{equation}
T^0_0=-\rho_{\phi}=-\Big(\frac{1}{2}\dot{\phi}^2+V(\phi)\Big),
\end{equation} 
and
\begin{equation}
T^1_1=T^2_2=T^3_3=p_{\phi}=\frac{1}{2}\dot{\phi}^2-V(\phi).
\end{equation}

Therefore the expression of the  dark energy equation of state parameter ($w_{DE}$) for a quintessence scalar field is give by,

\begin{equation}
w_{DE}=\frac{\dot{\phi}^2-2V(\phi)}{\dot{\phi}^2+2V(\phi)}.
\end{equation} 

It is also straightforward from the expressions of $\rho_{\phi}$ and $p_{\phi}$, that,

\begin{equation}
\rho_{\phi}+p_{\phi}=\dot{\phi}^2\geq0.
\end{equation} 

This indicates that, for $\rho_{\phi}\geq0$, $w_{DE}\geq-1$. Now the quintessence potential $V(\phi)$ can be expressed as,

\begin{equation}
V(\phi)=\left(\frac{1-w_{DE}}{1+w_{DE}}\right)\frac{\dot{\phi}^2}{2}.
\label{quintpot}
\end{equation} 

Equation (\ref{quintpot}) reveals that dark energy equation of state parameter ($w_{DE}$) for a quintessence model is time evolving.\\

Depending upon the nature of quintessence potential, quintessence models are classified in three different classes.

$\bullet$ If $V(\phi)<<\dot{\phi}^2$, then $w_{DE}\simeq1$ and the energy density varies as $\rho_{\phi}\propto a^{-6}$ which is equivalent to the stiff matter. This does not contribute like a dark energy. 

$\bullet$ If $V(\phi)>>\dot{\phi}^2$, then $w_{DE}\simeq-1$ and the energy density $\rho_{\phi}\approx constant$ which is equivalent to the cosmological constant.

$\bullet$ For the intermediate scenario $-1< w_{DE}<1$, the energy density $\rho_{\phi}\propto a^{-n}$. The quintessence field generates cosmic acceleration for $0\leq n\leq2$ \cite{copelandsami}.\\

 The idea of quintessence scalar field was first introduced in the context of inflation by Ratra and Peeble \cite{ratrapeeble} and by Wetterich \cite{wettequint}. There are ample amount of work in the literature in the context of late time cosmic acceleration with different type of quintessence potential. $V(\phi)\propto\frac{1}{\phi^{\alpha}}$ (where $\alpha$ is  a constant) type of potential was first introduced in  the context of dark energy.  An exponential potential can generate a power law expansion that is $a(t)\propto t^m$ ($m$ is a positive constant). But the exponential potential can not explain the transition from decelerated to accelerated  phase of expansion. A double exponential potential can successfully generate the present acceleration and the decelerated expansion phase that prevailed in the recent past \cite{sensethi}.  \\
The quintessence scalar models with some fine-tuning can address the problem of cosmic coincidence. Observations suggested that at present the energy density of the scalar field ($\rho_{\phi}$) and matter energy density ($\rho_m$) are comparable in order of magnitude. As these two energy densities decay at different rates, it requires a fine tuning the initial conditions to make them comparable at present time .  The idea  is to introduce the quintessence potential in such a way that the energy density ($\rho_{\phi}$) behaves in a very similar way as the dark matter density for a wide range of initial conditions. Actually the idea is that the dark energy density will follow the evolution which is very similar to the evolution of the matter density and at recent era, the dark energy would dominate the dynamics. This is known as the {\it tracking behaviour} of the quintessence dark energy. The idea of tracking quintessence was first introduced by Zlatev, Wang and Steinhardt \cite{trackquint}. Some important investigation regarding the tracking scalar field can be found in reference \cite{trackquint2}. Sahlen, Liddle and Parkinson have reconstructed quintessence potential and checked the viability of tracking behaviour \cite{shalenLidtrac}. On the other hand, the {\it thawing scalar field models} behave in the opposite way. The dark energy equation of state parameter ($w_{DE}$) for a thawing model is close to $-1$ at earlier epoch and it increases with time. This is opposite to the scenario of the tracking quintessence where $w_{DE}$ decreases with time. The behaviour of the dark energy equation of state depends upon the relative shape of the quintessence potential. Comprehensive discussion of thawing models of dark energy can be found in references \cite{scherrersen,thawingquint}. Observational constraints on tracking and thawing models are discussed by Chiba, Felice and Tsujikawa \cite{chibafetsu}. Comperative study of thawing and freezing models of dark energy has been presented by Pantazis, Nesseris and Perivolaropoulos \cite{pantazisnessper}.  Stability  analysis  of tracking quintessence has been discussed by Roy and Banerjee \cite{roybantrckquin}. Carvalho {\it et al.} showed that in a scalar field dominated cosmology, it is possible to have a transient accelerating expansion \cite{carvalholima}.

\subsubsection{K-essence model}

This is a scalar field model where the kinetic part has the dominating contribution in the energy density. This is the basic difference between K-essence model and quintessence model where the quintessence potential take the leading role in the contribution to the dark energy density. The idea of accelerated expansion driven by the kinetic part of the scalar field was first introduced in the context of inflation \cite{armengarr}. In the context of late time acceleration, it was introduced by Chiba {\it et al.} \cite{chibaT} and further generalization was done by Armendariz-Picon {\it et al.} \cite{armenda}

The action for the K-essence models is written as,
\begin{equation}
S= \int d^4 x \sqrt{-g} P(\phi, X),
\end{equation} 
where $X= - \frac{1}{2} (\bigtriangledown \phi)^2$  is the kinetic energy (KE) and the Lagrangian density $P(\phi, X)$ is in the form of pressure density. For a KE dominated scalar field, the Lagrangian density $P(\phi, X)\rightarrow0$ for $X\rightarrow0$. A series expansion of $P(\phi, X)$ around $X=0$ is given as,

\begin{equation}
P(\phi, X) = K(\phi) X + L(\phi) X^2 + ...,
\end{equation}

and the higher order terms of $X$ can be neglected for $X\rightarrow0$. The scalar field now can be redefined as $ \phi_{new} = \int^{\phi_{old}} d\phi \sqrt{\frac{L}{|K|}}$, and the Lagrangian density

$P (\phi, X) = f(\phi) (-X + X^2)$,

where $\phi \equiv \phi_{new}$, $X \equiv X_{new} = (\frac{L}{|K|}) X_{old} $ and $f(\phi) = \frac{K^2 (\phi_{old})}{L (\phi_{old})}$. In a flat homogeneous and isotropic Universe, the pressure ($p_{\phi}$) and energy density ($\rho_{\phi}$) of the scalar field for this model is written as,

\begin{equation}
p_{\phi}=P(\phi, X)=f(\phi)(-X + X^2),
\end{equation}
and
\begin{equation}
\rho_{\phi}=f(\phi)(-X + 3X^2).
\end{equation}

Thus the dark energy equation of state parameter is obtained as,

\begin{equation}
w_{\phi} = \frac{p}{\rho} = \frac{1-X}{1-3X}.
\end{equation}

For accelerated expansion, $w_{DE}<-\frac{1}{3}$, that means $X<\frac{2}{3}$. For $X=\frac{1}{2}$, $w_{DE}=-1$ that means this model can recover the cosmological constant model for a particular condition. 

\par For a viable cosmological model, $f(\phi)$ is to be fine tuned to be of the order of the energy density of the Universe. General discussions on the K-essence model are given  in \cite{kessencegen}.

\subsubsection{Tachyon field model}

The idea of tachyon field model is inspired from the string theory. A tachyon has a negative squared mass and the speed is greater than the speed of light. It is produced at the time of decay of D-brane \cite{dbrane} and has an equation of state parameter that varies between $-1$ to $0$ and thus it can be chosen as a viable candidate of dark energy \cite{tachdark}. The tachyonic dark energy models that  can successfully generate late time acceleration, are discussed in reference \cite{tachmodel}. 

\par The state of tachyon field rests at the maxima of the associated potential and with a small perturbation, it rolls down to the real mass. The relevant action for a tachyon field is given as,

\begin{equation}
 S= - \int d^4 x ~ V(\phi) \sqrt{-det(g_{ab} + \partial_{\alpha} \phi \partial_{\beta} \phi )},
\end{equation}
 where $V(\phi)$  is the tachyonnic potential. In FRW space time, the energy density $\rho_{\phi}$ and the pressure $p_{\phi}$ of the tachyon are written as, 

\begin{equation}
\rho_{\phi} = \frac{V(\phi)}{\sqrt{1 - \dot{\phi}^2}},
\end{equation}
and 
\begin{equation}
p_{\phi} = - V(\phi) \sqrt{1 - \dot{\phi} ^2}.
\end{equation}

Thus the dark energy equation of state parameter, $w_{DE}=\frac{p_{\phi}}{\rho_{\phi}}=\dot{\phi}^2-1$, and for accelerated expansion, $\dot{\phi}^2<\frac{2}{3}$. It can also be shown that  the saclar field energy density $\rho_{\phi}\propto a^{-m}$,  where $0<m<3$.

\subsubsection{Phantom field model}
In the context of dark energy, the phantom field was first discussed by Caldwell \cite{caldwellphantom}. The basic difference of phantom field from the quintessence field is that the phantom field has negative kinetic energy. The relevant action of the phantom field is written as,

\begin{equation}
S = \int d^4 x ~ L(\phi, X) ,
\end{equation}

where $L(\phi,X)=-X-V(\phi)$. The energy density ($\rho$) and the pressure ($p$) of the phantom field yield to be,

\begin{equation}
\rho_{\phi}=-\frac{\dot{\phi}^2}{2}+V(\phi),
\end{equation}

and

\begin{equation}
p_{\phi}=-\frac{\dot{\phi}^2}{2}-V(\phi).
\end{equation}

The equation of state parameter of dark energy for phantom field,

\begin{equation}
w_{DE}=\frac{\dot{\phi}^2+2V(\phi)}{\dot{\phi}^2-2V(\phi)},
\end{equation}

and for $V(\phi)>>\dot{\phi}^2$, $w_{DE}<-1$.  It leads to a very rapid expansion of the Universe up to infinite extent within a finite time. This scenario is called the Big Rip where both the volume and the expansion rate blow up to infinity. A phantom field potential with a maxima can avoid the Big Rip. For instance, $V(\phi)=V_0\Big[cosh(\frac{\alpha\phi}{m_{pl}})\Big]^{-1}$, where $\alpha$ is a constant, type of potential can avoid the Big Rip. The field rests at the maxima after a damped oscillation and thus the equation of state parameter $w_{DE}=-1$ as $\dot{\phi}=0$. Thus it can restore the scenario of cosmological constant.

\subsubsection{Chaplygin Gas model}

Chaplygin gas was introduced in the context of cosmic acceleration by Kamenshchik, Moschella and Pasquir \cite{kamenshchikchap}. The idea of Chaplygin gas is based on a special type of equation of state, $p=A/\rho$, where $A$  is a positive constant. A generalization of Chaplygin gas model has been discussed by Bento, Bertolami and Sen \cite{gcgbento}, where the equation of state is presented in a generalised form as, $p=-A/\rho^{\alpha}$,  (where $0<\alpha\leq1$). From the continuity equation, the energy density can be written as,

\begin{equation}
\rho=\Big[A+\frac{B}{a^{3(1+\alpha)}}\Big]^{\frac{1}{1+\alpha}},
\end{equation}

where $B$ is the integration constant. For $\alpha=1$, the asymptotic behaviour shows interesting results. When the scale factor $a$ is small, $\rho\sim\frac{\sqrt{B}}{a^3}$, that means it behaves like a pressureless dust matter. When $a >> (\frac{B}{A})^{\frac{1}{6}}$, $\rho \sim -p \sim \sqrt{A}$, so it resembles the cosmological constant. Thus the generalised Chaplygin gas model is an attempt towards the unification of dark energy and dark matter. Though the Chaplygin gas models have been ruled out by the CMB temperature anisotropy data \cite{amechap,bentochap2}, the generalised Chaplygin gas models are allowed within a narrow domain of the parameter, $0\leq\alpha\leq0.2$ \cite{amechap}.

\subsection{Modified gravity models}  

The other way to look for the plausible explanation of the phenomenon of cosmic acceleration is the modification of GR. There are various way of modification of the theory of gravity. Different $f(R)$ gravity models \cite{frmodels}, scalar tensor theory \cite{scalten}, higher dimensional gravity theories \cite{highgrav} etc. belong to the class of modified gravity theory.  Though modified gravity models are nice theoretical attempts and adequately account for the cosmic acceleration, these models are normally not very suitable to explain the local astronomical observations. Some of the modified gravity theories are discussed in the following.

\subsubsection{$f(R)$ gravity}

The simplest modification of GR is the $f(R)$ gravity where the modification is done in the space time action by replacing the Ricci scalar $R$ with an analytic function $f=f(R)$. Thus the $f(R)$ gravity action is given as,

\begin{equation}
S = \int d^4 x \sqrt{-g}\Big[f(R) + \mathcal{L}_m\Big],
\end{equation}

where $\mathcal{L}_m$ is the matter Lagrangian. The field equations are also modified as,

\begin{equation}
\Big(\frac{\partial f}{\partial R}\Big)^{-1} \Bigg[\frac{1}{2}g_{\mu \nu}\Big(f-\frac{\partial f}{\partial R}R\Big)+\triangledown_{\mu}   \triangledown_{\nu}\frac{\partial f}{\partial R}-g_{\mu \nu}\square\frac{\partial f}{\partial R}\Bigg]=-8\pi GT_{\mu \nu},
\end{equation}

where $T_{\mu\nu}$ is the energy momentum tensor of the matter distribution. For $f(R)=R$, the theory reduces to GR.

\par Depending on the form of $f(R)$, these models can generate the scenario of early inflation or late time acceleration. For instance, $f(R)\sim R^2$ type of models generate cosmic inflation and $f(R)\sim1/R^n$, with $n>0$ type of models are viable in case of late time acceleration. The $f(R)$ gravity in the context of inflation has been discussed by Satrobinsky \cite{starinffR}, Kerner \cite{kernerinffR} and by Duruisseau and Kerner \cite{durkerinffR}. As the curvature $R$  decreases with time, inverse power of $R$ in the expression of $f(R)$ can have significant contribution in generating the late time acceleration. In the context of late time cosmic acceleration, $f(R)$ gravity has been invoked by Capozziello {\it et al} \cite{capofR}, Nojiri and Odintsov \cite{nojodfR}, Carroll {\it et al.} \cite{carrollfR}, Das, Banerjee and Dadhich \cite{dasbandfR}; $f(R)\sim\frac{1}{R}$ type model has been emphasized by Carroll {\it et al.} \cite{carrollfR} and by Vollick \cite{vollickfR}. An exponential $f(R)$ model has been investigated by Das, Banerjee and Dadhich \cite{dasbandfR}. Some more discussions on cosmological dynamics of $f(R)$ gravity models are in reference \cite{frdis}. Amendola {\it et al.} \cite{frviableaman} have discussed the viability conditions of different $f(R)$ gravity models; $f(R)\sim\frac{1}{R}$ type  of models suffer from the problem of singularity at $R\rightarrow 0$. An exponential $f(R)$ model can resolve this singularity problem, but it conflicts the viability conditions discussed in \cite{frviableaman}. Reconstruction of $f(R)$ model from observational data has been discussed by Felice, Mukherjee and Wang \cite{frrecon}. The recent results of observational test of $f(R)$ gravity have been discussed by Capozziello and Salzano \cite{capozz}.

\par

\subsubsection{Scalar tensor theories}

The basic idea of a scalar tensor theory  is the nonminimal coupling between the scalar field and the geometry. The simplest approach in this direction is the Brans-Dicke theory \cite{bransdikke}, where the scalar field is coupled to the Ricci scalar ($R$) and thus the Lagrangian is given as,

\begin{equation}
\mathcal{L}=\frac{\phi R}{2}-\frac{\omega_{BD}}{2\phi}(\triangledown \phi)^2,
\end{equation}

where $\omega_{BD}$ is the Brans-Dicke parameter. This is a varying $G$ theory. In the weak field limit, the Brans-Dicke theory resembles the GR. But it is different from GR in non linear regime \cite{nbssbrans}. The generalised Brans-Dicke theory behaves as a coupled quintessence in conformal frame \cite{bransconf}. It has been shown by Banerjee and Pavon \cite{banpav} that the cosmic acceleration be generated in Brans-Dicke theory without introducing any exotic component in the matter sector. But it can hardly explain the smooth transition from decelerated to accelerated phase of expansion. It also requires a low $\omega_{BD}$ in cosmological scenario, but local astronomy demands high $\omega_{BD}$.

\section{Cosmological observations and the observational data sets}
\label{obsdata}

Cosmology today is a subject based on observations. The basic endeavour is to model different observational phenomena. The observations are mainly of the electromagnetic wave of different wavelength. Earlier, the observations were limited within the optical wavelength. But at present, the astronomical observations are going in the wavelength regime starting from microwave to gamma ray. 

\par In the following, some of the observations which are important in the context of cosmological modelling, mainly in the context of late time cosmic acceleration, have been discussed. These are the observations of type Ia supernova, observational measurement of Hubble parameter, the baryon acoustic oscillation and the cosmic microwave background (CMB) radiation.

\subsection{Observation of type Ia Supernova}

In the context of late time cosmic acceleration, the pioneering observation is the type Ia supernova. As already mentioned, the phenomenon of cosmic acceleration was first discovered during the observation of type Ia supernova by two different groups \cite{riess1,perlmutter}. The type Ia supernova is a {\it standard candle} for astronomical observation. A standard candle  is defined as a class of distinguishable objects of known intrinsic brightness and it can be distinguished in a wide range of distance. As the Universe expands, the distance between the object and the observer increases and the wavelength of the radiated photons also increases. Thus the radiated photon get redshifted.  The observed brightness of the objects and the redshift of the observed photons provide  a measurement of the expansion of the Universe. 

\par Distance measurements of supernova are presented at different redshift ($z$) in the form of distance modulus ($\mu_B(z)$) which is actually the difference between the {\it apparent magnitude} ($m_B$) and the {\it absolute magnitude} ($M_B$) of the B-band (wavelength band of blue line) of the observed spectrum of the supernova.  The apparent magnitude is related to the observed brightness or the observed photon flux ($F_B$). If two objects in the sky have apparent magnitude $m_1$ and $m_2$ with observed photon flux $F_1$ and $F_2$ respectively, then,

\begin{equation}
m_1-m_2=-2.5\log{\Big(\frac{F_1}{F_2}\Big)},
\end{equation}
\cite{araichou}. Now if the two objects have same luminosity $L$, then one can write,
\begin{equation}
m_1-m_2=5\log{\Big(\frac{d_1}{d_2}\Big)}
\end{equation}
where $d_1$ and $d_2$ are the luminosity distance of the respective objects (luminosity distance is defined in equation (\ref{lumdis})). It is imperative to note that the value of apparent magnitude ($m$) decreases if the observed brightness increases. The absolute magnitude ($M$) is defined as the apparent magnitude of the object if it is at 10 parsecs (pc) away from the observer. In case of a particular standard candle, the absolute magnitude is same for all observations. The distance of the object is related to the difference of apparent  and absolute magnitude as,

\begin{equation}
Distance/pc=10^{(m-M+5)/5},
\end{equation}

where the distance is scaled by the unit parsec (pc). Finally the distance modulus for the B-band of the observed supernova ($\mu_B(z)$) is written as, 

\begin{equation}
\mu_B(z)=m_B-M_B=5\log{\Big(\frac{d_L(z)}{Mpc}\Big)}+25,
\end{equation}

where $d_L(z)$ is the luminosity distance of the object in the unit of Mpc. This $d_L(z)$ can be expressed in terms of the present values of the Hubble parameter ($H_0$) and the deceleration parameter ($q_0$) as,

\begin{equation}
d_L(z)=\frac{z}{H_0}\left[1+(1-q_0)\frac{z}{2}+...\right],
\end{equation}
\cite{akrthcos}. The higher order terms can be neglected for small values of $z$. As the observed luminosity distance for nearby supernova is higher than the expected value, it can be concluded that the value of $q_0$ is negative. That means the present Universe is going through a phase of accelerated expansion. 
\par In the context of statistical analysis of different cosmological models discussed in the following chapters, the distance modulus data of 580  supernovae of Union 2.1 compilation \cite{snsuzuki} or the recent data of 740 supernovae sample of joint lightcurve analysis (jla) of the Sloan Digital Sky Survey II (SDSS II) and Supernova Legacy Survey (SNLS)  \cite{snbetoule} have been utilized.

\subsection{Observational measurement of Hubble parameter}

The Hubble parameter, which is the fractional rate of the expansion, is defined as, $H=\dot{a}/a$, where $a$ is the scale factor. The Hubble parameter can be estimated at different redshift ($z$), from observation. In the statistical analysis, discussed in the following chapters, the measurement of Hubble parameter by different groups have been used.

\par The estimation of the value of $H(z)$ can be obtained from the measurement of differential of redshift $z$ with respect to cosmic time $t$ as,

\begin{equation}
H(z)=-\frac{1}{(1+z)}\frac{dz}{dt}.
\end{equation}

The differential age of galaxies have been used as an estimator of $dz/dt$ by Simon {\it et al.} \cite{simonohd}. Measurement of cosmic expansion
history using red-enveloped galaxies was done by Stern {\it et al.} \cite{sternohd} and by Chuang and Wang \cite{chuangohd}. Measurement of expansion history from WiggleZ Dark Energy Survey has been discussed by Blake {\it et al.} \cite{blakeohd}. Measurement of Hubble parameter at low redshift using the differential age method with the Sloan Digital Sky Survey data have been presented by Zhang {\it et al.} \cite{zhangohd}. Compilation of
observational Hubble parameter measurement has been presented by Moresco {\it et al.} \cite{morescoohd}. Finally, the measurement of Hubble parameter at $z= 2.3$ by Busca {\it et al.} \cite{buscaohd} or the improved estimation  at $z= 2.34$ by Delubac {\it et al.} \cite{delubacohd} has also been incorporated in the data set, used in the present analysis. Table \ref{hubbtable} presents the $H(z)$ measurements which have been used in the present analysis. The measurement of $H_0=67.80\pm0.77 km~ s^{-1}~Mpc^{-1}$ from Planck combining   Planck+WP+highL+BAO data (Planck temperature data (Planck), WMAP polarization data (WP), other high  resolution CMB observation data (highL) and baryon acoustic oscillation data (BAO)) \cite{planck2013} has also been used in the analysis.

%%%%%%%%%%%%%%%%%%%%%%%%%
\begin{table}
\caption{{\small $H(z)$ data table (in unit [$km$ $s^{-1}$$Mpc^{-1}$])
}}
\begin{center}
\resizebox{0.47 \textwidth}{!}{  
\begin{tabular}{ c   c   c  c  c } 
 \hline
 \hline
 $z$  & H & $\sigma_H$ & References \\ 

 \hline
0.07      & 69      & 19.6  & Zhang {\it et al.}\cite{zhangohd}\\ 
0.1       & 69      & 12    & Simon {\it et al.}\cite{simonohd}\\ 
0.12      & 68.6    & 26.2  & Zhang {\it et al.}\cite{zhangohd}\\ 
0.17      & 83      & 8     & Simon {\it et al.}\cite{simonohd}\\ 
0.179     & 75      & 4     & Moresco {\it et al.}\cite{morescoohd}\\ 
0.199     & 75      & 5     & Moresco {\it et al.}\cite{morescoohd}\\  
0.2       & 72.9    & 29.6  & Zhang {\it et al.}\cite{zhangohd}\\  
0.27      & 77      & 14    & Simon {\it et al.}\cite{simonohd}\\ 
0.28      & 88.8    & 36.6  & Zhang {\it et al.}\cite{zhangohd}\\  
0.35      & 76.3    & 5.6   & Chuang and Wang\cite{chuangohd}\\ 
0.352     & 83      & 14    & Moresco {\it et al.}\cite{morescoohd}\\ 
0.4       & 95      & 17    & Simon {\it et al.}\cite{simonohd}\\ 
0.44      & 82.6    & 7.8   & Blake {\it et al.}\cite{blakeohd}\\ 
0.48      & 97      & 62    & Stern {\it et al.}\cite{sternohd}\\ 
0.593     & 104     & 13    & Moresco {\it et al.}\cite{morescoohd}\\ 
0.6       & 87.9    & 6.1   & Blake {\it et al.}\cite{blakeohd}\\ 
0.68      & 92      & 8     & Moresco {\it et al.}\cite{morescoohd}\\ 
0.73      & 97.3    & 7     & Blake {\it et al.}\cite{blakeohd}\\ 
0.781     & 105     & 12    & Moresco {\it et al.}\cite{morescoohd}\\ 
0.875     & 125     & 17    & Moresco {\it et al.}\cite{morescoohd}\\  
0.88      & 90      & 40    & Stern {\it et al.}\cite{sternohd}\\ 
0.9       & 117     & 23    & Simon {\it et al.}\cite{simonohd}\\ 
1.037     & 154     & 20    & Moresco {\it et al.}\cite{morescoohd}\\ 
1.3       & 168     & 17    & Simon {\it et al.}\cite{simonohd}\\ 
1.43      & 177     & 18    & Simon {\it et al.}\cite{simonohd}\\  
1.53      & 140     & 14    & Simon {\it et al.}\cite{simonohd}\\ 
1.75      & 202     & 40    & Simon {\it et al.}\cite{simonohd}\\ 
2.34      & 222     & 7     & Delubac {\it et al.}\cite{delubacohd}\\

 \hline
 \hline
\end{tabular}
}
\end{center}
\label{hubbtable}
\end{table}
%%%%%%%%%%%%%%%%%%%%%%%%%%%%%%%%%%%

\subsection{Baryon acoustic oscillation}

Baryon acoustic oscillation (BAO) is  the oscillation of the tightly coupled photon-baryon plasma before the recombination. The idea of statistical standard ruler with baryon acoustic oscillation is based on the fact that  the clustering of galaxies may have  a preferred scale. This preferred scale of galaxy clustering can be used to constrain the angular diameter distance. The phenomenon behind this distance correlation of galaxy clustering is actually the baryon acoustic oscillation. 

\par Before {\it recombination}, the formation of neutral hydrogen atom and the decoupling of photon from baryon matter, the mixture of baryons and photons was in the form of a hot plasma due to tight coupling between photons and baryons via Thomson scattering. The radiation pressure and the gravitational attraction acted as two competing forces and thus set up oscillations in the plasma. A single spherical over-density  in the tightly coupled photon-baryon plasma would propagate with a speed $c_s=c/\sqrt{3(1+3\rho_b/4\rho_\gamma)}$, where $\rho_b$ is the baryon density and $\rho_\gamma$ is the photon density \cite{eisensbao}.

\par At recombination, the photon decoupled from the baryon and propagated freely, forming the cosmic microwave background (CMB). The baryon became neutral at recombination and the spherical shells formed due to the baryon oscillation remain imprinted on the distribution of the baryonic matter in the Universe. The {\it comoving sound horizon} ($r_s$) at a given redshift ($z$) is the distance travelled by the acoustic wave in the time interval starting from the beginning of matter formation to that given redshift $z$. The acoustic scale is the distance scale at which the galaxy clusters are correlated.  

\par In  the statistical analysis, discussed in the following chapters, the BAO data along with the measurement of acoustic scale ($l_A$) and the sound horizon ($r_s$) at {\it photon-electron decoupling ($z_*$)} and at {\it photon drag epoch ($z_d$)} have been used. The comoving sound horizon is defined as,

\begin{equation}
r_s(z)=\frac{c}{\sqrt{3}}\int_0^{1/(1+z)}\frac{da}{a^2H^2(a)\sqrt{1+a(3\Omega_{b0}/4\Omega_{\gamma 0})}},
\end{equation}

where $\Omega_{b0}$ is the present value of the baryon density parameter and $\Omega_{\gamma 0}$ is the present value of the photon density parameter. The acoustic scale at photon decoupling is defined as,

\begin{equation}
l_A(z_*)=\pi\frac{d_A(z_*)}{r_s(z_*)},
\end{equation}

where $d_A(z_*)=c\int_0^{z_*}\frac{dz'}{H(z')}$ is the comoving angular diameter distance at decoupling. Another important definition is of dilation  scale ,

\begin{equation}
D_V(z)=\Big[czd_A^2(z)/H(z)\Big]^{1/3}.
\label{dVscale}
\end{equation}
This is actually a geometric mean of two transverse and one radial directions measurements of the distance for BAO. According to the Planck results, the redshift of photon decoupling is $z_*\approx1091$ and the redshift of photon drag epoch is $z_d\approx1059$ \cite{planck2013}. The Planck measurement of the acoustic scale at photon decoupling $l_A(z_*)=301.74\pm0.19$ and ratio of comoving sound horizon at drag epoch ($z_d$) and at decoupling ($z_*$) is $r_s(z_d)/r_s(z_*)=1.019\pm0.009$ \cite{planck2013,wangwangcmb}.

\par The BAO data are normally given in the from $(r_s(z_d)/D_V(z_{BAO}))$ and it can be scaled to $(d_A(z_*)/D_V(z_{BAO}))$ for convenience in the context of studying the dark energy models. Detail discussion regarding the statistical analysis of cosmological models using BAO data is presented by Giostri {\it et al.} \cite{giostribao}. Different measurement of BAO data have been utilized in the present context. This are from the 6dF Galaxy Survey at $z=0.106$ \cite{beutlerbao}, measurements of WiggleZ team at $z=0.44$, $z=0.60$ and $z=0.73$ \cite{blakebaowiggleZ}, measurements of BAO from Sloan Digital Sky Survey (SDSS) at $z=0.2$ and $z=0.35$ \cite{percilavSDSSbao} and the Baryon Oscillation Spectroscopic Survey (BOSS) at $z=0.32$ (BOSS LOWZ) and at $z=0.57$ (BOSS CMASS) \cite{bossandersonbao}. It is worth mentioning that all the data points referred to here have not been incorporated in a single analysis so as to avoid the effects of complicated correlations between the measurements. Different combinations of the BAO data points have been utilized in different analysis discussed in the following chapters.

\subsection{Cosmic Microwave Background Radiation}

The discovery of the cosmic microwave background (CMB) radiation has opened a new window in observational cosmology \cite{penziascmb}. Cosmic microwave background (CMB) radiation is the relic photon that decoupled from the baryon at recombination. The temperature of the Universe at recombination was $T\approx3000$K. The radiated photon temperature decreases due to redshift and the observed CMB temperature at present epoch is $T=2.728\pm0.004$K \cite{fixsencmb}. The redshift of recombination, when the photons decoupled from the baryons, is $z_*=1091$ \cite{planck2013}.

\par Though the distribution of temperature of the CMB photon is highly isotropic, fluctuation of the order of $10^{-5}$ has been observed by Cosmic Background Explorer (COBE) \cite{cobe}, Differential Microwave Radiometer (DMR) \cite{cobedmr} and later by Wilkinson Microwave Anisotropy Probe (WMAP) \cite{komatsuWMAP} and Planck satellite \cite{planck2013,planck2015}. The temperature fluctuation is presented as $\Theta(\hat{n})=\Delta T/T$, where $\hat{n}$ is the direction vector, $T$ is the average CMB temperature and $\Delta T=(T(\hat{n})-T)$. For a Gaussian fluctuation, the multipole moments of the temperature fluctuation field is written as,

\begin{equation}
\Theta_{lm}=\int d\hat{n}Y^*_{lm}\Theta(\hat{n}).
\end{equation}

where $Y^*_{lm}$ is the complex conjugate of the spherical harmonics. The power spectrum is written as,

\begin{equation}
<\Theta^*_{lm}\Theta_{l'm'}>=\delta_{ll'}\delta_{mm'}C_l.
\end{equation}

The multipole moment $l$ is related to the angular separation $\theta$ as $\theta=2\pi/l$; thus large multipole moment corresponds to small angular separation. Finally the power spectrum is expressed as

\begin{equation}
\Delta_T^2=\frac{l(l+1)}{2\pi}C_lT^2.
\end{equation}

A comprehensive review on the CMB temperature and polarization spectrum is presented by Hu and Dodelson \cite{hudodcmb}.

\par In the statistical analysis of different cosmological models, discussed in the following chapters, the CMB shift parameter data has been used. The CMB shift parameter is related to the position of the first acoustic peak in the power spectrum of CMB temperature anisotropy. The CMB shift parameter is defined as,

\begin{equation}
R_{CMB}=\sqrt{\Omega_{m0}}H_0\int_0^{z_*}\frac{dz}{H(z)},
\end{equation}

where $\Omega_{m0}$ is  the present value of the matter density parameter, $H_0$ is the present value of the Hubble parameter and $z_*$ is the redshift of photon decoupling. The value of the CMB shift parameter is estimated from the CMB data with some fiducial assumption about the cosmological model. In the present analysis, the estimation of the CMB shift parameter by Wang and Wang \cite{wangwangcmb} from the combined analysis with Planck+lensing+WP data has been utilized, where the estimated value of the shift parameter is $R_{CMB}=1.7407\pm0.0094$ at 1$\sigma$ confidence level.

\vskip 2.0 cm

\section{Statistical analysis of cosmological models}
\label{statanalysis}

Now that a large amount of observational data are available, the relevance of a cosmological model can be tested against the available data. These tests crucially depend on the  statistical analysis of the observed data and estimation of the values of the relevant cosmological parameters. In the context of late time acceleration, the relevant physical parameters are the dark energy equation of state parameter ($w_{DE}$), the present value of the matter density parameter ($\Omega_{m0}$), the dark energy density parameter ($\Omega_{DE0}$) etc. There are also other parameters which are introduced through the corresponding models and they are related to some physical quantity. The estimation of the parameters and the statistical comparison of different models have been discussed in brief in the following section.

\subsection{The $\chi^2$ and the likelihood}

To estimate the parameter values of cosmological models from the observational data, a $\chi^2$-statistics has been adopted. The $\chi^2$ is defined as,

\begin{equation}
\chi^2=\sum_{i}\frac{[\epsilon _{obs}(z_i)-\epsilon_{th}(z_i,\{\theta\})]^2}{\sigma_i^2},
\end{equation}

where $\epsilon_{obs}$ is the observationally estimated value of the observable at redshift $z_i$, $\epsilon_{th}(z_i,\{\theta\})$ is the form of the observable as given by the model as a function of the set of model parameters $\{\theta\}$ and $\sigma_i$ is the uncertainty associated to the measurement at $z_i$. 

\par The $\chi^2$ for the observational Hubble parameter data (OHD) is written as,

\begin{equation}
\chi^2_{{\tiny OHD}}=\sum_{i}\frac{[H_{obs}(z_i)-H_{th}(z_i,\{\theta\})]^2}{\sigma_i^2},
\label{chiOHD}
\end{equation}
where $H_{obs}$ is the observed value of the Hubble parameter, $H_{th}$ is theoretical one and $\sigma_i$ is the uncertainty  associated to the $i^{th}$ measurement. The $\chi^2$ is a function of the set of model parameters $\{\theta\}$.

\par The $\chi^2$ of the supernova distance modulus data is defined in a slightly complicated way so as to marginalize the nuisance parameter $H_0$ and to incorporate the systematics of the distance modulus measurements. The method discussed by Farooq, Mania and Ratra \cite{farooqmaniarartasne} has been adopted. The $\chi_{SNe}^2$ has been defined as
\begin{equation}
\chi_{SNe}^2=A(\{\theta\})-\frac{B^2(\{\theta\})}{C}-\frac{2\ln{10}}{5C}B(\{\theta\})-Q,
\end{equation}  
where 
\begin{equation}
A(\{\theta\})=\sum_{\alpha,\beta}(\mu_{th}-\mu_{obs})_{\alpha}(Cov)^{-1}_{\alpha\beta}(\mu_{th}-\mu_{obs})_{\beta},
\end{equation}
\begin{equation}
B(\{\theta\})=\sum_{\alpha}(\mu_{th}-\mu_{obs})_{\alpha}\sum_{\beta}(Cov)^{-1}_{\alpha\beta},
\end{equation}
\begin{equation}
C=\sum_{\alpha,\beta}(Cov)^{-1}_{\alpha\beta},
\end{equation}
and the $Cov$ is the covariance matrix of the data. Here the $Q$ is a constant which does not depend upon the parameters and hence has been ignored.

\par The relevant $\chi^2$ for baryon acoustic oscillation (BAO) data, namely $\chi^2_{BAO}$, is defined as:
\begin{equation}
\chi^2_{BAO}={\bf X^{t}C^{-1}X},
\label{chibao}
\end{equation}

where ${\bf X}=\left(\left(\frac{d_A(z_*)}{D_V(z_{BAO})}\right)_{th}-\left(\frac{d_A(z_*)}{D_V(z_{BAO})}\right)_{obs}\right)$ in the form of a column matrix and ${\bf C^{-1}}$ is the inverse of the covariance matrix.

\par The $\chi^2_{\tiny CMBShift}$ is defined as 

\begin{equation}
\chi^2_{\tiny CMBShift}=\frac{(R_{obs}-R_{th}(z_*,\{\theta\}))^2}{\sigma^2},
\end{equation}

where $R_{obs}$ is the value of CMB shift parameter estimated from observational data, $R_{th}(z_*,\{\theta\})$ is from the theoretical model  and $\sigma$ is the corresponding uncertainty.

\par For statistical analysis with combinations of different data sets, the $\chi^2$ associated to different data sets are added up to define the combined $\chi^2$ as,

\begin{equation}
\chi^2_{\tiny combined}=\sum_d\chi^2_{\tiny d},
\end{equation}
 where $d$ denotes the data sets taken into account for that particular combination.

\par The likelihood function is defined as,

\begin{equation}
L(\{\theta\})=\exp{(-\chi^2/2)}.
\end{equation}

The likelihood is also a function of the model parameters. In Bayes' theorem, the posterior probability distribution of parameter ($\theta$) is expressed as,

\begin{equation}
p(\theta|D,I)=\frac{p(\theta|I)p(D|\theta,I)}{p(D|I)},
\end{equation}  

where $I$ is the proposition representing the prior information, $D$ represents the data; $p(D|\theta,I)$ is the probability of obtaining the data if parameter $\theta$ is given according to $I$; $p(D|\theta,I)$ is  the likelihood; $p(\theta|I)$ is the prior probability; $p(D|I)$ is called the global likelihood which is actually the normalization factor $p(D|I)=\int p(\theta|I)p(D|\theta,I)d\theta$ ensuring $\int p(\theta|D,I)d\theta=1$. Thus the $\chi^2$ and the likelihood are connected to the Bayesian approach and the prime endeavour of statistical analysis of a cosmological model is to figure out the posterior probability distribution of model parameters. Comprehensive discussions on statistical analysis with the Bayesian approach is presented by  Gergory \cite{gregorybayesian} and by Hobson {\it et al.} \cite{hobsonbayesian}.

\subsection{Estimation of the parameter values and propagation of error}

\subsubsection{Maximum likelihood analysis}

\begin{figure}[t]
\begin{center}
\includegraphics[angle=0, width=0.95\textwidth]{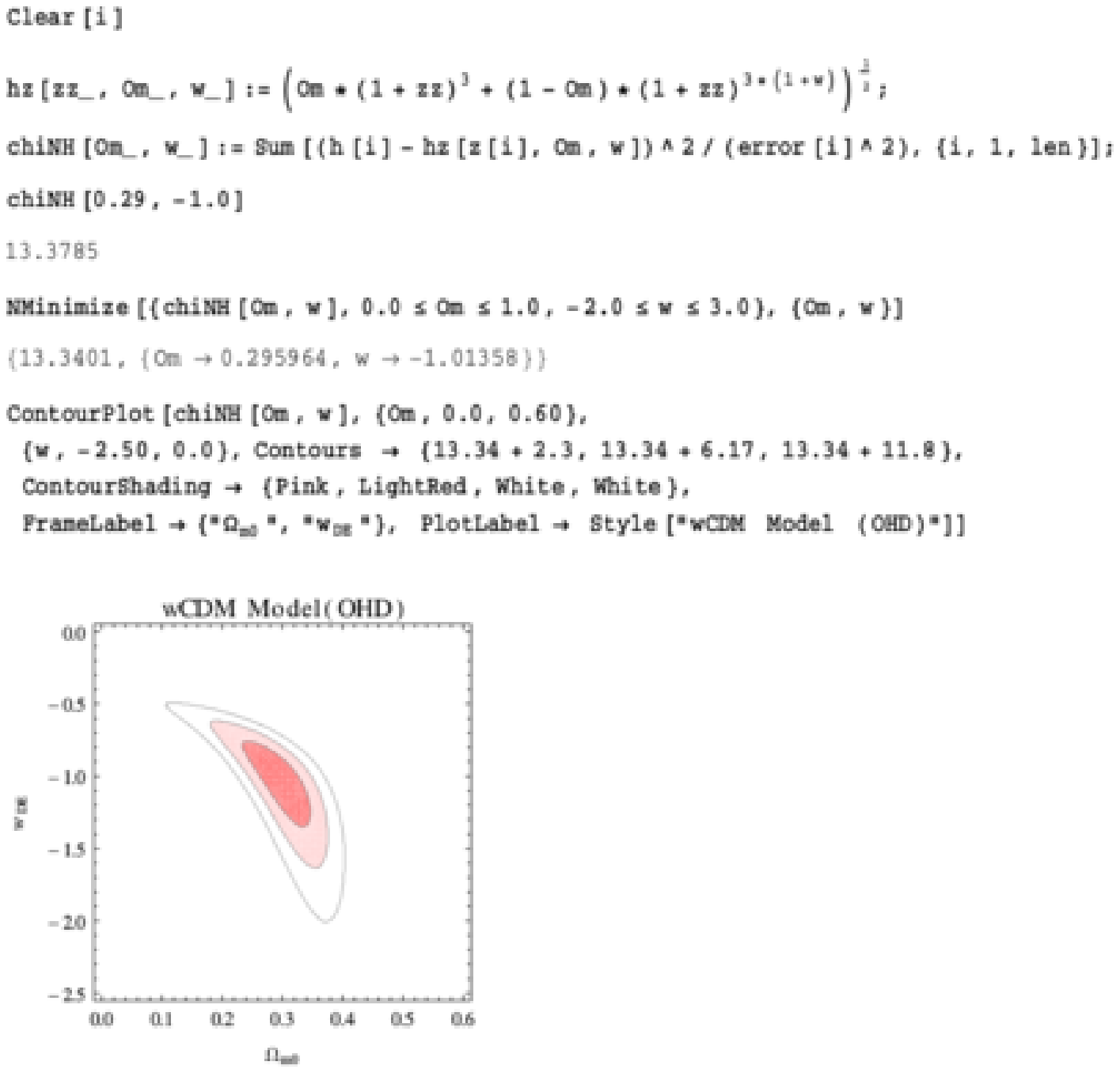}
\caption{{\small A part of the code that shows how the contours on the parameter space are obtained for $w$CDM model using OHD data.}}
\label{OHDCodech1}
\end{center}
\end{figure}

\begin{figure}[t]
\begin{center}
\includegraphics[angle=0, width=0.95\textwidth]{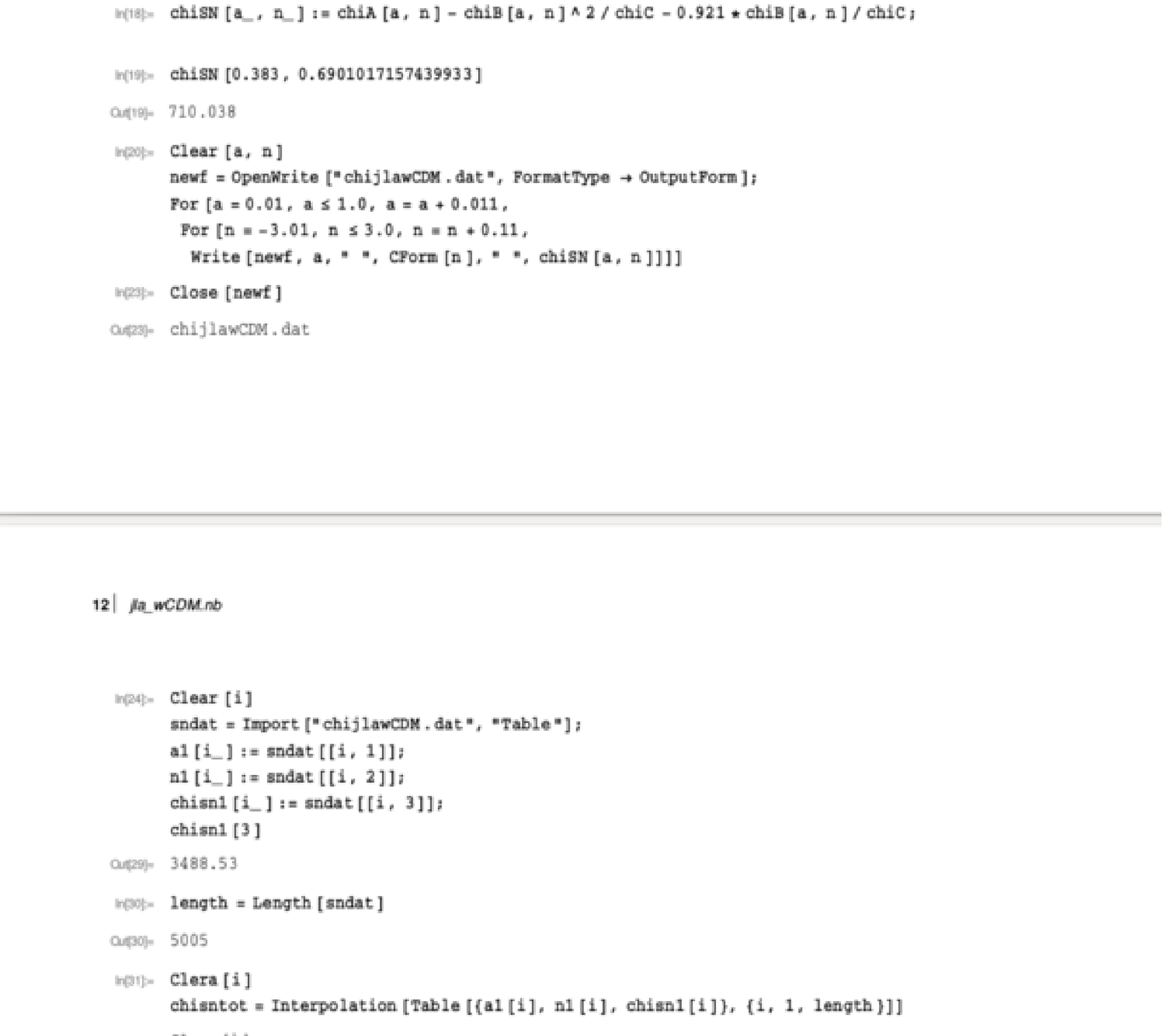}
\caption{{\small A part of code for $w$CDM model using SNe data. It shows how the relevant $\chi^2$ is defined for the SNe data.}}
\label{SNeCodech1}
\end{center}
\end{figure}

To obtain the best fit values of the model parameters, the maximum likelihood analysis has been adopted. It is clear from the definition of likelihood that the minimum value of the corresponding $\chi^2$ would maximise the value of the likelihood function. Thus the $\chi^2$ minimization is equivalent to the maximum likelihood analysis. Analysis for different cosmological models, discussed in the following chapters, have been done numerically using the basic grid searching of likelihood where the range of the parameters are divided into grids and all possible combinations are evaluated to obtain the maximum likelihood and the corresponding best fit values of the parameters $\{\hat{\theta}\}$. Numerical analysis have been done in Mathematica. The code to obtain the contour plots for $w$CDM model with  OHD data has been shown in figure \ref{OHDCodech1} and some part of the code for SNe data are sohwn in the figure \ref{SNeCodech1}. The best fit values of the parameters are obtained by minimizing the corresponding $\chi^2$. The confidence contours are obtained by adding the $\Delta\chi^2$ with the $\chi^2_{min}$ where the value of $\chi^2$ at the boundary of the the contour on the parameter space is $\chi^2_{min}+\Delta\chi^2$. The value of $\Delta\chi^2$ for different confidence level on 2 dimensional parameter space are,  $\Delta\chi^2=2.3$ (at 1$\sigma$), $\Delta\chi^2=6.17$ (at 2$\sigma$) and $\Delta\chi^2=11.8$ (at 3$\sigma$).

%%%%%%%%%%%%%%%%%%%%%%%%%%%%%%%%%%%%%%%%%%%%%%%%%%%
%%%%%%%%%%%%%%%%%%%%%%%%%%%%%%%%%%%%%%%%%%%%%%

%%%%%%%%%%%%%%%%%%%%%%%%%%%%%%%%%%%%%%%%%%%%%%%%%%%
%%%%%%%%%%%%%%%%%%%%%%%%%%%%%%%%%%%%%%%%%%%%%%

\par To estimate the error associated to the best fit values of the model parameters obtained from the maximum likelihood analysis, it is required to calculate the parameter covariance matrix from the $\chi^2$. The parameter covariance matrix is defined as,

\begin{equation}
Cov=\left(\frac{\partial^2\chi^2}{\partial\theta_i\partial\theta_j}\right)^{-1}\Bigg|_{\{\hat{\theta}\}}.
\end{equation}

The diagonal terms of the matrix give the variance ($\sigma^2_{\theta_i}$) of the corresponding parameter and the off-diagonal terms are the covariance ($cov(\theta_i,\theta_j)$) of the two parameters associated to that term. In this way the parameter values are estimated from the observational data. 

\subsubsection{Propagation of errors}

Sometimes it is required to figure out the evolution of different quantities which are functions of the model parameters. As  there are uncertainties associated with the parameter values, the quantities which are the functions of of the model parameters, would also have errors along with the best fit curves. The Taylor series expansion of $y(\{\theta\})$ around the best fit value is written as,

\begin{equation}
y(\{\theta\})=y(\{\theta\})\Big|_{\{\hat{\theta}\}}+\sum_i\left(\frac{\partial y}{\partial\theta_i}\right)\Bigg|_{\{\hat{\theta}\}}(\theta_i-\hat{\theta}_i)+higher~order~terms,
\end{equation}
where $y=y(\{\theta\})$ is a function of the model parameters. If the likelihood of the parameters are Gaussian function, then the higher order terms can be ignored as for a Gaussian distribution, the probability falls rapidly for higher deviation from the best fit point. Thus only very small values of $(\theta_i-\hat{\theta}_i)$ are probabilistically significant. The variance of $y(\{\theta\})$ thus looks like,

\begin{equation}
\sigma_y^2=\sum_i\sigma_{\theta_i}^2\left(\frac{\partial y}{\partial\theta_i}\right)^2\Bigg|_{\{\hat{\theta}\}}+\sum_{ij,i\neq j}cov(\theta_i,\theta_j)\left(\frac{\partial y}{\partial\theta_i}\frac{\partial y}{\partial\theta_j}\right)\Bigg|_{\{\hat{\theta}\}}.
\end{equation}

\subsection{Bayesian evidence and model selection}
\label{beiv}

Statistical comparison of different models can be done with different model selection criteria like the Akaike Information Criterion (AIC) and Bayesian Information Criterion (BIC).

\par The Akaike Information Criterion (AIC) \cite{aic} is defined as,

\begin{equation}
AIC=-2\ln{L_{max}}+2\kappa,
\end{equation}

where $L_{max}$ is the maximum likelihood and $\kappa$ is  the number of free parameter in the model. 

The Bayesian Information Criterion (BIC)  \cite{bic}, which is actually based on the Bayesian evidence, is defined as,

\begin{equation}
BIC=-2\ln{L_{max}}+2\kappa\ln{N},
\end{equation}

where $N$ is the number of data point used in the analysis. 

\par The definitions show that the AIC and BIC  are quite close to each other. For comparison between two models, the difference between the value of AIC of  the two models, i.e. $\Delta AIC$, and  the difference between the value of BIC of  the two models, i.e. $\Delta BIC$ are important. If the two models have same number of free parameters and same number of data points used in the statistical analysis, then the $\Delta AIC$ and $\Delta BIC$ are same. If $\Delta AIC$(or $\Delta BIC$)$<1$, then the models are in close proximity of each other according to the statistical model selection. If $1<\Delta AIC(or ~ \Delta BIC)<5$, the models are not very close according to preference. If $5<\Delta AIC(or ~ \Delta BIC)$ then the models are significantly different.

\par In a Bayesian analysis, calculation of Bayesian evidence is a more powerful approach for the model selection as it more general. The information criteria are approximated from Bayesian evidence. The Bayesian evidence is defined as,

\begin{equation}
Evidence=\int (Prior\times Likelihood)d\theta_1d\theta_2...d\theta_n,
\end{equation}

where $\theta_1,\theta_2,...,\theta_n$ are the model parameters. The ratio of Bayesian evidence of two models is called the Bayes' factor \cite{kassbayesfac}. If the Bayes' factor is greater than $10$, then the evidence is said to be strong regarding the model selection. The model with higher value  of the evidence is more preferred.

\section{Reconstruction of cosmological models and the present work}

As there is no theoretically compelling reason to pick any model for the late time dynamics of the universe from other branches of physics, such as particle physics  theory, the present trend  is to find the some viable model right from the observations. The basic idea is to choose a viable evolution scenario, choosing $a=a(t)$ or $H=H(t)$, and using that finding out the cosmological parameters like the dark energy equation of state parameter, the dark energy density parameter, the quintessence scalar field or the potential etc. using statistical techniques with the observational data. This kind of reverse engineering is called the {\it reconstruction}.  Pioneering work in this direction was by Ellis and Madsen \cite{ellismadsenreconst}, where the scalar field potential was found from a given evolution scenario, i.e., $a=a(t)$, through Einstein's equations.  Starobinsky showed that the potential associated to the scalar field can be reconstructed using the density perturbation \cite{staroreconst}. The data of distance measurement of supernova has been utilized by Huterer and Turner \cite{huterreconst} and by Saini {\it et al.} \cite{sainireconst}.

\par Reconstruction of dark energy that drives the cosmic acceleration, normally involves the finding out the dark energy equation of state parameter $w_{DE}$ as a function of redshift, i.e. $w_{DE}=w_{DE}(z)$ \cite{huterreconst,sainireconst,vsahniaasrarore}. Reconstruction is mainly done in two different ways. One is called  {\it parametric reconstruction} where a suitable ansatz for $w_{DE}(z)$ is chosen and the value of the model parameters are estimated from the observational data \cite{reparam,cplparam,scherrersen}. Parametric reconstruction with other cosmological quantities like deceleration parameter, total equation of state parameter are also there in literature \cite{GongWang2007}. The other way of reconstruction is a non parametric approach which is an attempt to reconstruct the evolution of $w_{DE}(z)$ directly from observational data without any prior assumption about the functional form \cite{nonparreconst}.

\par Cosmological quantities, that only consist of the scale factor and its time derivatives, are called  the kinematical parameters. Kinematic approach in the study of cosmic evolution is independent of any particular gravity theory. The Hubble parameter, deceleration parameter, the jerk parameter etc., belong to the set of kinematic quantities. Reconstruction of different kinematical quantities using the observational data depicts the nature of cosmic evolution without any apriori assumption regarding the dark energy. A kinematic approach was discussed by Riess et al. \cite{riess2}, where a linear parametrization of deceleration parameter $q(z)$ has been used to estimate the value of redshift at which the transition from decelerated to accelerated expansion happened. The cosmological jerk parameter, which is a dimensionless representation of the third-order time derivative of the scale factor, has been used as a diagnostic of dark energy models by Sahni et al. \cite{sahnistatefinder} and Alam et al. \cite{alamstatefinder}. The jerk parameter and a combination of jerk and deceleration parameter together have been identified as the statefinder parameter in these two investigations. Reconstruction of dark energy equation of state through the parametrization of cosmological jerk has been discussed by Luongo \cite{luongojerk}. Kinematic approach to the modelling of accelerating Universe has been discussed by Rapetti et al. \cite{rapettijerk}, where a constant jerk parameter model has been invoked. Evolving jerk parameter models has been investigated by Zhai et al. \cite{zhaijerk}.

\par The present work, which has been discussed in the following chapters, is devoted to the reconstruction of the dark energy models using diverse observational data sets. The parametric approach has been adopted for the reconstruction of cosmological models. The reconstruction of kinematic quantities and the possibility of interaction between dark energy and dark matter have also been emphasised.

\par In the second chapter, a parametric reconstruction of the dark energy equation of state for a quintessence scalar field model has been discussed. A new parametric form of the dark energy equation of state parameter ($w_{DE}$) has been introduced. It is written as,

\begin{equation} 
w_{DE}(z)=-\frac{3}{\alpha (1+z)^3+ 3}, 
\label{wzch1}
\end{equation}   

where $\alpha$ is the model parameter. Effectively there are two parameters, one is the $\alpha$ and the other one is the matter density parameter $\Omega_{m0}$. The values of  the model parameter have been estimated by $\chi^2$ minimization technique using different combinations of the data sets, namely the OHD, SNe, BAO and CMB shift parameter data. The values obtained in the combined analysis with these four data sets are, $\Omega_{m0}=0.284\pm0.007$ and $\alpha=-0.0009\pm0.0117$ at 1$\sigma$ error bar. The best fit value of $\alpha$ is negative and  thus the model shows a preference towards the phantom nature of dark energy. The corresponding scalar field potential has also been studied. The potential comes out to be a double exponential potential and it is a generalization of the potential discussed by Sen and Sethi \cite{sensethi}.

\par In the third chapter, a parametric reconstruction of the effective or total equation of state parameter ($w_{eff}(z)=p_{tot}/\rho_{tot}$) has been discussed. The form of $w_{eff}$ is chosen as,

\begin{equation}
w_{eff}=-\frac{1}{1+\alpha(1+z)^n},
\end{equation}

and the expression of the Hubble parameter has been obtained for this model as,

\begin{equation}
h^2(z)=H^2(z)/H^2_0=\Bigg(\frac{1+\alpha(1+z)^n}{1+\alpha}\Bigg)^{\frac{3}{2n}},
\end{equation}
where $H_0$ is the present value of the Hubble parameter. The model contains two parameter $\alpha$ and $n$. The series expansion of $h^2(z)$ will show that there is a term with $(1+z)^3$ which represents the matter density and the corresponding coefficient $\left(\frac{\alpha}{1+\alpha}\right)^{3/n}$ is the matter density parameter. The reconstructed mode mimics the $\Lambda$CDM model for $n=3$, thus the value of the model parameter $n$, obtained in the statistical analysis, would indicate the deviation of the model form $\Lambda$CDM. The values of the model parameters, obtained in the statistical analysis with SNe, OHD, BAO and CMB shift parameter data, are $\alpha=0.444\pm0.042$ and  $n=2.907\pm0.136$ at 1$\sigma$ confidence level.  A comparative study between the reconstructed model and the $w$CDM dark energy model has been done by representing both the models on ($q_0,j_0$) parameter space where $q_0$ is the present value of the deceleration parameter and $j_0$ is the present value of cosmological jerk parameter. It has been shown that the reconstructed model ensures tighter constraints on the kinematic quantities like the deceleration parameter and the jerk parameter.

\par The fourth chapter is devoted to the reconstruction of cosmological jerk parameter $j(z)$. This is a kinematic approach which is independent of any prior assumption about the gravity theory. The jerk parameter is defined as,

\begin{equation}
j(z)=-\frac{1}{H^3a}\frac{d^3a}{dt^3}=-1+(1+z)\frac{[H^2(z)]'}{H^2(z)}-\frac{1}{2}(1+z)^2\frac{[H^2(z)]''}{H^2(z)},
\label{jerkch1}
\end{equation}
where a prime denotes the derivative with respect  to the  redshift $z$.  The formalism discussed by Zhai {\it et al} \cite{zhaijerk} has been adopted in the present work. The form of jerk parameter is assumed as, 
\begin{equation}
j(z)=-1+j_1\frac{f(z)}{h^2(z)},
\end{equation} 
where $j_1$ is a constant and $f(z)$ is an analytic function of $z$. Four ansatz for $j(z)$ have been chosen in the present work. 
The four ansatz chosen are given below,
\begin{equation}
Model ~I.~~~~   j(z)=-1+j_1\frac{1}{h^2(z)},
\label{jmodel1}
\end{equation}
\begin{equation}
Model ~II.~~~~~  j(z)=-1+j_1\frac{(1+z)}{h^2(z)},
\label{jmodel2}
\end{equation}
\begin{equation}
Model ~III.~~~~~  j(z)=-1+j_1\frac{(1+z)^2}{h^2(z)},
\label{jmodel3}
\end{equation}
\begin{equation}
~~Model ~IV.~~~~~   j(z)=-1+j_1\frac{1}{(1+z)h^2(z)}.
\label{jmodel4}
\end{equation}
The models effectively have two parameters. One is coming in the expression of the Hubble parameter as integration constant ($c_1$), and other one is the $j_1$. The parameter $c_1$ is equivalent to the matter density parameter $\Omega_{m0}$. The value of the model parameters obtained from the combined analysis using OHD, SNe, BAO and CMBShift data are, ~~{\it Model I.} $c_1=0.298\pm0.010$, $j_1=0.078\pm0.140$;~~ {\it Model II.} $c_1=0.299\pm0.008$, $j_1=0.045\pm0.050$;~~ {\it Model III.} $c_1=0.300\pm0.008$, $j_1=0.017\pm0.015$;~~  {\it Model IV.} $c_1=0.298\pm0.008$, $j_1=0.112\pm0.176$. The nature of evolution of deceleration parameter,  jerk parameter and dark energy equation of state parameter have also been studied. A Bayesian analysis, mainly the calculation of the Bayesian evidence shows that the models are at close proximity of each other according to statistical preference and it can only be concluded that Model IV is marginally preferred to the other models.

\par  The fifth chapter also about a kinematic approach to the reconstruction of dark energy. The possibility of the interaction between dark energy and dark matter has been investigated in this reconstruction. The reconstruction has been done with an assumption of a slowly varying jerk parameter which is effectively a constant in the low redshift regime. The expression of Hubble parameter obtained in this case is,

\begin{equation}
h^2(z)=A(1+z)^{\frac{3+\sqrt{9-8(1+j)}}{2}}+(1-A)(1+z)^{\frac{3-\sqrt{9-8(1+j)}}{2}}.
\label{h2zAjch1}
\end{equation}
The model has two parameters $A$ and constant jerk parameter $j$. The parameter $A$ which comes as an integration constant, is actually the matter density parameter. For $j=-1$, the present model mimics the $\Lambda$CDM. Any deviation of the the value of parameter $j$ from $-1$ does not allow the dark matter and the dark energy components to have independent conservation. Thus it invokes the possibility of interaction between dark matter and the dark energy. The total conservation equation can be written into two parts,  

\begin{equation}
\dot{\rho}_m+3H\rho_m=\eta,
\label{mattconch1} 
\end{equation} 
and 
\begin{equation} 
\dot{\rho}_{DE}+3H(1+w_{DE})\rho_{DE}=-\eta.
\label{deconch1}
\end{equation}
The interaction term $\eta$ can be represented in a dimensionless way as, $Q=\frac{8\pi G}{3H_0^3}\eta$. The values of the model parameters, obtained in the combined analysis with OHD, SNe and BAO  are, $A=0.286\pm0.0015$ and $j=-1.027\pm0.037$. The interaction term has also been reconstructed and it shows that the possibility of interaction is high in the past, but it is very low at present. The value of parameter $j$ obtained from the statistical analysis shows that the model remains at close proximity of $\Lambda$CDM. The dark energy equation of state parameter shows a slight inclination towards the non-phantom nature.

\par In the sixth chapter, the interaction rate of dark energy and dark matter has been reconstructed for holographic dark energy model. The idea of holographic dark energy is based on fundamental thermodynamic consideration, namely the {\it holographic principle} \cite{hooft,susskind}. It relates a  short distance cut-off (ultraviolet cut-off) to a long distance cut-off (infrared (IR) cut-off).  Finally the holographic dark energy density is written as,
\begin{equation}
\rho_{H}=3C^2M_P^2/L^2,
\label{rhoh}
\end{equation} 
where $C^2$ is a dimensionless constant, $M_p^2=(8\pi G)^{-1}$ and $L$ is the associated IR cut-off.  In the present work, the Hubble horizon is taken as the IR cut-off, i.e. $L=H^{-1}$. The interaction rate between dark energy and dark matter has been reconstructed for two different parameterization of the deceleration parameter, Model I. $q(z)=q_1+\frac{q_2}{(1+z)^2}$ and Model II. $q(z)=\frac{1}{2}+\frac{q_1+q_2z}{(1+z)^2}$. The total conservation equation has been written into two parts, similar to that in equation (\ref{mattconch1}) and (\ref{deconch1}), and the interaction rate is defined as $\Gamma=\eta/\rho_H$. The interaction rate can be expressed as,

\begin{equation}
\Gamma=-3Hrw_{eff},
\end{equation}
where $w_{eff}$ is the effective or total equation of state parameter and the $r=\rho_m/\rho_H$ is called the {\it coincidence parameter}. The evolution of the interaction rate, the nature of dark energy equation of state parameter have been studied. The cosmic coincidence problem in the context of holographic dark energy has also been discussed. In a spatially flat universe, holographic dark energy model with Hubble horizon as the IR cut-off  has a constant value of coincident parameter $r$. The dark energy equation of state parameter shows a phantom nature at present and at high redshift it tends to zero, thus the dark energy component is indistinguishable from the dark matter at high redshift.

\par Finally, chapter seven contains the concluding remark and relevant discussions regarding the overall work presented in the thesis.    

%%%%%%%%%%%%%%%%%%%%%%%%%%%%%%%%%%%%%%%%%%%%%%%%%%%%%%%%%%%%%%%%%%%%%%%%%%%%%%

%*******************************************************************************
%****************************** Second Chapter *********************************
%*******************************************************************************

\chapter{A reconstruction of dark energy equation of state}

\ifpdf
    \graphicspath{{Chapter2/Figs/Raster/}{Chapter2/Figs/PDF/}{Chapter2/Figs/}}
\else
    \graphicspath{{Chapter2/Figs/Vector/}{Chapter2/Figs/}}
\fi
\section{Introduction}

To explain the cosmic acceleration, the exotic component introduced in the energy budget of the Universe is called the dark energy. The attempt towards finding a viable model using the observational data is the {\it reconstruction}, which is actually the reverse way of building the model. The idea is to assume a viable evolution scenario and then to find the suitable matter distribution consistent with the evolution scenario. The reconstruction of dark energy models mainly hovers around finding the nature and evolution of the dark energy equation of state. In the present chapter, a new parametric dark energy equation of state has been discussed.  A scaler field ($\phi$) model, associated with the reconstructed dark energy equation of state, has been picked up and the nature and evolution of the scalar field potential ($V(\phi)$) has been investigated. The reconstruction is based upon the parametrization of the dark energy equation of state parameter $w_{DE}(z)$. In this chapter, a new parametric dark energy equation of state parameter has been proposed. Eventually the equation of state leads to a particular form of the quintessence potential, $V(\phi)=V_1e^{\lambda\phi}+V_2e^{-\lambda\phi}+V_0$, where $V_1$, $V_2$, $V_0$ and $\lambda$ are constants. Statistical analysis of this model is carried out using the type Ia supernova distance modulus data (SNe), observational Hubble parameter data (OHD), Baryon Acoustic oscillation data (BAO) and the CMB shift parameter data (CMBShift). This model also shows a slight inclination towards the phantom behaviour across the limit $w_{DE}=-1$.

\section{Reconstruction of the scalar field potential from the equation of state parameter}
\label{sec2ch2}

The field equations for a spatially flat FRW universe with cold dark matter (given by a pressureless fluid) and a scalar field ($\phi$) are 
\begin{equation}
3H^2=8\pi G(\rho_m + \rho_{\phi}),
\label{friedmann1ch2}
\end{equation}
\begin{equation}
2\dot{H}+3H^2=-8\pi G p_{\phi},
\label{friedmann2ch2}
\end{equation}
where $H$ is the Hubble parameter given by $H=\frac{\dot{a}}{a}$, $\rho_m $ is the matter energy density and $ \rho_{\phi}$  and $p_{\phi}$ are the contributions of the scalar field to the energy density and pressure sectors respectively. The latter two are given by 

\begin{equation}
\rho_{\phi}=\frac{\dot{\phi}^2}{2}+V(\phi),
\label{rhophich2}
\end{equation}
\begin{equation}
p_{\phi}=\frac{\dot{\phi}^2}{2}-V(\phi),
\label{pphich2}
\end{equation}
where $V(\phi)$ is the scalar potential. An overhead dot indicates a differentiation with respect to the cosmic time $t$. The pressureless cold dark matter satisfies its own conservation equation which leads to 
\begin{equation}
\rho_m= \rho_{m0}(1+z)^3,
\label{rhomch2}
\end{equation}
where $z$ is the redshift.  The constant, $\rho_{m0}$, is the present value of the dark matter density. With equations (\ref{friedmann1ch2}),(\ref{friedmann2ch2}) and (\ref{rhomch2}), the wave equation for the scalar field,
\begin{equation}
\Box\phi+\frac{dV}{d\phi}=0,
\label{waveequnch2}
\end{equation} 
is a consequence of the Bianchi identity and does not lead to an independent equation. From equations (\ref{friedmann1ch2}) and (\ref{friedmann2ch2}), the equation of state parameter $w_{DE}$ can be written as
\begin{equation}
w_{DE}=\frac{p_{\phi}}{{\rho}_{\phi}}=-\frac{2\dot{H}+3H^2}{3H^2-8\pi G\rho_m}.
\label{wDEch2}
\end{equation}
One can replace the argument `$t$' by the redshift $z$ in this equation. With the aid of the equation (\ref{rhomch2}), the equation (\ref{wDEch2}) would look like 
\begin{equation}
2(1+z)H\frac{dH}{dz}=3(1+w_{DE})H^2-8\pi G\rho_{m0}(1+z)^3w_{DE}.
\label{differentialequnch2}
\end{equation}  
As we have three unknown quantities  $a$, $\phi$ and $V(\phi)$ against only two equations, namely equation (\ref{friedmann1ch2}) and (\ref{friedmann2ch2}) to solve for them, we can choose an ansatz so as to close the system of equations. In what follows, a one parameter equation of state, given by 
\begin{equation}
w_{DE}(z)=-\frac{3}{\alpha (1+z)^3+ 3}, 
\label{wzch2}
\end{equation}  
 is chosen where $\alpha$ is  a constant parameter. The reason for choosing this kind of $w_{DE}$  is that for high $z$, i.e. at the early stage of evolution, $w_{DE}$ is almost zero so that it is hardly distinguishable from the equation of state parameter of a pressureless fluid, but  gradually decreases to more and more negative values so as to yield an increasing negative pressure. For $\alpha = 0$, $w_{DE}$ reduces to $-1$, i.e., that of a cosmological constant. For $\alpha<0$, the model leads to a phantom behaviour i.e. $w_{DE}<-1$. Normally a dark energy model is chosen such that it remains subdued at an early stage, i.e., for high value of $z$ and evolves to dominate over the dark matter only at the later stage of evolution. The present choice is qualitatively different from such models as at the early stage with $w_{DE}$ approaching zero, the dark energy is indistinguishable from the pressureless dark matter rather than being subdued.

\par With equation (\ref{wzch2}), one can integrate equation (\ref{differentialequnch2}) to obtain 
\begin{equation} 
H^2(z)=H_0^2\left[\frac{(\alpha+3\Omega_{m0})}{(\alpha+3)}(1+z)^3+\frac{3(1-\Omega_{m0})}{(\alpha +3)}\right], 
\label{Hzch2}
\end{equation}
where $H_0$ is the present value of Hubble parameter and $\Omega_{m0}$ is the present matter density parameter defined as $\Omega_{m0}=\frac{8\pi G\rho_{m0}}{3H_0^2}$. From the expression of Hubble parameter (equation \ref{Hzch2}), the model looks like the $\Lambda$CDM a bit in disguise. But this is not exactly the case as we can not determine parameter $\alpha$ in terms of $\Omega_{m0}$ from the flatness constraint.

 The deceleration parameter $q$, defined as ($-\frac{a\ddot{a}}{{\dot{a}^{2}}}$), can be written in terms of $z$ as
\begin{equation}
q(z)=-1+\frac{3(\alpha+3\Omega_{m0})}{2(\alpha+3)}\frac{H_0^2(1+z)^3}{H^2(z)}.
\label{qzch2}
\end{equation}

The nature of evolution  of $q(z)$ can be investigated utilizing the values of the parameters $\alpha$ and $\Omega_{m0}$, constrained by observation. Now from equation (\ref{friedmann1ch2}) and (\ref{friedmann2ch2}), one can write (using the expression for $ \rho_{\phi}$  and $p_{\phi}$) 
\begin{equation}
2\dot{H}=-8\pi G(\rho_{m}+\dot{\phi}^2),
\label{hdotch2}
\end{equation}
which can be written as
\begin{equation}
8\pi G(1+z)^2H^2\Bigg(\frac{d\phi}{dz}\Bigg)^2=2(1+z)H\frac{dH}{dz}-3H_0^2\Omega_{m0}(1+z)^3,
\label{phiequationch2}
\end{equation}
if $z$ is used as the argument instead of $t$. This can be integrated to yield (using equation (\ref{Hzch2})) the result

\begin{align}
\nonumber
\sqrt{8\pi G}\phi(z)=&\frac{2}{3}\sqrt{\frac{3\alpha(1-\Omega_{m0})}{\alpha+3\Omega_{m0}}}
\ln{}\Big[2(\alpha+3\Omega_{m0})(1+z)^{\frac{3}{2}}\\
&+2\sqrt{(\alpha+3\Omega_{m0})^2(1+z)^3+3(1-\Omega_{m0})(\alpha+3\Omega_{m0})}
\Big].
\label{phizch2}
\end{align}

An addition of the field equations (\ref{friedmann1ch2}) and (\ref{friedmann2ch2}) will now yield 
\begin{equation}
8\pi G V(z)=\frac{3\alpha H_0^2(1-\Omega_{m0})}{2(\alpha+3)}(1+z)^3+\frac{9H_0^2(1-\Omega_{m0})}{(\alpha+3)}.
\label{Vzch3}
\end{equation} 
In this expression $z$ can be replaced by $\phi$  using equation (\ref{phizch2}) to obtain the potential as a function of $\phi$ as,

\begin{align}
\nonumber
8\pi G V(\phi)=&\frac{3H_0^2(1-\Omega_{m0})exp(\Phi)}{128(\alpha+3)(\alpha+3\Omega_{m0})^2}+\frac{27H_0^2(1-\Omega_{m0})^3exp(-\Phi)}{2(\alpha+3)}\\
&+\frac{9H_0^2(1-\Omega_{m0})(3\alpha+\alpha\Omega_{m0}+12\Omega_{m0})}{4\alpha(\alpha+3)(\alpha+3\Omega_{m0})},
\label{Vphich2}
\end{align}

where $\Phi=3\sqrt{8\pi G}\sqrt{\frac{\alpha+3\Omega_{m0}}{3\alpha(1-\Omega_{m0})}}\phi$. 
\vskip 1.0cm

\section{Observational constraints on the parameters}
\label{sec3ch2}

The essential part of parametric reconstruction is the estimation of the parameter values from the observational data. There are two parameters in the model, the matter density parameter  $\Omega_{m0}$ and the parameter $\alpha$ which is introduced through the expression of $w_{DE}$. Here the observational Hubble parameter data (OHD), type Ia supernova distance modulus data (SNe), baryon acoustic oscillation data (BAO) and the CMB shift parameter (CMBShift) data have been used for the statistical analysis. 

\par The observational Hubble data set (OHD) is obtained from the measurement by different groups. Hubble parameter is measured directly from cosmic chromometres and differential age of galaxies in the redshift range $0<z<1.8$ \cite{simonohd,sternohd,chuangohd,blakeohd,zhangohd,morescoohd}. Measurement of Hubble parameter at $z=2.3$ \cite{buscaohd} has also been incorporated in the data set. The measurement of $H_0$ from Planck \cite{planck2013} has also been utilized. 

\par The distance modulus ($\mu_{B}(z)$) data set from type Ia supernova observations is very widely used one for the analysis of dark energy models. In the present work, the SNe data set of Union 2.1 compilation \cite{snsuzuki} has been utilized.

\par Baryon Acoustic Oscillation data (BAO) \cite{beutlerbao,blakebaowiggleZ,percilavSDSSbao} along with the measurement of {\it comoving sound horizon} at photon decoupling epoch ($z_*=1090.43\pm0.65$) and at photon drag epoch ($z_d=1059.29\pm0.65$) and the estimation of the value of {\it acoustic scale} at decoupling obtained from Planck results \cite{planck2013,wangwangcmb} have been incorporated in the statistical analysis.

\par Cosmic Microwave Background (CMB) data, in  the form of a distance prior, namely the CMB shift parameter $R_{\tiny CMB}$, estimated from Planck data in \cite{wangwangcmb}, has also been utilized here.

\par $\chi^2$-minimization (which is equivalent to the maximum likelihood analysis) technique has been adopted in  the present work for the statistical analysis.  Detail discussion about the observational data sets and the statistical techniques are presented in previous chapter (section \ref{obsdata} and \ref{statanalysis}).

\par
\begin{figure}[H]
\begin{center}
\includegraphics[angle=0, width=0.35\textwidth]{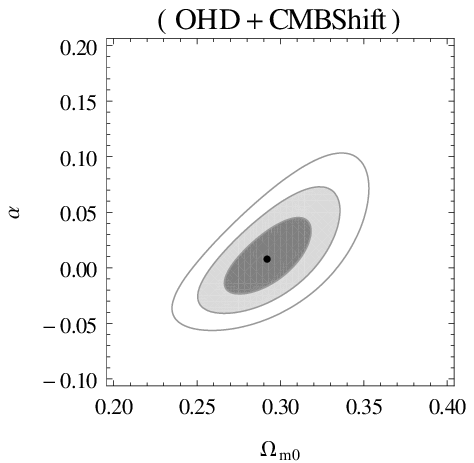} 
\includegraphics[angle=0, width=0.35\textwidth]{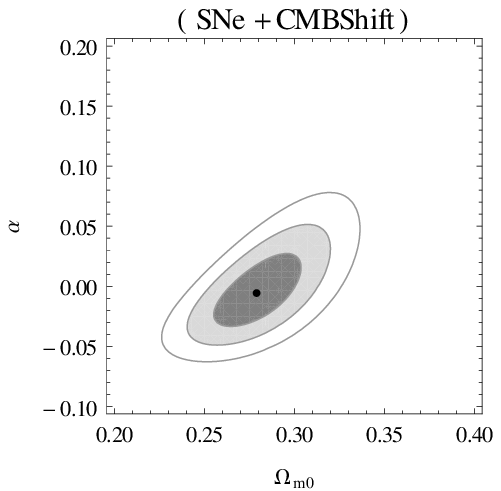}\\
\includegraphics[angle=0, width=0.35\textwidth]{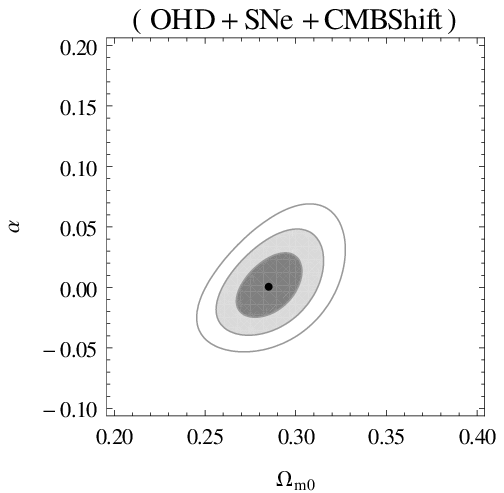} 
\includegraphics[angle=0, width=0.35\textwidth]{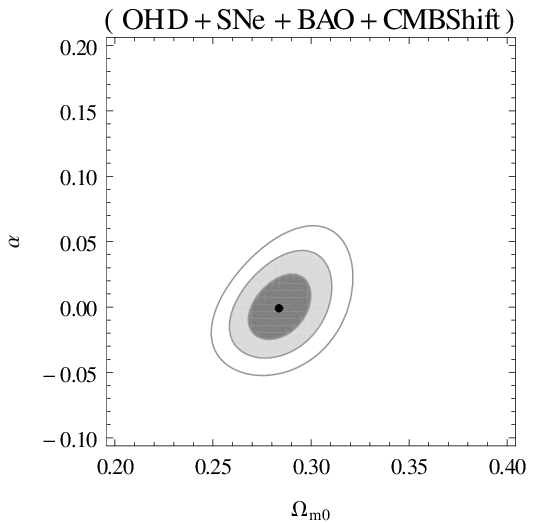} 
\caption{{\small Confidence contours on the 2D parameter space obtained for different combinations of the data sets. The 1$\sigma$, 2$\sigma$ and 3$\sigma$ confidence regions are presented from inner to outer area. The central black dots represent the corresponding best fit points. The upper left one is obtained for (OHD+CMBShitf), upper right is for (SNe+CMBShift), lower left is for (OHD+SNe+CMBShift) and lower right is for                  (OHD+SNe+BAO +CMBShift).}}
\label{contourplotch2}
\end{center}
\end{figure}

\begin{figure}[H]
\begin{center}
\includegraphics[angle=0, width=0.35\textwidth]{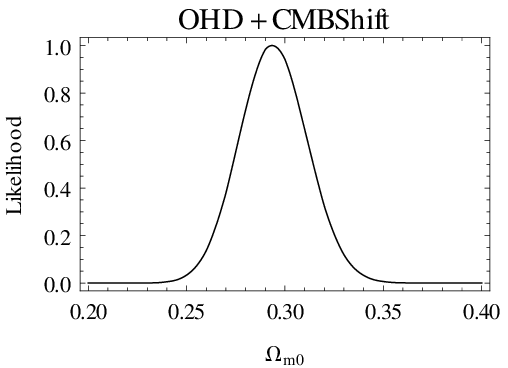}
\includegraphics[angle=0, width=0.35\textwidth]{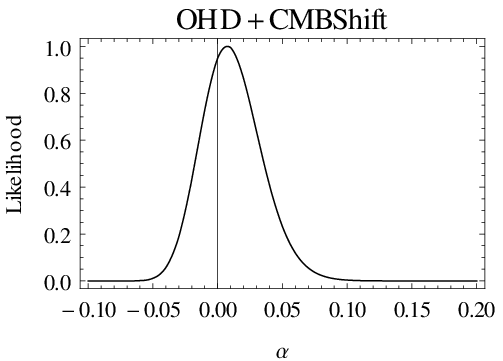}\\  
\includegraphics[angle=0, width=0.35\textwidth]{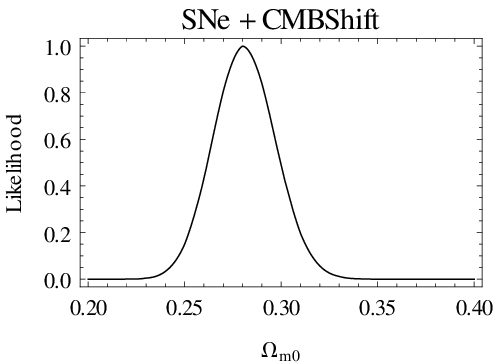}
\includegraphics[angle=0, width=0.35\textwidth]{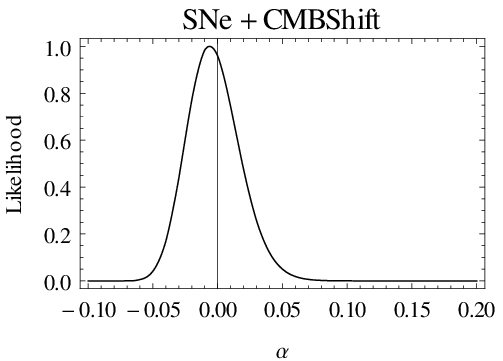}\\
\includegraphics[angle=0, width=0.35\textwidth]{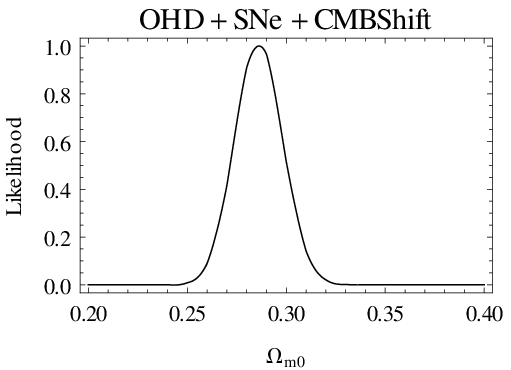} 
\includegraphics[angle=0, width=0.35\textwidth]{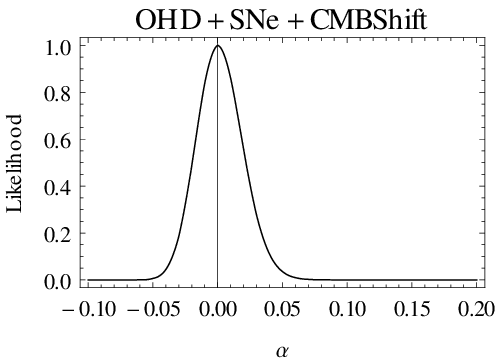}\\ 
\includegraphics[angle=0, width=0.35\textwidth]{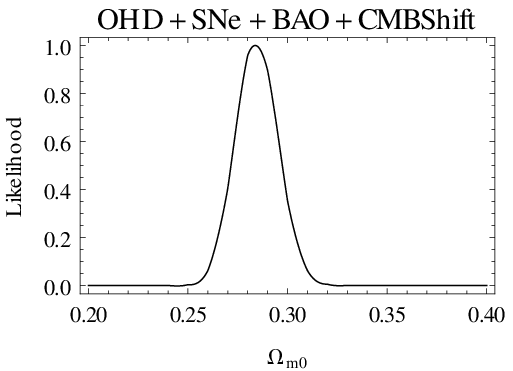} 
\includegraphics[angle=0, width=0.35\textwidth]{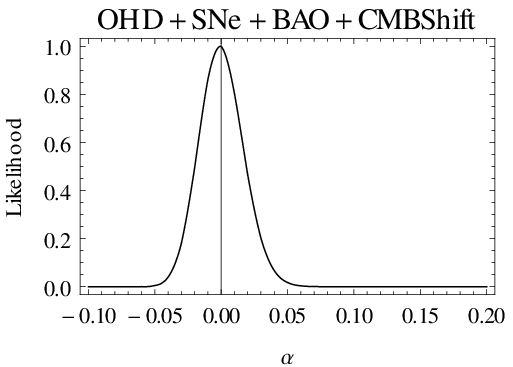} 
\caption{{\small Plots of marginalised likelihood functions for different combinations of the data sets. Left panels show the likelihood as a function of $\Omega_{m0}$ and the right panels show the likelihood as function of $\alpha$.}}
\label{likelihoodch2}
\end{center}
\end{figure}

 Figure \ref{contourplotch2} presents the confidence contours on 2D parameter space of the model obtained for different combination of the data sets. It is important to note that the CMB shift parameter data  is common to all combinations taken into account in the analysis. Actually the addition of the shift parameter data leads to substantial improvement of the constraints on the model parameters. Table \ref{tableCombinedanalysisch2} contains the best fit values of the parameters $\alpha$ and $\Omega_{m0}$ along with the allowed variation in the 1$\sigma$ error bar. The best fit values are obtained by the usual $\chi^2$ minimization technique.   Figure \ref{likelihoodch2} presents the marginalised likelihood functions. The likelihood function plots are well fitted to Gaussian distribution for both the parameters as arguments. 
 
%%%%%%%%%%%%%%%%%%%%%%%%%
\begin{table}[h!]
\caption{{\small Results of the statistical analysis. The reduced $\chi^2$ i.e. $\chi^2_{min}/d.o.f.$ where $d.o.f.$ is the number of degrees of freedom of that $\chi^2$ distribution, the best fit values of the parameters along with the 1$\sigma$ error bar obtained for different combinations of the data sets are presented.}}
\begin{center}
\resizebox{0.85\textwidth}{!}{  
\begin{tabular}{ c |c |c c } 
 \hline
  \hline
  Data  & $\chi^2_{min}/d.o.f.$ & Parameters \\ 
 \hline
 \hline
  {\small OHD+CMBShift} &  13.34/26 & $\Omega_{m0}=0.292\pm0.012$, $\alpha=0.0078\pm0.0160$\\ 
 \hline
  {\small SNe+CMBShift} &  562.23/577 & $\Omega_{m0}=0.279\pm0.012$, $\alpha=-0.0056\pm0.0156$\\ 
  \hline
  {\small SNe+OHD+CMBShift} & 575.89/603 &  $\Omega_{m0}=0.285\pm0.008$, $\alpha=0.0005\pm0.0124$\\ 
  \hline
  {\small SNe+OHD+BAO+CMBShift} &  578.04/606 & $\Omega_{m0}=0.284\pm0.007$, $\alpha=-0.0009\pm0.0117$\\ 
  \hline
  \hline
\end{tabular}
}
\end{center}
\label{tableCombinedanalysisch2}
\end{table}
%%%%%%%%%%%%%%%%%%%%%%%%%%%%%%%%%%%
%%%%%%%%%%%%%%%%%%%%%%%%%%%%%%%%%%%%%%%%%%%%%%%%%%%%%%%%%%%%%%%%%%%%

%%%%%%%%%%%%%%%%%%%%%%%%%%%%%%%%%%%%%%%%%%%%%%%%%%%%%%%%%%%%%%%%%%%%%%%%%

Contour plots on 2D parameter space (figure \ref{contourplotch2}), obtained from different combinations of the data sets, shows that the parameters ($\Omega_{m0}$ and $\alpha$) have a positive correlation between them. The likelihood plots (figure \ref{likelihoodch2}) show that the result obtained from the analysis with OHD+SNe+CMBShift and OHD+SNe+BAO+CMBSHift are very close to $\Lambda$CDM model as the bets fit values of the model parameter $\alpha$ is very close to zero. The results obtained for OHD+CMBShift and SNe+CMBShift have slightly higher deviation from $\Lambda$CDM limit. Though the $\Lambda$CDM always remains within 1$\sigma$ confidence region of the reconstructed model.

%%%%%%%%%%%%%%%%%%%%%%%%%%%%%%%%%%%%%%%%%%%%%%%%%%%
%%%%%%%%%%%%%%%%%%%%%%%%%%%%%%%%%%%%%%%%%%%%%%

%%%%%%%%%%%%%%%%%%%%%%%%%%%%%%%%%%%%%%%%%%%%%%%%%%%
\begin{figure}[t]
\begin{center}
\includegraphics[angle=0, width=0.35\textwidth]{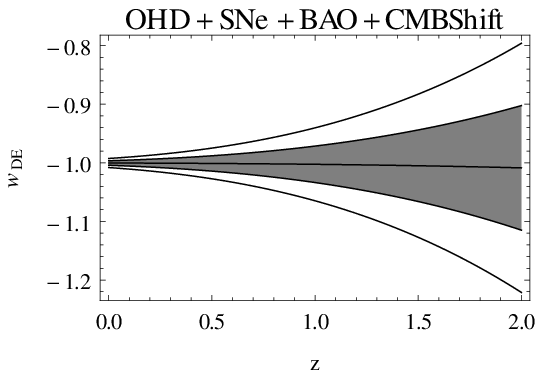}
\includegraphics[angle=0, width=0.35\textwidth]{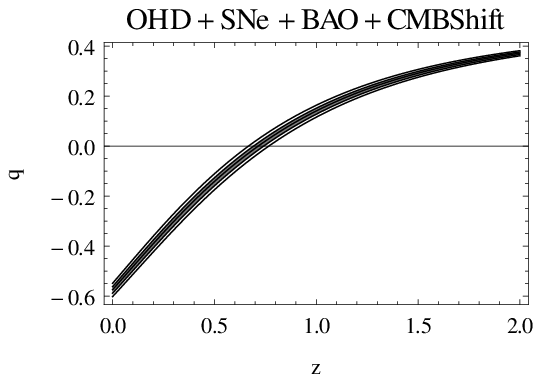}
\caption{{\small Plots show the behaviour of dark energy equation of state parameter $w_{DE}(z)$ (left) and the deceleration parameter $q(z)$ (right) as a function of redshift $z$ within 1$\sigma$ and 2$\sigma$ confidence levels with the central black line representing the best fit curve.}}
\label{wzqzch2}
\end{center}
\end{figure}
%%%%%%%%%%%%%%%%%%%%%%%%%%%%%%%%%%%%%%%%%%%%%%
%%%%%%%%%%%%%%%%%%%%%%%%%%%%%%%%%%%%%%%%%%%%%%%%%%%
\begin{figure}[t]
\begin{center}
\includegraphics[angle=0, width=0.355\textwidth]{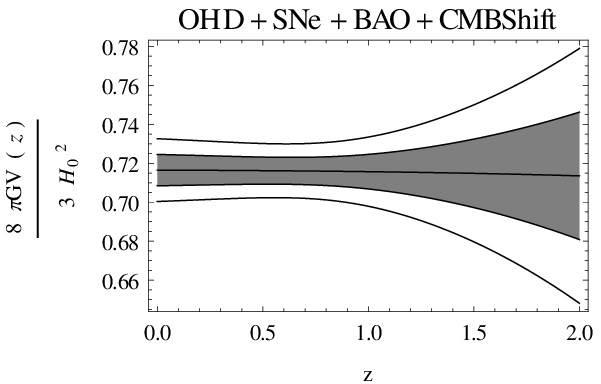} 
\includegraphics[angle=0, width=0.36\textwidth]{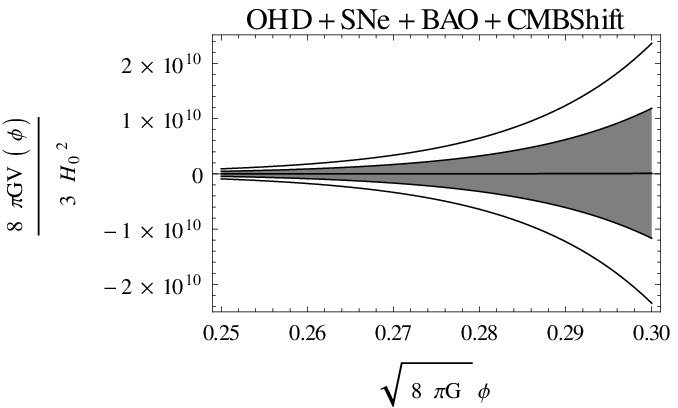} 
\caption{{\small The plots show the quintessence potential as a function of redshift $z$ (left) and also as a function of the quintessence scalar field (right) for 1$\sigma$ and 2$\sigma$ confidence level with the best fit curves represented by the central dark line.}}
\label{Vzphich2}
\end{center}
\end{figure}
%%%%%%%%%%%%%%%%%%%%%%%%%%%%%%%%%%%%%%%%%%%%%%

\par The plot of dark energy equation of state parameter ($w_{DE}(z)$) (left panel of figure \ref{wzqzch2}) shows that it is almost constant in the low redshift regime and has a slight inclination towards the phantom nature. The associated uncertainty increases with increase in redshift.  Right panel of figure \ref{wzqzch2} shows the plot of deceleration parameter.  The corresponding scalar field potential shows a slowly varying nature (figure \ref{Vzphich2})

\section{Discussion}
\label{sec4ch2}

This chapter presents a scalar field model of dark energy where a dark energy equation of state parameter $w_{DE}$, which is chosen as a one parameter function of $z$, is reconstructed from the observational data.  Tighter constraints on the parameters can be obtained using proper combination of the data sets (figure \ref{contourplotch2}). The last row in table \ref{tableCombinedanalysisch2} shows the best fit values of the model parameter $\alpha$ (which describes the equation of state given by equation (\ref{wzch2})) and the matter density parameter $\Omega_{m0}$,  obtained for various combination of SNe, OHD, BAO and CMB shift parameter data. The values, when all the four data sets are combined, are given as $\alpha=-0.0009\pm0.0117$ and  $\Omega_{m0}=0.284\pm0.007$ in 1$\sigma$ confidence region. The best fit value of $\alpha$ is negative, but very close to zero. That means the proposed model is very  close to $\Lambda$CDM with a tendency towards favouring the phantom nature of dark energy (left panel of figure \ref{wzqzch2}). The present value of dark energy equation of state parameter ($w_0$) is constrained to be $ w_0=-1.000\pm0.004$ at 1$\sigma$ confidence level by the present reconstruction.

\par It has already been mentioned that one characteristic feature of  the parametrization, discussed in the present chapter, is that the dark energy equation of state parameter tends to zero at high redshift. Thus the dark energy behaves like dust matter at high redshift causing an extra dust like contribution but  it is significant only at asymptotic limit.  In addition to that,  high-redshift data, namely the CMB shift parameter, constraints the value of the model parameter $\alpha\sim10^{-3}$ (in order of magnitude). Thus the extra dust like contribution does not have any significant effect on the cosmic structure formation.

\par The right panel of figure \ref{wzqzch2} shows that the deceleration parameter $q$ starts positive for a higher $z$, and attains negative value near $z=0$ with a signature flip between $z=0.6$ and $0.8$. This is consistent with the recent analysis by Farooq and Ratra \cite{farooqratra}.

\par The reconstructed quintessence potential is shown in the figure \ref{Vzphich2}. The left panel, which depicts $V=V(z)$, clearly indicates  that $V(z)$ remains almost flat. So one can say that the potential is a {\it freezing} potential as opposed to a {\it thawing} one (see for example, the work of Caldwell and Linder \cite{caldwelllinder} and that of Scherrer and Sen \cite{scherrersen}). 

\par The potential is shown as a function of the scalar field ($\phi$) in the right panel of figure \ref{Vzphich2} and the analytic form is given in equation (\ref{Vphich2}).  A similar potential had already been discussed by Sen and Sethi \cite{sensethi}. The potential obtained in the present work in the form $V(\phi)=V_1e^{\lambda\phi}+V_2e^{-\lambda\phi}+V_0$ where $V_1$, $V_2$, $V_0$ and $\lambda$ are constants, is a generalization of the potential given by Sen and Sethi, where $V_1=V_2$.  The requirement of $V_{1}= V_{2}$ would yield, from equation (\ref{Vphich2}), the condition
\begin{equation}
{\Omega}_{m0} = \frac{-(\alpha - 3) + \sqrt{(\alpha - 3)^{2} + 12(\alpha \mp \frac{1}{24})}}{6}.
\label{v1v2ch2}
\end{equation}
If we take realistic values of ${\Omega}_{m0}$ between $\frac{1}{4}$ and $\frac{1}{3}$, the value of $\alpha$ will lie between $-\frac{25}{36}$ and $-\frac{17}{16}$ leading to the values of $w_{DE}$ at $z=0$ between -1.301 and -1.548 respectively, well into the phantom regime of $w_{DE} < -1$. But this is out of 2$\sigma$ error bar of $w_{DE}$ at $z=0$ of the model presented in this work. Thus the Sen and Sethi model is allowed but statistically not favoured.

\par A recent analysis \cite{xiaLCDMwCDM} using CMB temperature anisotropy and polarization data, along with other non-CMB data, estimates the values of the parameters as $\Omega_{m0}=0.293\pm0.013$ at 1$\sigma$ confidence level for the $\Lambda$CDM model and $\Omega_{m0}=0.270\pm0.014$, $w_{DE}=-1.167\pm0.061$ at 1$\sigma$ confidence level for $w$CDM model. The present model is inclined towards the $\Lambda$CDM model, and the value of $\Omega_{m0}$ remains in between the values obtained for $\Lambda$CDM and $w$CDM (within 1$\sigma$ confidence level of both the models). 

\par A study of different parameterizations of dark energy equation of state by Hazra {\it et al} \cite{hazra} has  obtained the parameter values in various cases. For example, Chevallier-Polarski-Linder (CPL) parametrization \cite{cplparam} yields $\Omega_{m0}=0.307_{-0.046}^{+0.041}$, $w_0=-1.005_{-0.17}^{+0.15}$, Scherrer and Sen (SS) parameterization \cite{scherrersen} yields $\Omega_{m0}=0.283_{-0.030}^{+0.028}$, $w_0=-1.14_{-0.09}^{+0.08}$ and generalized Chaplygin gas (GCG) parameterization \cite{gcgbento,bentochap2} shows  $\Omega_{m0}=0.32_{-0.012}^{+0.013}$, $w_0=-0.957_{\tiny non-phantom}^{+0.007}$. Hence the present reconstruction is consistent with the CPL parameterization at 1$\sigma$ confidence level. The SS parameterization requires a slightly  lower value of $w_0$ (out of 1$\sigma$ error bar) but the value of $\Omega_{m0}$ is highly consistent. The non phantom prior assumption of GCG parameterization is not in agreement with the present model, but the lower bound of 1$\sigma$ error bar for GCG parameterization is within the 1$\sigma$ confidence region of the present model.

\par A reconstruction of quintessence potential described by a polynomial series constrains the present value of dark energy equation of state $w_0=-0.978_{-0.031}^{+0.032}$ \cite{hupeiris}. This is within the 1$\sigma$ error bar of the present model. 

\par It deserves mention that systematic uncertainties of observations might have its imprints on the results of these analyses. For instance, the colour-luminosity parameter might depend on the redshift, and hence affect the magnitude in the analysis of Supernova data \cite{wangwangSNeSys}. We also refer to the analyses of Rubin {\it et al} \cite{rubinSNeSys} and Shafer and Huterer \cite{shaferSNeSys} for some very recent development in connection with the systematics.

\chapter{A reconstruction of the effective equation of state} 

% **************************** Define Graphics Path**************************
\ifpdf
    \graphicspath{{Chapter3/Figs/Raster/}{Chapter3/Figs/PDF/}{Chapter3/Figs/}}
\else
    \graphicspath{{Chapter3/Figs/Vector/}{Chapter3/Figs/}}
\fi

\section{Introduction}

In this chapter, a parametric reconstruction of the effective or total equation of state is presented. The functional form of  effective equation of state parameter is chosen in such a way that it tends to zero at high value of redshift which is the signature of matter dominated universe. The present value of the effective equation of state parameter depends on the model parameters which have been constrained from the observational data.

\par The present work is not based on a purely kinematical approach, it rather assumes GR as the theory of gravity, but there is hardly any prior assumption about the distribution of the components in the matter sector. The prime endeavour of this reconstruction is to figure out the distribution of the matter components instead of any prior assumption about them.  The possibility of interaction between the components can also be investigated in this framework.  For comparison, a standard dark energy model, namely the $w$CDM, has also been explored using the same data sets.  For the $w$CDM model, the dark energy equation of state parameter ($w_{DE}$) is assumed to be a constant throughout the evolution. At the same time, the dark matter is allowed to have an independent conservation. The cosmological constant model or the $\Lambda$CDM and the $w$CDM model are at present the most popular dark energy models as they are well consistent with most of the observational data. For these reasons, the $w$CDM model has been chosen in the present work as an example for a comparison with the reconstructed model. Different model selection criteria unambiguously  show the consistency of this model with the standard $w$CDM dark energy model. For direct comparison, both the models have also been presented through ($q_0$,$j_0$) parameter space, where $q_0$ is the present value of deceleration parameter and $j_0$ be the present value of jerk parameter.

\section{Reconstruction of the model} 
\label{reconstch3}

The Friedmann equations for spatially flat universe are obtained as,

\begin{equation}
3H^2=8\pi G\rho,
\label{friedmann1ch3}
\end{equation}

\begin{equation}
2\dot{H}+3H^2=-8\pi Gp,
\label{friedmann2ch3}
\end{equation}
where $H$ is the Hubble parameter defined as $H=\frac{\dot{a}}{a}$ (an over-headed dot denotes the derivative with respect to cosmic time $t$), the $\rho$ is the total energy density and the $p$ is the pressure. Now the effective or total equation of state parameter ($w_{eff}$) is defined as
\begin{equation}
w_{eff}=\frac{p}{\rho}.
\end{equation}
This $\rho$ and $p$ take care of the density and the pressure respectively for all the forms of the matter present in the universe taken together. Now using equations (\ref{friedmann1ch3}) and (\ref{friedmann2ch3}), the effective equation of state parameter is written as
\begin{equation}
w_{eff}=-\frac{2\dot{H}+3H^2}{3H^2}.
\label{weffch3}
\end{equation}
 It is convenient to use redshift (z) as the argument instead of cosmic time $t$ as $z$ is a dimensionless quantity. If the argument of differentiation of $H$ is changed from cosmic time $t$ to redshift $z$, defined as $z+1=\frac{a_0}{a}$, where $a_0$ is the present value of the scale factor. Now one can write, 
\begin{equation}
\dot{H}=-(1+z)H\frac{dH}{dz}.
\label{dHdzch3}
\end{equation}
In the present work,  a parametric form of the effective equation of state $w_{eff}$, as a function of redshift $z$, is assumed as
\begin{equation}
w_{eff}(z)=-\frac{1}{1+\alpha(1+z)^n},
\label{weffzch3}
\end{equation}
where $\alpha$ and $n$ are two model parameters. It is now clear from the observation of large scale structure and the existing models of structure formation that the contribution to the energy budget of the universe was dominated by dark matter at high redshift. In the recent era, the prime contribution is coming from the exotic component dubbed as dark energy. As the dark matter is pressureless, the effective equation of state at high redshift was effectively zero. At the epoch of recent acceleration, it has a negative value which is less than $-\frac{1}{3}$. The functional form of the effective equation of state (equation (\ref{weffzch3})) assumed for the present reconstruction can easily accommodate these two phases of evolution. For positive values of the model parameter $\alpha$ and $n$, the values of $w_{eff}(z)$ tends to zero a high value of the redshift $z$ whereas at $z=0$, its value depends  upon the model parameter $\alpha$. It is also clear from the expression of $w_{eff}(z)$ (equation (\ref{weffzch3})) that a positive value of the model parameter $\alpha$ always fixes a lower bound to the value of $w_{eff}(z)$ and keeps it in the non-phantom regime. It is interesting to note that for $\Lambda$CDM model, the effective equation of state parameter can be expressed as, 

\begin{equation}
w_{eff}^{\Lambda CDM}=-\frac{1}{[1+(\Omega_{m0}/\Omega_{\Lambda 0})(1+z)^3]}.
\end{equation}
Thus the parameter $\alpha$ is equivalent to the ratio of dark mattre to dark energy density at $\Lambda$CDM limit.

\par Introducing the assumed ansatz of $w_{eff}(z)$ (equation (\ref{weffzch3})) to equation (\ref{weffch3}) and (\ref{dHdzch3})), the differential equation for $H$ reads as
\begin{equation}
\frac{2}{3}(1+z)\frac{1}{H}\frac{dH}{dz}-1=-\frac{1}{1+\alpha(1+z)^n}.
\end{equation}
The solution obtained for the Hubble parameter as a function of redshift is 
\begin{equation}
H(z)=H_0\Bigg(\frac{1+\alpha(1+z)^n}{1+\alpha}\Bigg)^{\frac{3}{2n}},
\label{hubbleparamch3}
\end{equation}
where $H_0$ is the value of Hubble parameter at $z=0$. One interesting point regarding this expression of Hubble parameter is that for $n=3$, this reproduces the $\Lambda$CDM model. Hence the estimated value of the model parameter $n$ will clearly indicate whether a $\Lambda$CDM or a time evolving dark energy is preferred by observations. 

\par It is important to note at this point that in the series expansion of $h^2(z)$ (where $h(z)=H(z)/H_0$), which can be obtained from equation (\ref{hubbleparamch3}), there will be a term with $(1+z)^3$. This corresponds to the dark matter density. The coefficient of $(1+z)^3$ is $\left(\frac{\alpha}{1+\alpha}\right)^{3/n}$. It is equivalent to the matter density parameter $\Omega_{m0}$ which is the ratio of present matter density and the present critical density ($3H_0^2/8\pi G$). Thus the contribution of the dark energy can be obtained by subtracting this term from $h^2(z)$,
\begin{equation} 
\Omega_{DE}(z)=h^2(z)-\Big(\frac{\alpha}{1+\alpha}\Big)^{\frac{3}{n}}(1+z)^3,
\end{equation}
($\Omega_{DE}$ is the dark energy density scaled by the present critical density). Similarly the pressure contribution of the dark energy can be obtained using equation (\ref{friedmann1ch3}) and (\ref{friedmann2ch3}) along with the expression of Hubble parameter $H(z)$ obtained in equation (\ref{hubbleparamch3}) . Finally the dark energy equation of state parameter can be written as a function of redshift and the associated model parameters as

\begin{equation}
{\small
w_{DE}(z)= -\frac{\Bigg(\frac{1+\alpha(1+z)^n}{1+\alpha}\Bigg)^{\frac{3}{n}}-\Big(\frac{\alpha}{1+\alpha}\Big)(1+z)^n\Bigg(\frac{1+\alpha(1+z)^n}{1+\alpha}\Bigg)^{\frac{3}{n}-1}}{\Bigg(\frac{1+\alpha(1+z)^n}{1+\alpha}\Bigg)^{\frac{3}{n}}-\Big(\frac{\alpha}{1+\alpha}\Big)^{\frac{3}{n}}(1+z)^3}.
}
\end{equation}
It is clear from the expression of $w_{DE}(z)$ that for $n=3$, the value $w_{DE}=-1$, which is the $\Lambda$CDM.

\section{Results of statistical analysis}
\label{resultch3}

As already mentioned, four observational data sets, namely OHD, SNe, BAO and CMBSift, are used in the statistical analysis.

\par The observational Hubble data set (OHD) is obtained from the measurement by different groups. Hubble parameter is measured directly from cosmic chromometres and differential age of galaxies in the redshift range $0<z<1.8$ \cite{simonohd,sternohd,chuangohd,blakeohd,zhangohd,morescoohd}. Measurement of Hubble parameter at $z=2.34$ by Delubac {\it et al} \cite{delubacohd} has also been incorporated in the data set. 

\par The distance modulus ($\mu(z)$) data set from type Ia supernova observations is very widely used for the analysis of dark energy models. In this work, the 31 binned distance modulus data sample of the recent joint light-curve analysis \cite{sn3} has been utilized.

\par Baryon Acoustic Oscillation data (result of 6dF Galaxy Survey at redshift $z=0.106$ (Beutler {\it et al} \cite{beutlerbao}), and the results of Baryon Oscillation Spectroscopic Survey (BOSS) at redshift $z=0.32$ (BOSS LOWZ) and at redshift z = 0.57 (BOSS CMASS)(Anderson {\it et al} \cite{bossandersonbao})), along with the measurement of {\it comoving sound horizon} at photon decoupling epoch ($z_*=1090.43\pm0.65$) and at photon drag epoch ($z_d=1059.29\pm0.65$) and the estimation of the value of {\it acoustic scale} at decoupling obtained from Planck results \cite{planck2013,wangwangcmb} have been incorporated in the statistical analysis.

\par Cosmic Microwave Background (CMB) data, in  the form of a distance prior, namely the CMB shift parameter $R_{\tiny CMB}$, estimated from Planck data in \cite{wangwangcmb}, has also been utilized here.

\par $\chi^2$-minimization technique has been adopted in  the present work for the estimation of the model parameters. Detail discussion about the observational data sets and the statistical techniques are presented in the first chapter (section \ref{obsdata} and \ref{statanalysis}).

\begin{figure}[H]
\begin{center}
\includegraphics[angle=0, width=0.27\textwidth]{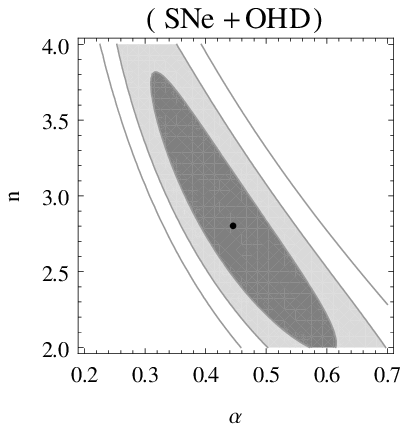}
\includegraphics[angle=0, width=0.27\textwidth]{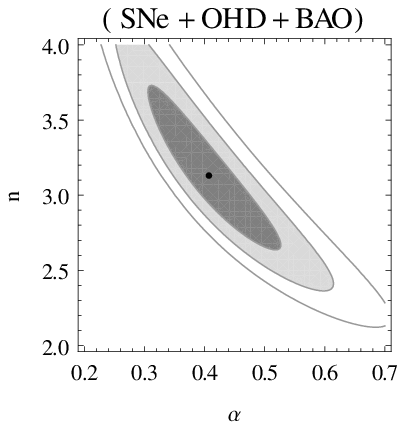}
\includegraphics[angle=0, width=0.262\textwidth]{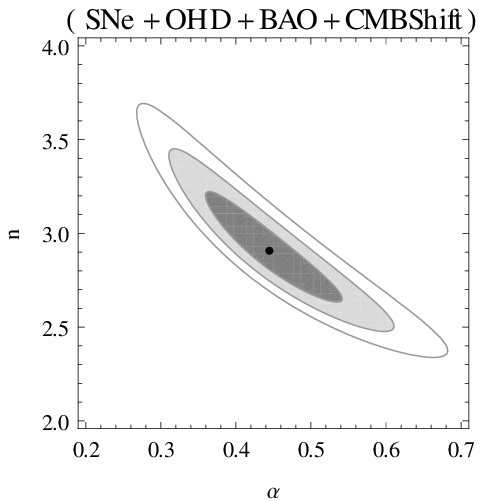}
\end{center}
\caption{{\small Confidence contours on the 2D parameter space of the $w_{eff}(z)$ model obtained for obtained for different combinations of the data sets. 1$\sigma$, 2$\sigma$ and 3$\sigma$ confidence regions are presented from inner to outer portion and the central black dots represent the corresponding  best fit points. The left panel shows the confidence contours obtained for SNe+OHD, the middle panel shows the confidence contours obtained for SNe+OHD+BAO and the right panel shows confidence contours for SNe+OHD+BAO +CMBShift.}}
\label{weffcontourch3}
\end{figure}

\begin{figure}[H]
\begin{center}
\includegraphics[angle=0, width=0.273\textwidth]{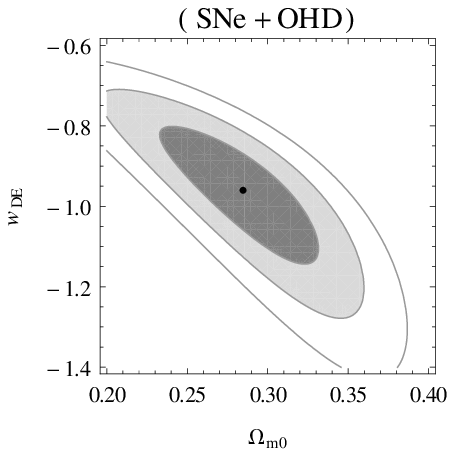}
\includegraphics[angle=0, width=0.273\textwidth]{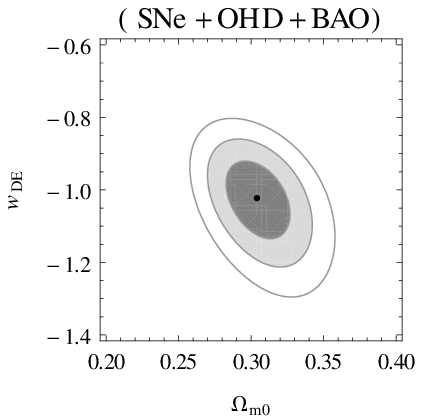}
\includegraphics[angle=0, width=0.262\textwidth]{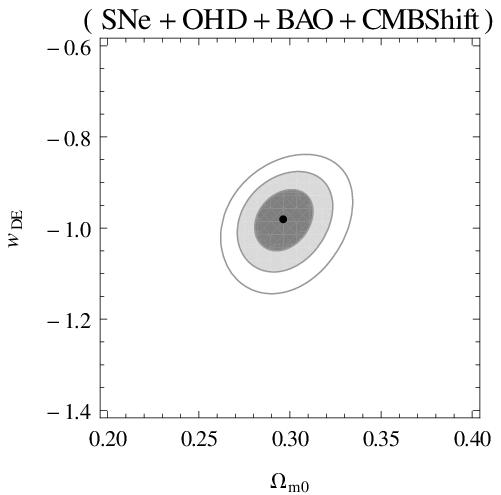}
\end{center}
\caption{{\small Confidence contours on the 2D parameter space of the $w$CDM model obtained for obtained for different combinations of the data sets. 1$\sigma$, 2$\sigma$ and 3$\sigma$ confidence regions are presented from inner to outer portion and the central black dots represent the corresponding  best fit points. The left panel shows the confidence contours obtained for SNe+OHD, the middle panel shows the confidence contours obtained for SNe+OHD+BAO and the right panel shows confidence contours for SNe+OHD+BAO +CMBShift.}}
\label{wCDMcontourch3}
\end{figure}

\begin{figure}[H]
\begin{center}
\includegraphics[angle=0, width=0.32\textwidth]{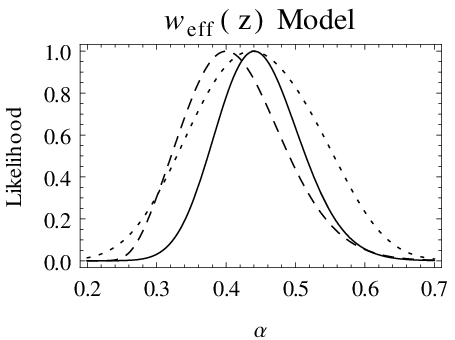}
\includegraphics[angle=0, width=0.32\textwidth]{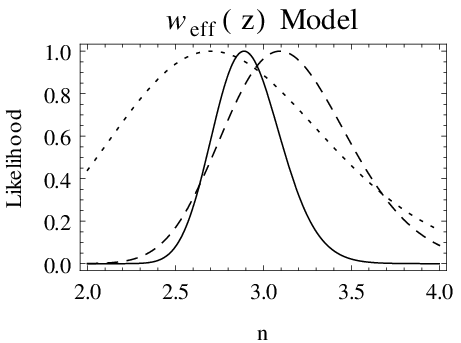}\\
\includegraphics[angle=0, width=0.32\textwidth]{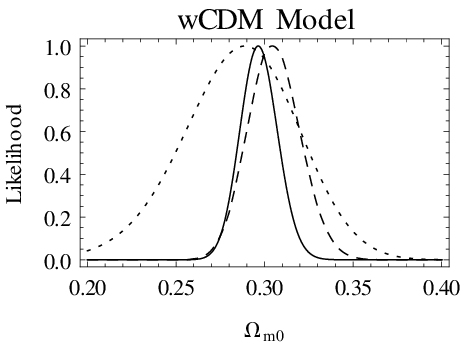}
\includegraphics[angle=0, width=0.32\textwidth]{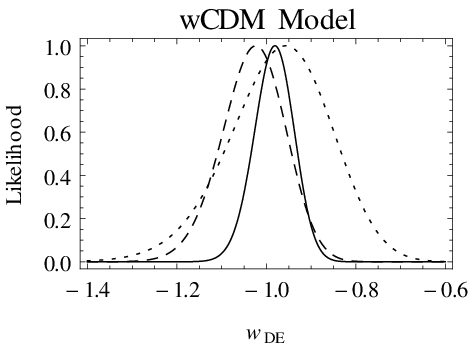}
\end{center}
\caption{{\small Plots  of marginalised likelihood as functions of model parameters. The upper panels show the likelihood for  the $w_{eff}(z)$ model and the lower panels show likelihood of $w$CDM model obtained from the statistical analysis with different combinations of the data sets. The dotted curves  represent the likelihood obtained for SNe+OHD, the dashed curves are obtained for SNe+OHD+BAO and the solid curves show the likelihood for  SNe+OHD+BAO+CMBShift.}}
\label{wefflikelihoodch3}
\end{figure}

Figure \ref{weffcontourch3} shows the confidence contours on the two dimensional (2D) parameter space for the $w_{eff}(z)$ model for different combinations of the data sets. Similarly figure \ref{wCDMcontourch3} presents the confidence contours on the 2D parameter space of the $w$CDM model, which has two parameters, namely the matter density parameter $\Omega_{m0}$ and the constant dark energy equation of state parameter $w_{DE}$. Marginalised likelihoods for the $w_{eff}(z)$ model and $w$CDM model are presented in figure \ref{wefflikelihoodch3}. It is clear from the likelihood plots that the likelihood functions are well fitted to a Gaussian distribution function when all four data sets are take into account. It is also apparent from the confidence contours (figure \ref{weffcontourch3} and figure \ref{wCDMcontourch3}) and the likelihood function plots (figure \ref{wefflikelihoodch3}) that the addition of CMB shift parameter data leads to substantially tighter constraints on the model parameters for both the models.

%%%%%%%%%%%%%%%%%%%%%%%%%
\begin{table}[h!]
\caption{{\small Results of statistical analysis of the $w_{eff}(z)$ and $w$CDM  models combining OHD, SNe, BAO and CMBShift data}}
\begin{center}
\resizebox{0.69\textwidth}{!}{  
\begin{tabular}{ c |c |c c  } 
  \hline
 \hline
 Model  & $\chi^2_{min}/d.o.f.$ &  Parameters \\ 

 \hline
  $w_{eff}(z)$ & 48.21/52 & $\alpha=0.444\pm0.042$ ; $n=2.907\pm0.136$\\ 
 \hline
  $w$CDM & 48.24/52 & $\Omega_{m0}=0.296\pm0.007$ ;  $w_{DE}=-0.981\pm0.031$\\ 
 \hline
  \hline
\end{tabular}
}
\end{center}
\label{tableweffwCDMch3}
\end{table}
%%%%%%%%%%%%%%%%%%%%%%%%%%%%%%%%

%%%%%%%%%%%%%%%%%%%%%
%%%%%%%%%%%%%%%%%%%%%%
%%%%%%%%%%%%%%%%%%%%%
%%%%%%%%%%%%%%%%%%%%%%

%%%%%%%%%%%%%%%%%%%%%
%%%%%%%%%%%%%%%%%%%%%%
%%%%%%%%%%%%%%%%%%%%%%%%%%%%%%%%%%%%%%%%%%%%%%%%%%%%%%%%%%%%%%%%%%%%%%%%%%

Table \ref{tableweffwCDMch3} contains the results obtained from the statistical analysis combining OHD, SNe, BAO and CMB shift parameter data. The parameter values and the associated 1$\sigma$ uncertainty have been presented along with the reduced $\chi^2$ i.e. $\chi^2_{min}/d.o.f$, where $d.o.f.$ is the {\it degrees of freedom} of the associated $\chi^2$. The reduced $\chi^2$ is a measure of the goodness of the fitting. The fitting would be rated to be good if the value of reduced $\chi^2$ is close to one.

\par The model parameter $n$ is important to figure out the deviation of the reconstructed model from the $\Lambda$CDM. for $n=3$, the reconstructed model actually becomes the $\Lambda$CDM. The confidence contours obtained from different combinations of the data sets show that the $\Lambda$CDM is always within 1$\sigma$ confidence regions. The result obtained from the analysis using only the SNe and OHD shows a higher deviation of the best fit value of the parameter $n$ from the corresponding $\Lambda$CDM value than the  results obtained by introducing the BAO and CMB shift parameter data along with SNe and OHD (figure \ref{weffcontourch3} and the upper right panel of figure \ref{wefflikelihoodch3}). The associated uncertainty obtained from the analysis with SNe+OHD is very large (left panel of figure \ref{weffcontourch3}) and the constraints become tighter with the addition of other data sets, namely the BAO and CMB shift parameter (middle and right panels of figure \ref{weffcontourch3}). The best fit  value of $n$ obtained for SNe+OHD is less than 3 (left panel of figure \ref{weffcontourch3}), for SNe+OHD+BAO it is greater than 3 (middle panel of figure \ref{weffcontourch3}) and for SNe+HOD+BAO+CMBShift, it is slightly less that 3 (right panel of figure \ref{weffcontourch3}). So it is apparent that the nature of deviation from $\Lambda$CDM varies according to the combination of data sets used for the analysis. It also deserves mention that the addition of CMB shift parameter data keeps the model in close proximity of $\Lambda$CDM and also ensures much tighter constraints on the parameter values. Another interesting point to note is that the parameters of the reconstructed $w_{eff}$ model have a negative correlation between them for all the combinations of the data sets considered in the present analysis (figure \ref{weffcontourch3}).  But for $w$CDM model, the nature of correlation between the parameter changes for different combinations of the data sets (figure \ref{wCDMcontourch3}). For the analysis with SNe+OHD and SNe+OHD+BAO the parameters of $w$CDM have a negative correlation but the correlation is positive when the CMB shift parameter data is taken into account.

\par The deceleration parameter, a dimensionless representation of the second order time derivative of the scale factor, is defined as $q=-\frac{1}{H^2}\frac{\ddot{a}}{a}$. It can also be written in terms of Hubbele parameter and its derivative with respect to redshift as,
\begin{equation}
q(z)=-1+\frac{1}{2}(1+z)\frac{(h^2)'}{h^2}.
\end{equation}
For the present $w_{eff}$ model, the expression of the deceleration parameter is obtained as,
\begin{equation}
q(z)=-1+\frac{3\alpha(1+z)^n}{2(1+\alpha(1+z)^n)}.
\label{presentdecelch3}
\end{equation}

\begin{figure}[H]
\begin{center}
\includegraphics[angle=0, width=0.32\textwidth]{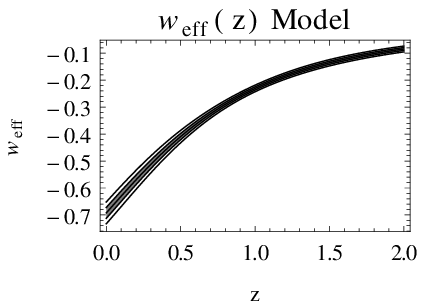}
\includegraphics[angle=0, width=0.32\textwidth]{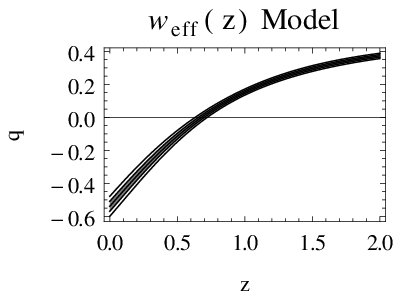}\\
\includegraphics[angle=0, width=0.32\textwidth]{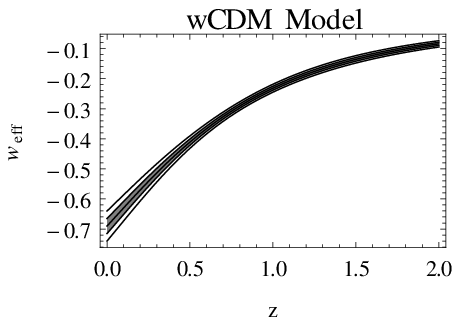}
\includegraphics[angle=0, width=0.32\textwidth]{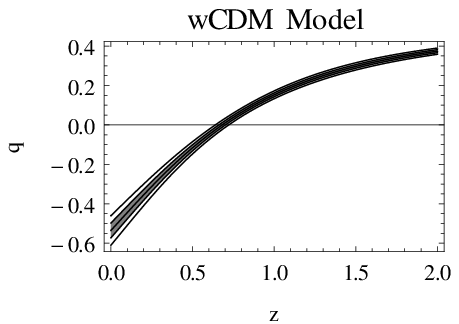}
\end{center}
\caption{{\small Plots of effective equation of state parameter and the deceleration parameter as functions of redshift $z$ for $w_{eff}(z)$ model (upper panels) and $w$CDM model (lower panels). The 1$\sigma$ and  2$\sigma$ confidence regions along with the central black line representing the corresponding the best fit curves obtained from the analysis combining the SNe, OHD, BAO and CMB shift parameter data are presented.}}
\label{weffzqzplotch3}
\end{figure}

In figure \ref{weffzqzplotch3}, the plot of effective equation of state parameter ($w_{eff}$) and the deceleration parameter ($q$) as functions of redshift $z$ for both  the $w_{eff}$ model and $w$CDM model have been presented. The central dark lines represent for the best fit curves and the 1$\sigma$ and 2$\sigma$ confidence region are given from inner to outer part.

%%%%%%%%%%%%%%%%%%%%%
%%%%%%%%%%%%%%%%%%%%%%
%%%%%%%%%%%%%%%%%%%%%
\begin{figure}[H]
\begin{center}
\includegraphics[angle=0, width=0.32\textwidth]{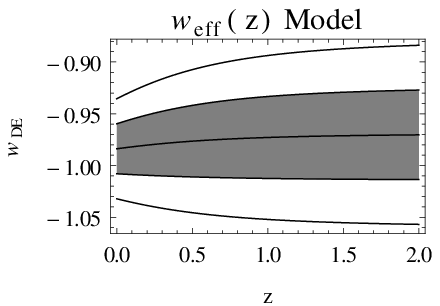}
\includegraphics[angle=0, width=0.32\textwidth]{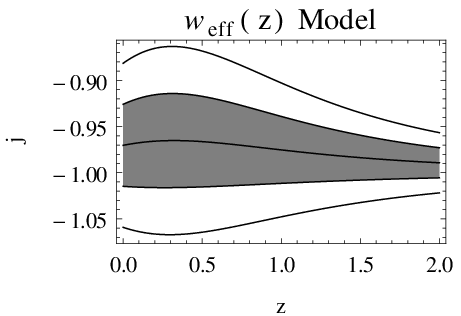}\\
\includegraphics[angle=0, width=0.32\textwidth]{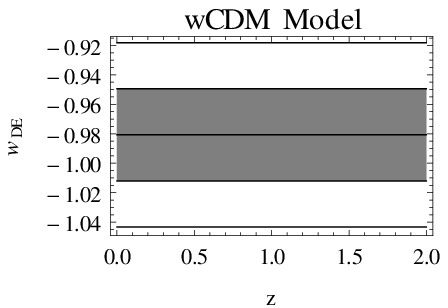}
\includegraphics[angle=0, width=0.32\textwidth]{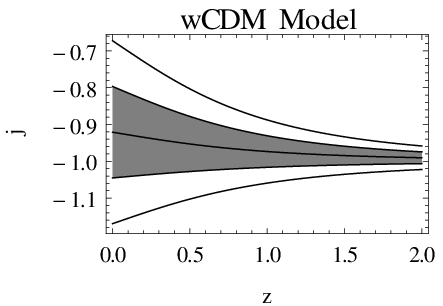}
\end{center}
\caption{{\small Plots of dark energy equation of state parameter ($w_{DE}$) and the cosmological jerk parameter ($j(z)$) parameter as functions of redshift $z$ for $w_{eff}(z)$ model (upper panels) and $w$CDM model (lower panels). The 1$\sigma$ and  2$\sigma$ confidence regions along with the central black line representing the corresponding the best fit curves obtained from the analysis combining the SNe, OHD, BAO and CMB shift parameter data are presented.}}
\label{wDEjzplotch3}
\end{figure}
%%%%%%%%%%%%%%%%%%%%%%

 Figure \ref{weffzqzplotch3} reveals the fact that the effective equation of state ($w_{eff}$) and deceleration parameter ($q$) evolve in very similar way for both the models. For the proposed model, the deceleration parameter shows a signature flip in between the redshift value $0.6$ to $0.8$, which is well consistent with the analysis of observational data by Farooq and Ratra \citep{farooqratra}. Similar behaviour has been obtained for $w$CDM model also.

\par Figure \ref{wDEjzplotch3} shows the plots of dark energy energy equation of state parameter $w_{DE}$ and cosmic jerk parameter $j$ for both the models. The jerk parameter $j$, which is the dimensionless representation of the 3rd order time derivative of the scale factor $a(t)$, is defined as
\begin{equation}
j=-\frac{1}{aH^3}\frac{d^3a}{dt^3}.
\label{jerkch3}
\end{equation}
It is sometimes defined without the negative sign. In the present work, this convention has been used similar to that in reference \citep{zhaijerk}. The jerk parameter can also be expressed in terms of Hubble parameter and its derivative with respect to redshift as,
\begin{equation}
j(z)=-1+(1+z)\frac{(h^2)'}{h^2}-\frac{1}{2}(1+z)^2\frac{(h^2)''}{h^2},
\end{equation}
and for the present model, the expression is
\begin{equation}
j(z)=-1-\frac{3\alpha(n-3)(1+z)^n}{2(1+\alpha(1+z)^n)}+\frac{3\alpha^2(n-3)(1+z)^{2n}}{2(1+\alpha(1+z)^n)^2}.
\label{presentjerkch3}
\end{equation}

 The jerk parameter is also important to understand the deviation of the model from $\Lambda$CDM as for a universe with cosmological constant and cold dark matter, the value of jerk parameter is always $-1$. The dark energy equation of state remains almost flat and shows the preference toward the non-phantom nature of dark energy for the reconstructed $w_{eff}$ model (upper left panel of figure \ref{wDEjzplotch3}). That means its behaviour is very similar to that of $w$CDM model. The plots of $w_{DE}(z)$ and cosmological jerk $j(z)$ show that tighter constraints on their present values are obtained for the reconstructed model than the $w$CDM. 
\par For the reconstructed $w_{eff}$ model, plots (upper panels of figure \ref{wDEjzplotch3}) show that the dark energy equation of state parameter $w_{DE}(z)$ is better constrained at low redshift but the jerk parameter $j(z)$ is better constrained at high redshift. The plots also show that the best fit value of $w_{DE}(z)$ has higher deviation from $-1$ at high redshift and on the other hand the best fit value of $j(z)$ has a higher deviation from the corresponding $\Lambda$CDM value at low redshift. It also indicates that the reconstructed model allows a wide variation of the value of $w_{DE}(z)$ at high redshift, but the value of the jerk parameter $j(z)$ is not allowed to have a wide variation at high redshift. The plot of jerk parameter $j(z)$ for the $w$CDM model (lower right panel of figure \ref{wDEjzplotch3}) also shows a similar behaviour. Actually the dark energy equation of state or the deceleration parameter are highly sensitive to the matter dark energy ratio through the parameter $\alpha$ and thus they are better constrained at low redshift. But the jerk parameter is more sensitive to the existence of the early matter dominated era through the parameter $n$ (equation (\ref{presentjerkch3})) and hence it is better constrained at high redshift.

\par For statistical comparison of the $w_{eff}(z)$ model to the $w$CDM model, two model selection criterion  have been invoked, the Akaike information criterion (AIC) and the Bayesian Information Criterion (BIC). These have been discussed in details in section \ref{beiv}. For the $w_{eff}(z)$ model in comparison with $w$CDM, the  $\Delta$AIC and  $\Delta$BIC vales are
\begin{equation}
\Delta AIC=\chi^2_{min}(w_{eff}(z))-\chi^2_{min}(wCDM)=-0.03,
\label{deltaAICch3}
\end{equation}
and
\begin{equation}
\Delta BIC=\chi^2_{min}(w_{eff}(z))-\chi^2_{min}(wCDM)=-0.03.
\label{deltaBICch3}
\end{equation}
Here $\Delta$AIC and $\Delta$BIC are equal as both the models have two free parameters and the number of data points used of the statistical analysis are same for both the models. The reconstruction of the parametric effective equation of state parameter $w_{eff}$, which is independent of any prior assumption regarding the nature of dark energy, is highly consistent with the  dynamical dark energy model, namely the $w$CDM model.

\section{Representation on $(q_0,j_0)$ parameter space}
\label{comwCDMch3}

For a direct comparison between these two models it would be convenient to look at them through the same parameter space. Now the present value of deceleration parameter $q_0$ and present value of the jerk parameter $j_0$ can be used as the parameters replacing the corresponding model parameters. The $q_0$ and $j_0$ can be obtained from equations (\ref{presentdecelch3}) and (\ref{presentjerkch3}) respectively, as 
\begin{equation}
q_0=-1+\frac{3\alpha}{2(1+\alpha)},
\label{q0ch3}
\end{equation}
and 
\begin{equation}
j_0=-1-\frac{3\alpha(n-3)}{2(1+\alpha)}+\frac{3\alpha^2(n-3)}{2(1+\alpha)^2}.
\label{j0ch3}
\end{equation}
From these two equations, namely equation (\ref{q0ch3}) and (\ref{j0ch3}), the model parameter $\alpha$ and $n$ can be expressed in terms of $q_0$ and $j_0$.  Substituting those expressions of $\alpha$ and $n$ in equation (\ref{hubbleparamch3}), $h^2(z)$ for the reconstructed model can be written in terms of $q_0$ and $j_0$ as,
\begin{equation}
h^2(z)=\Bigg(\frac{(1-2q_0)}{3}+\frac{2(1+q_0)}{3}(1+z)^{\frac{3(1+j_0)+3(1+q_0)(2q_0-1)}{(1+q_0)(2q_0-1)}}\Bigg)^{\frac{(1+q_0)(2q_0-1)}{(1+j_0)+(1+q_0)(2q_0-1)}}.
\label{HubbleWeffq0j0ch3}
\end{equation}
In the same way, the Hubble parameter for $w$CDM model can be expressed in terms of parameters $q_0$ and $j_0$ as
\begin{equation} 
h^2(z)=\Bigg(1-\frac{(1-2q_0)^2}{3(1-2q_0)-2(1+j_0)}\Bigg)(1+z)^3+\Bigg(\frac{(1-2q_0)^2}{3(1-2q_0)-2(1+j_0)}\Bigg)(1+z)^{\frac{2(1+j_0)}{3-2(1+q_0)}}.
\label{HubblewCDMq0j0ch3}
\end{equation}

A similar statistical analysis has been carried out to estimates the values of the kinematical parameters $q_0$ and $j_0$ for both the models. Different model selection criteria and the Bayesian evidence are obviously important to judge the consistency between the models. But the representation of the models on the same parameter space is important to understand whether the models allow any common region on the parameter space. It also shows whether the nature of correlation between the parameters are same for both the models.  

\par
\begin{figure}[H]
\begin{center}
\includegraphics[angle=0, width=0.28\textwidth]{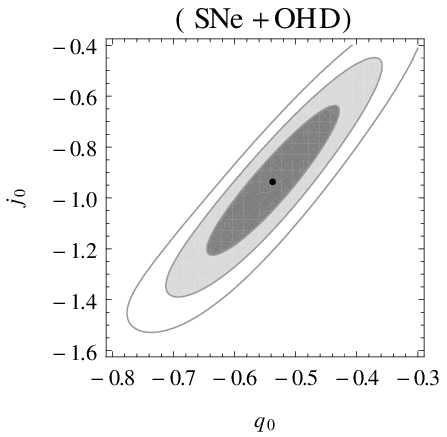}
\includegraphics[angle=0, width=0.28\textwidth]{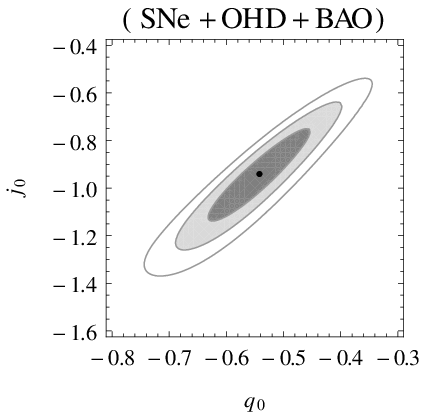}
\includegraphics[angle=0, width=0.28\textwidth]{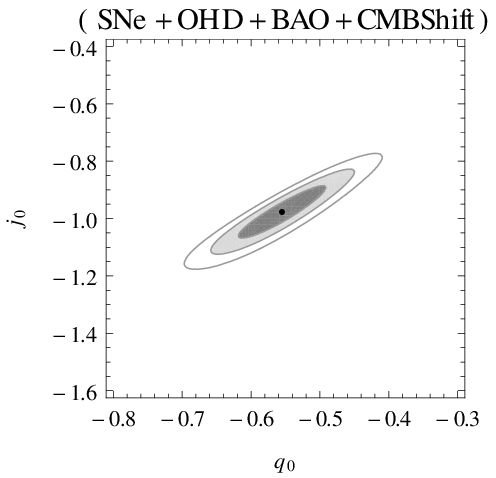}
\end{center}
\caption{{\small Confidence contours on the 2D parameter space ($q_0$,$j_0$) for the $w_{eff}(z)$ model obtained for different combinations of the data sets. 1$\sigma$, 2$\sigma$ and 3$\sigma$ confidence regions are presented from inner to outer portion and the central black dots represent the corresponding  best fit points. The left panel shows the confidence contours obtained for SNe+OHD, the middle panel shows the confidence contours obtained for SNe+OHD+BAO and the right panel shows confidence contours for SNe+OHD+BAO+ CMBShift.}}
\label{weffq0j0Contourplotsch3}
\end{figure}

\begin{figure}[H]
\begin{center}
\includegraphics[angle=0, width=0.28\textwidth]{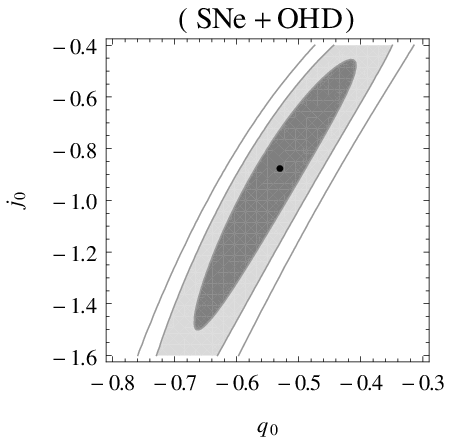}
\includegraphics[angle=0, width=0.28\textwidth]{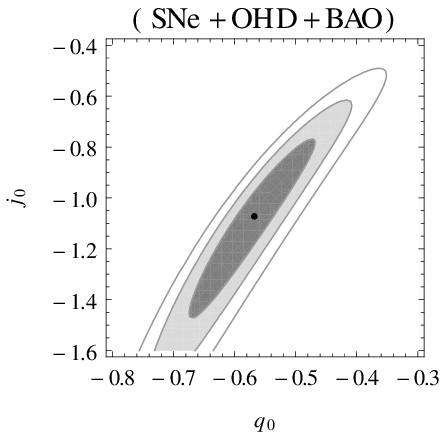}
\includegraphics[angle=0, width=0.28\textwidth]{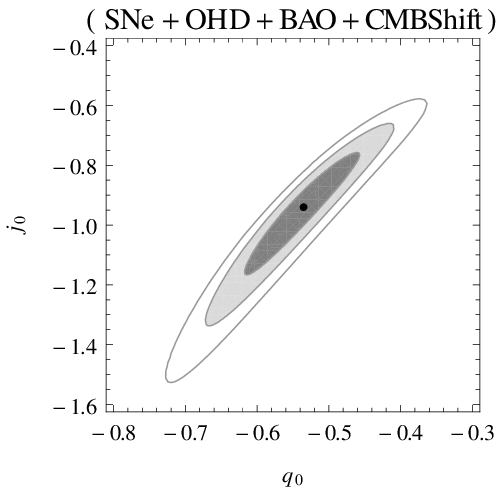}
\end{center}
\caption{{\small Confidence contours on the 2D parameter space ($q_0$,$j_0$) for the $w$CDM model obtained for different combinations of the data sets. 1$\sigma$, 2$\sigma$ and 3$\sigma$ confidence regions are presented from inner to outer portion and the central black dots represent the corresponding  best fit points. The left panel shows the confidence contours obtained for SNe+OHD, the middle panel shows the confidence contours obtained for SNe+OHD+BAO and the right panel shows confidence contours for SNe+OHD+BAO+ CMBShift.}}
\label{wCDMq0j0Contourplotsch3}
\end{figure}

\begin{figure}[H]
\begin{center}
\includegraphics[angle=0, width=0.32\textwidth]{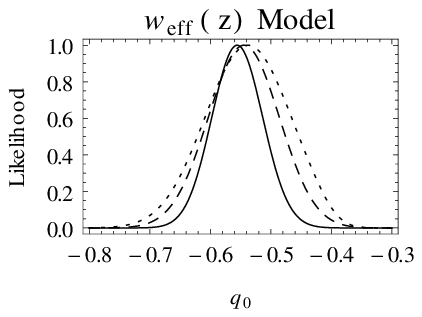}
\includegraphics[angle=0, width=0.32\textwidth]{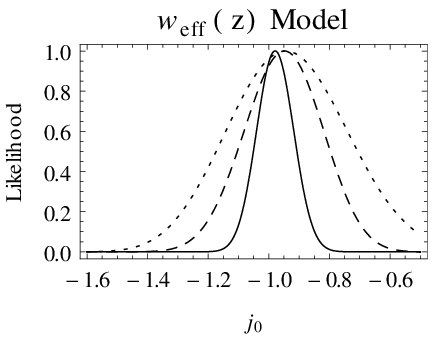}\\
\includegraphics[angle=0, width=0.32\textwidth]{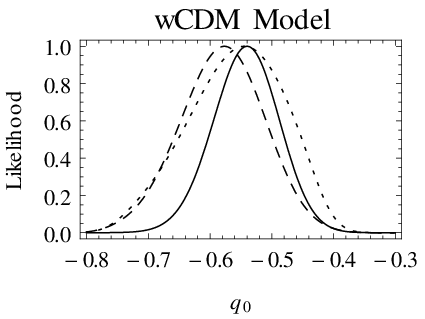}
\includegraphics[angle=0, width=0.32\textwidth]{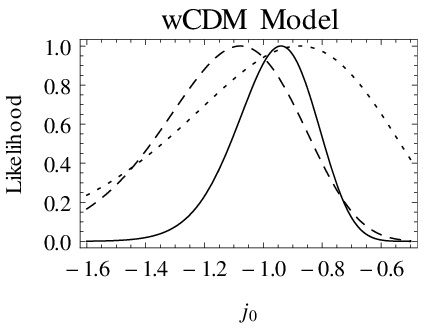}
\end{center}
\caption{{\small The marginalised likelihood as functions of $q_0$ and $j_0$ for the $w_{eff}(z)$ model (upper panels) and for the $w$CDM model (lower panels) obtained for different combinations of the datasets. The dotted curves  represent the likelihood obtained for SNe+OHD, the dashed curves are obtained for SNe+OHD+BAO and the solid curves show the likelihood for  SNe+OHD+BAO+CMBShift.}}
\label{q0j0Likelihoodch3}
\end{figure}

 Table \ref{tableweffwCDMq0j0ch3} presents the results of statistical analysis, obtained from the statistical analysis combining SNe, OHD, BAO and CMB shift parameter data, for the reconstructed $w_{eff}$ model and $w$CDM models on ($q_0$,$j_0$) parameter space. Figure \ref{weffq0j0Contourplotsch3} and figure \ref{wCDMq0j0Contourplotsch3} show the 2D confidence contours on ($q_0$, $j_0$) parameter space.  Figure \ref{q0j0Likelihoodch3} shows the marginalised likelihood with $q_0$ and $j_0$ as the arguments for both the models. 

%%%%%%%%%%%%%%%%%%%%%%%%%
\begin{table}[h!]
\caption{{\small Results of statistical analysis of the $w_{eff}(z)$ and $wCDM$ models with $q_0$ and $j_0$ as the parameters using the combination of OHD, SNe, BAO and CMBShift data}}
\begin{center}
\resizebox{0.63\textwidth}{!}{  
\begin{tabular}{ c |c |c c c } 
 \hline
 \hline
 Model  & $\chi^2_{min}/d.o.f.$ & $q_0$ & $j_0$ \\ 

 \hline
  $w_{eff}(z)$ & 49.45/52 & $-0.555\pm0.030$ & $-0.977\pm0.043$\\ 
 \hline
  $w$CDM & 48.24/52 & $-0.535\pm0.037$ &  $-0.940\pm0.094$\\ 
 \hline
 \hline
\end{tabular}
}
\end{center}
\label{tableweffwCDMq0j0ch3}
\end{table}
%%%%%%%%%%%%%%%%%%%%%%%%%%%%%%%%%%%

%%%%%%%%%%%%%%%%%%%%%
%%%%%%%%%%%%%%%%%%%%%%
%%%%%%%%%%%%%%%%%%%%%
%%%%%%%%%%%%%%%%%%%%%%

%%%%%%%%%%%%%%%%%%%%%
%%%%%%%%%%%%%%%%%%%%%%

Though the best fit value of the parameters $q_0$ and $j_0$ obtained for both the models are close enough, the present $w_{eff}(z)$ model minimizes the uncertainty of the parameter values. This is also consistent with the results obtained from figure \ref{wDEjzplotch3}.

The representation of the reconstructed model and the $w$CDM model on the ($q_0$,$j_0$) parameter space (figure \ref{weffq0j0Contourplotsch3} and figure \ref{wCDMq0j0Contourplotsch3}) clearly show that the confidence contours of the models are consistent with each other and the correlations between the parameters are very much similar for both the models. It is also clear that substantially tighter constraint on the present value of cosmological jerk parameter ($j_0$) has been achieved for the model reconstructed in the present work than the $w$CDM dark energy model.

\section{Discussion}
\label{disch3}

In the present work, a model is built up by considering a parametric form of the effective equation of state parameter. The constraints on the model parameters of the $w$CDM model have also been obtained using the same sets of data. The idea is to draw a direct comparison between these two models. Different model selection criteria clearly indicate the consistency between the $w$CDM and the model reconstructed in the present work.

\par The model parameter $n$ is an indicator of deviation of the model from cosmological constant. Now the value of $n$ obtained from the likelihood analysis is very close to $3$ which indicates that the reconstructed $w_{eff}$ is in close proximity of $\Lambda$CDM. As already mentioned that the the contour plots along with the best fit points (figure \ref{weffcontourch3}) and the likelihood plots (upper right panel of figure \ref{wefflikelihoodch3}) show that the amount of deviations of the model from $\Lambda$CDM vary for different combinations of  the data sets. The addition of CMB shift brings the best fit value of the model parameter $n$ very close to the corresponding $\Lambda$CDM value and much tighter constraints have also been achieved.   In the series expansion expansion of $h^2(z)$, there is a term evolving as $(1+z)^3$. This is equivalent to the matter density and the constant coefficient of this term is the present matter density parameter ($\Omega_{m0}$). For the reconstructed $w_{eff}(z)$ model, the value of the model parameters obtained are $\alpha=0.444\pm0.042$ and $n=2.907\pm0.136$ at 1$\sigma$ confidence level. As the parameter $\alpha$ is connected to the matter density parameter ($\Omega_{m0}$), the value of $\Omega_{m0}$ obtained for this model is $0.296\pm0.011$, which is consistent with the value obtained from the same analysis for $w$CDM model.

\par A recent analysis of $\Lambda$CDM and $w$CDM model by Xia, Li and Zhang \citep{xiaLCDMwCDM}, using the CMB temperature anisotropy and polarization data along with other non-CMB data, estimates the value of the matter density parameter $\Omega_{m0}=0.293\pm0.013$ at 1$\sigma$ confidence level for $\Lambda$CDM and $\Omega_{m0}=0.270\pm0.014$ at 1$\sigma$ confidence level for $w$CDM. Hence the value of the matter density parameter obtained in the present work is very close to the value obtained for $\Lambda$CDM  by Xia, Li and Zhang \citep{xiaLCDMwCDM}. A recent analysis by Hazra {\it et al} \citep{hazra} has presented the analysis of different parameterizations of dark energy using various recent observational data sets. The parameter values obtained are $\Omega_{m0}=0.307_{-0.046}^{+0.041}$, $w_{DE}(z=0)=-1.005_{-0.15}^{+0.17}$ for Chevallier-Polarski-Linder (CPL) parametrization \citep{cplparam}, $\Omega_{m0}=0.283_{-0.030}^{+0.028}$, $w_{DE}(z=0)=-1.14_{-0.09}^{+0.08}$ for Scherrer and Sen (SS) parameterization \citep{scherrersen} and $\Omega_{m0}=0.32_{-0.012}^{+0.013}$, $w_{DE}(z=0)=-0.95_{non-phantom}^{+0.007}$ for generalized Chaplygin gas (GCG) model \citep{gcgbento}. It is clear that the CPL parameterization is in good agreement with the presently reconstructed $w_{eff}$ model. But the SS parameterization has a preference towards a lower value of the dark energy equation of state though the value of matter density parameter is within 1$\sigma$ confidence region of the reconstructed model. The non-phantom prior assumption of GCG parameterization is in good agreement with the present $w_{eff}$ model, though the present model also allows the phantom behaviour within 1$\sigma$ confidence level. The GCG has a clear preference towards a higher value of the matter density.

\par The plot of deceleration parameter $q(z)$ (upper left panel of figure \ref{weffzqzplotch3}) shows that the reconstructed model successfully generates the late time acceleration along with the decelerated expansion phase which prevailed before the accelerated expansion phase. The redshift of transition from decelerated to accelerated phase of expansion lies in between the redshift range 0.6 to 0.8 which is consistent with the recent analysis by Farooq and Ratra \cite{farooqratra}.

\par The equation of state parameter of dark energy achieved for the model presented here remains almost constant. The nature of effective equation of state ($w_{eff}(z)$) and the deceleration parameter $q(z)$ are also very much similar to that of $w$CDM model (figure \ref{weffzqzplotch3}). Figure \ref{wDEjzplotch3} presents the plots of $w_{DE}$ and $j(z)$ for the reconstructed $w_{eff}$ model and the $w$CMD model. It is clear from the plots that the reconstructed model puts tighter constraints on the present values of dark energy equation of state parameter ($w_{DE}(z=0)$) and cosmological jerk ($j_0$) than the $w$CDM model. Another interesting point is that the uncertainties associated to the value  of $w_{DE}(z)$ and $j(z)$ vary according to its deviations from the $\Lambda$CDM. A higher deviation of the best fit value from the corresponding $\Lambda$CDM value increases the associated uncertainty. It is also apparently clear from  figure \ref{wDEjzplotch3} that  the reconstructed $w_{eff}$ model allows a wide variation for the value of dark energy equation of state parameter $w_{DE}(z)$ at high redshift, but the value of the jerk parameter $j(z)$ at high redshift  is not allowed to have a wide variation.    

\par It deserves mention that the systematics of supernova data has been taken into account as it might have its signature on the results. The effect of redshift dependence of colour-luminosity parameter of distance modulus measurement has been discussed by Wang and Wang \cite{wangwangSNeSys}. There are some recent discussion on the impact of supernova systematics which can also be referred to in this context \cite{rubinSNeSys,shaferSNeSys}.

\chapter{Reconstruction of the cosmological jerk parameter}

% **************************** Define Graphics Path **************************
\ifpdf
    \graphicspath{{Chapter4/Figs/Raster/}{Chapter4/Figs/PDF/}{Chapter4/Figs/}}
\else
    \graphicspath{{Chapter4/Figs/Vector/}{Chapter4/Figs/}}
\fi

\section{Introduction}
This chapter is devoted to the discussion of a kinematic reconstruction of cosmological models. Reconstruction of a dark energy (DE), that drives the current acceleration of the universe, normally involves finding the dark energy equation of state parameter $w_{DE}$, given by $ w_{DE}=\frac{p_{DE}}{\rho_{DE}}$, where $p_{\phi}$ and ${\rho}_{\phi}$ represent the contribution to the  pressure and the density respectively by the DE \cite{sainireconst,vsahniaasrarore}. There are only a few attempts to find a suitable model by a reconstruction of the kinematical quantities like the deceleration parameter or higher order derivatives of the scale factor. In spite of its being the natural choice amongst the kinematical quantities, as discussed in the introduction, the jerk parameter $j$, given by $j =-\frac{1}{a H^{3}}\frac{d^{3}a}{dt^{3}}$, has only a limited application until now in the context of the accelerated expansion. Cosmological jerk parameter has been used as a diagnostic of dark energy models by Sahni et al. \cite{sahnistatefinder} and Alam et al. \cite{alamstatefinder}. The jerk parameter and a combination of jerk and deceleration parameter together have been identified as the statefinder parameter in these two investigations. Visser \cite{visser} initiated a serious discussion on the jerk parameter, albeit from a different motivation. The idea was to look at the equation of state of the cosmic fluid as a Taylor expansion in terms of a background density. Analytical expressions for kinematical parameters, constructed by involving higher order derivatives of the scale factor in various forms of matter in a Friedmann universe, had been presented by Dunajski and Gibbons \cite{dunajskigibbons}. 

\par A reconstruction of the dark energy equation of state through a parametrization of the jerk parameter has been very recently given by Luongo \cite{luongojerk}.  A systematic study of jerk as the way towards building up a model for the accelerated expansion was given by Rapetti {\it et al}\cite{rapettijerk}. In an  exhaustive recent work by Zhai {\it et al} \cite{zhaijerk}, a reconstruction of $j$ has been attempted. For four different forms of $j=j(z)$, they fixed the parameters in $j$ by using observational Hubble parameter data (OHD) and supernovae (SNe) data. The present chapter is based on the reconstruction of cosmological jerk parameter from diverse observational data sets. 

\par It should be clearly mentioned that the equation of state parameter $w_{DE}$, the energy density of the dark energy, the potential of the quintessence field etc. are all part of the theoretical input, and hence constitute the fundamental ingredients of the model. The deceleration parameter, the jerk etc. are kinematical quantities, and thus are the outcome of the solution of the system of equations of the model. So no wonder that a reconstruction of $w_{DE}$ holds the centre-stage in attempts towards building up a model for the accelerating universe. However, it should be mentioned that the basic advantage of this kind of reconstruction through kinematical quantities neither assumes any theory of gravity (like GR, $f(R)$ gravity, scalar-tensor theory etc.) nor does it assume a given matter distribution like a quintessence field or any other exotic matter through any equation of state to start with. Thus, this reconstruction may lead to some understanding of the basic matter distribution and the possible interaction amongst themselves without any a priori assumptions on them.

\par The motivation of the present work is to reconstruct a dark energy model through the jerk parameter $j$. If $j$ is known as $j(t)$ or $j(z)$, the third order differential equation for the scale factor is known and hence one can find the evolution, at least in principle. It is wellknown that the $\Lambda$CDM model does very well in explaining the present acceleration of the universe but fails to match the theoretically predicted value of $\Lambda$. Thus the attempts to build up dark energy models hover around finding one which in the limit of $z=0$ (the present epoch) resembles a $\Lambda$CDM model. The work by Zhai {\it et al} \cite{zhaijerk} is also along that line. For a $\Lambda$CDM model, one has a constant $j$ as $j=-1$. Zhai  {\it et al} wrote their ansatz in such a way that the present value of $j$ is actually $-1$, i.e., $j(z=0) = -1$. All the four different forms of $j=j(z)$, they wrote, have this feature. 

\par The present work is more general in two different ways. First, the functional form of $j=j(z)$ in this work is more relaxed in the sense that its present value depends on a parameter to be fixed by observations, and is allowed to have a value widely different from $-1$ at $z=0$, if so required by the data sets. In this work, different ansatzs for $j=j(z)$ are taken, all of which are quite free to take values very much different from $-1$. The second point is related to sophistication, rather than principle, in the sense that where Zhai {\it et al} used a combination of two data sets namely the OHD and the SNe, we have a combination of four data sets. In addition to the two used by  Zhai {\it et al}, the baryon acoustic oscillation (BAO) data and the very recent CMB shift parameter (CMBShift) data have also been incorporated in the present work. As a result, the parameters of the theory are constrained to narrower limits. The results obtained by the present work also show that the observations very strongly indicate an inclination towards a $\Lambda$CDM behaviour of the present distribution of matter with a marginal preference towards a non-phantom behaviour. \\

\section{Kinematical quantities} 
\label{kich4}
                                         
The kinematical quantities in cosmology are constructed from the {\it scale factor} $a$ and its time derivatives.  The Hubble parameter ($H$) is defined as $H(t)=\frac{\dot{a}}{a}$, where a dot indicates derivative with respect to $t$. Similarly the deceleration parameter $q$ and the jerk parameter $j$ are defined as,
\begin{equation}
q(t)=-\frac{1}{aH^2}\frac{d^2a}{dt^2},
\end{equation}
and
\begin{equation}
j(t)=-\frac{1}{aH^3}\frac{d^3a}{dt^3}.
\end{equation}

One can extend this chain of derivatives, such as the fourth order derivative, called the snap parameter $s$ can be defined as $s = \frac{1}{a H^{4}}\frac{d^{4} a}{d t^{4}}$ and so on \cite{visser}. We shall however, restrict to the third derivative, namely the jerk parameter, as the evolution of $q$ is of physical interest now. It is to be noted that in defining $j$, there are two conventions of using a positive or a negative sign. We have adopted the convention as used by Zhai {\it et al} \cite{zhaijerk} and not the one used by Visser \cite{visser} and Luongo \cite{luongojerk}. This preference is only because we shall compare our results with that of Zhai {\it et al} \cite{zhaijerk}.

\par Hubble parameter $H(t)$ can also be written as a function of the redshift $z$, as $H=H(z)$. Redshift $z$ is defined as $(1+z)=\frac{a_0}{a(t)}$, where $a_0$ is the present value of the scale factor. The deceleration parameter $q(z)$ and the jerk parameter $j(z)$ can be written in terms of $H(z)$ with $z$ as the argument as

\begin{equation}
q(z)=-1+\frac{1}{2}(1+z)\frac{[H^2(z)]'}{H^2(z)},
\label{decelerationparameterch4}
\end{equation}
\begin{equation}
j(z)=-1+(1+z)\frac{[H^2(z)]'}{H^2(z)}-\frac{1}{2}(1+z)^2\frac{[H^2(z)]''}{H^2(z)},
\label{jerkparameterch4}
\end{equation}
where a prime denotes the derivative with respect  to the  redshift $z$.

\section{Reconstruction of jerk parameter}
\label{reconstch4}

Einstein's field equations for a spatially flat homogeneous and isotropic universe are,
\begin{equation}
3H^2=8\pi G(\rho_m+\rho_{DE}),
\label{friedmann1ch4}
\end{equation}
\begin{equation}
2\dot{H}+3H^2=-8\pi Gp_{DE},
\label{friedmann2ch4}
\end{equation}
where $\rho_m$ is the density of the pressureless dust matter, $\rho_{DE}$ and $p_{DE}$ are the contribution of the dark energy to the energy density and pressure sector. An overhead dot represents the differentiation with respect to the time. 

\par For $\Lambda$CDM cosmology, jerk parameter (defined in \ref{jerkparameterch4}) is a constant with the value $j_{\Lambda CDM}=-1$. We follow the parametric reconstruction of $j(z)$ in the similar line as discussed by Zhai {\it et al} \cite{zhaijerk}. The major difference, as discussed before, is that we do not restrict to $j=-1$ for $z=0$. So we do not assume a $\Lambda$CDM for the present universe {\it a priori}, but rather allow the model to behave in a more general way. The present value of $j$ is allowed to be fixed by the observational data.  We write the jerk parameter as
\begin{equation}
j(z)=-1+j_1\frac{f(z)}{h^2(z)},
\end{equation} 
where $j_1$ is a constant, $h(z)=\frac{H(z)}{H_0}$ is the Hubble parameter scaled by its present value and $f(z)$ is an analytic function of $z$. Four ansatz for $j(z)$ have been chosen in the present work, which will be discussed separately. Here $j_1$ is the model parameter to be fixed by observational data. Model I is the one where the evolution of $j$ is taken care of solely by $h^{2}(z)$. For the other three, the redshift $z$ also contributes explicitly and not through $h^2(z)$ alone. The four ansatz chosen are given below,
\begin{equation}
Model ~I.~~~~   j(z)=-1+j_1\frac{1}{h^2(z)},
\label{jmodel1ch4}
\end{equation}
\begin{equation}
Model ~II.~~~~~  j(z)=-1+j_1\frac{(1+z)}{h^2(z)},
\label{jmodel2ch4}
\end{equation}
\begin{equation}
Model ~III.~~~~~  j(z)=-1+j_1\frac{(1+z)^2}{h^2(z)},
\label{jmodel3ch4}
\end{equation}
\begin{equation}
~~Model ~IV.~~~~~   j(z)=-1+j_1\frac{1}{(1+z)h^2(z)}.
\label{jmodel4ch4}
\end{equation}

By substituting these expressions in the definition of $j$, given by equation (\ref{jerkparameterch4}), one can get the solutions for $h^{2}$ as, 
\begin{equation}
Model~I. ~~~~h^2(z)=\frac{H^2(z)}{H_0^2}=c_1(1+z)^3+c_2+\frac{2}{3}j_1ln(1+z),
\label{jmodelh1ch4}
\end{equation}  
\begin{equation}
Model~II.~~~~h^2(z)=\frac{H^2(z)}{H_0^2}=c_1(1+z)^3+c_2+j_1(1+z),
\label{jmodelh2ch4}
\end{equation}
\begin{equation}
Model~III.~~~~h^2(z)=\frac{H^2(z)}{H_0^2}=c_1(1+z)^3+c_2+j_1(1+z)^2,
\label{jmodelh3ch4}
\end{equation}
\begin{equation}
Model~IV.~~~~h^2(z)=\frac{H^2(z)}{H_0^2}=c_1(1+z)^3+c_2-j_1\frac{1}{2(1+z)}.
\label{jmodelh4ch4}
\end{equation}
Here $c_1$ and $c_2$ are integration constants. Now from the scaling $h^2(z)=\frac{H^2(z)}{H_0^2}$, one has the constraint $h(z=0)=1$, which connects the constants as
\begin{equation}
Model~I. ~~~~~~c_2=1-c_1,
\label{c2model1ch4}
\end{equation}  
\begin{equation}
Model~II.~~~~~~~~c_2=1-j_1-c_1,
\label{c2model2ch4}
\end{equation}
\begin{equation}
Model~III.~~~~~~~c_2=1-j_1-c_1,
\label{c2model3ch4}
\end{equation}
\begin{equation}
Model~IV.~~~~~~~~c_2=1+\frac{j_1}{2}-c_1.
\label{c2model4ch4}
\end{equation}
Thus each of the models have two independent parameters, $(c_1,j_1)$. It is quite apparent from the expression of $h^2(z)$, through the Einstein's field equation, $3H^{2} = 8\pi G \rho$, that the parameter $c_1$ is equivalent to the matter density parameter $\Omega_{m0}$ which is the ratio of present matter density ($\rho_{m0}$) and present critical density ($\rho_c=3H_0^2/8\pi G$). Equations (\ref{jmodelh2ch4}), (\ref{jmodelh3ch4}) and (\ref{jmodelh4ch4}) show that effectively one has a cold dark matter, a cosmological constant and another component of dark energy of the form $p = w\rho$ where $w$ is a negative constant. For the first choice of the model in equation (\ref{jmodelh1ch4}), the dark energy is not given by such a simple equation of state.

\par The properties of dark energy can also be ascertained to a certain extent from the analytic expressions of the Hubble parameter for the models. As the dust matter is minimally coupled to the dark energy for these models, the dark energy density scaled by critical density can be expressed as
\begin{equation}
\frac{\rho_{DE}}{\rho_c}=h^2(z)-c_1(1+z)^3.
\label{rhoDEch4}
\end{equation}
The contribution of dark energy to the pressure sector can be obtained from equation (\ref{friedmann2ch4}) as 
\begin{equation}
\frac{p_{DE}}{\rho_c}=\frac{2}{3}(1+z)hh'-h^2(z).
\label{pDEch4}  
\end{equation}
Finally the dark energy equation of state parameter can be expressed as 
\begin{equation}
w_{DE}(z)=\frac{\frac{2}{3}(1+z)hh'-h^2}{h^2(z)-c_1(1+z)^3}.
\label{wDEch4}
\end{equation}
The analytical expressions of $w_{DE}$ obtained for the models are 
\begin{equation}
Model~I. ~~~w_{DE}=-1+\frac{\frac{2}{9}j_1}{\frac{2}{3}j_1\ln{(1+z)}+(1-c_1)},
\label{wDEmodel1ch4}
\end{equation}  
\begin{equation}
Model~II.~~~w_{DE}=-1+\frac{\frac{1}{3}j_1(1+z)}{j_1(1+z)+(1-c_1-j_1)},
\label{wDEmodel2ch4}
\end{equation}
\begin{equation}
Model~III.~~~w_{DE}=-1+\frac{\frac{2}{3}j_1(1+z)^2}{j_1(1+z)^2+(1-c_1-j_1)},
\label{wDEmodel3ch4}
\end{equation}
\begin{equation}
Model~IV.~~~w_{DE}=-1+\frac{\frac{1}{6}\frac{j_1}{(1+z)}}{(1-c_1+\frac{1}{2}j_1)-\frac{j_1}{2(1+z)}} .
\label{wDEmodel4ch4}
\end{equation}

\par The statistical analysis have been carried out using various combinations of the data sets. Four different observational data sets have been used for the statistical analysis of the model in the present work. These are the observational Hubble data (OHD), distance modulus data from type Ia supernove (SNe), baryon acoustic oscillation (BAO) data along with the value of acoustic scale at photon electron decoupling and the ratio of comoving sound horizon at decoupling and at drag epoch estimated from CMB radiation power spectrum and the CMB shift parameter (CMBShift) data.

\par The observational Hubble data set (OHD) is obtained from the measurement by different groups. Hubble parameter is measured directly from cosmic chromometres and differential age of galaxies in the redshift range $0<z<1.8$ \cite{simonohd,sternohd,chuangohd,blakeohd,zhangohd,morescoohd}. Measurement of Hubble parameter at $z=2.34$ by Delubac {\it et al} \cite{delubacohd} has also been incorporated in the data set. 

\par The distance modulus ($\mu(z)$) data set from type Ia supernova observations is very widely used one for the analysis of dark energy models. In this work, the 31 binned distance modulus data sample of the recent joint light-curve analysis \cite{sn3} has been utilized.

\par Baryon Acoustic Oscillation data (result of 6dF Galaxy Survey at redshift $z=0.106$ (Beutler {\it et al} \cite{beutlerbao}), and the results of Baryon Oscillation Spectroscopic Survey (BOSS) at redshift $z=0.32$ (BOSS LOWZ) and at redshift z = 0.57 (BOSS CMASS)(Anderson {\it et al} \cite{bossandersonbao})), along with the measurement of {\it comoving sound horizon} at photon decoupling epoch ($z_*=1090.43\pm0.65$) and at photon drag epoch ($z_d=1059.29\pm0.65$) and the estimation of the value of {\it acoustic scale} at decoupling obtained from Planck results \cite{planck2013,wangwangcmb} have been incorporated in the statistical analysis.

\par Cosmic Microwave Background (CMB) data, in  the form of a distance prior, namely the CMB shift parameter, estimated from Planck data in \cite{wangwangcmb}, has also been utilized here.  
 
\par The constraints on the parameters are obtained by the $\chi^2$ minimization or equivalently the maximum likelihood analysis. The statistical techniques regarding the parameter estimation has been discussed in detail in section \ref{statanalysis}. 

\par
\begin{figure}[H]
\begin{center}
\includegraphics[angle=0, width=0.32\textwidth]{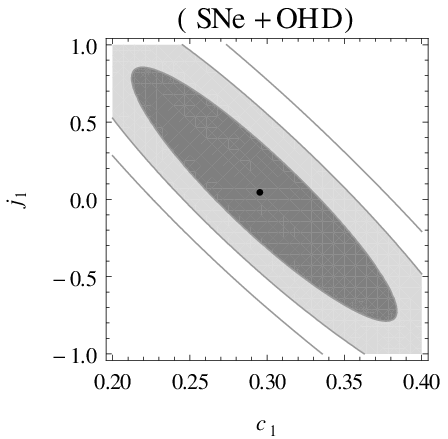}
\includegraphics[angle=0, width=0.32\textwidth]{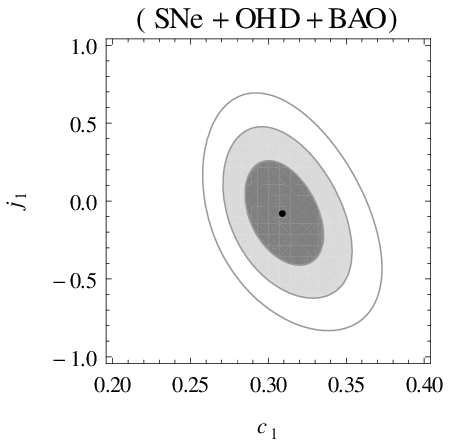}\\
\includegraphics[angle=0, width=0.32\textwidth]{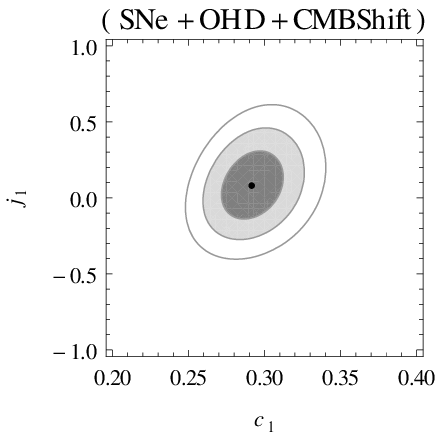}
\includegraphics[angle=0, width=0.32\textwidth]{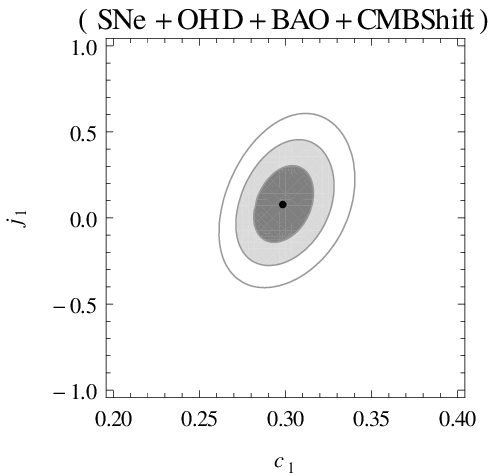}
\end{center}
\caption{{\small The confidence contours on 2D parameter space of Model I. The 1$\sigma$, 2$\sigma$ and 3$\sigma$ confidence contours are presented from inner to outer regions and the central black dots represent the corresponding best fit points. The upper left panel is obtained for (SNe+OHD), upper right panel is for (SNe+OHD+BAO), lower left panel is for (SNe+OHD+CMBShift) and lower right panel is for (SNe+OHD+BAO +CMBShift).}}
\label{jz1contoursch4}
\end{figure}

\begin{figure}[H]
\begin{center}
\includegraphics[angle=0, width=0.32\textwidth]{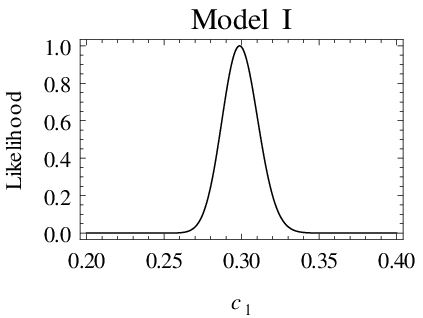}
\includegraphics[angle=0, width=0.32\textwidth]{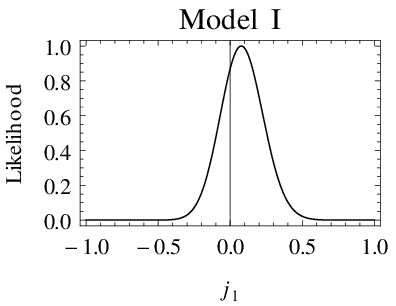}
\end{center}
\caption{{\small The marginalised likelihood functions of Model I obtained for SNe+OHD+BAO +CMBShift.}}
\label{jz1likelihoodch4}
\end{figure}

 Figure \ref{jz1contoursch4} shows the confidence contours on the 2D parameter space ($c_1$,$j_1$) of Model I obtained for various combinations of the data sets. Figure \ref{jz1likelihoodch4} presents the marginalised likelihood as functions of the model parameters $c_1$ and $j_1$ for Model I obtained for SNe+OHD+BAO +CMBShift data. The likelihood functions are well fitted to Gaussian distribution function with the best-fit parameter values $c_1=0.298\pm0.010$ and $j_1=0.078\pm0.140$. Table \ref{tablejzM1ch4} presents the results of statistical analysis for Model I. It is clear from the results that the addition of CMB shift parameter data leads to a substantial improvement of the parameter constraints.

\par
\begin{figure}[H]
\begin{center}
\includegraphics[angle=0, width=0.32\textwidth]{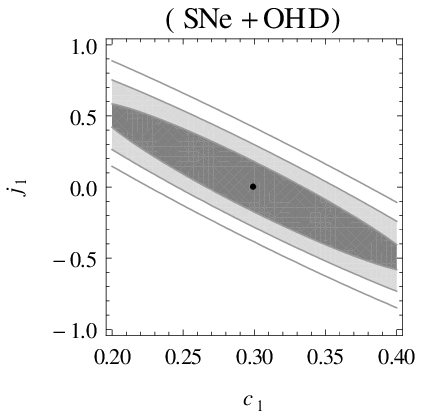}
\includegraphics[angle=0, width=0.32\textwidth]{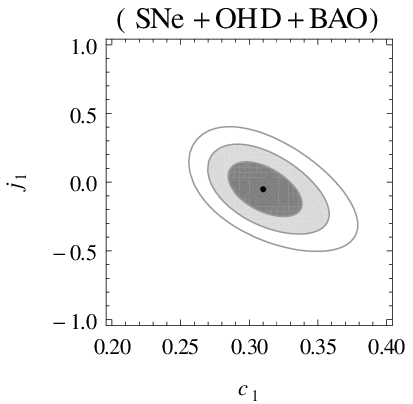}\\
\includegraphics[angle=0, width=0.32\textwidth]{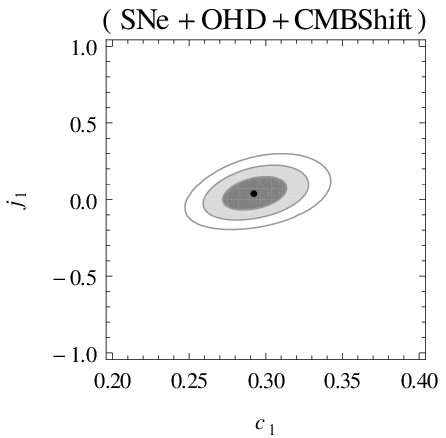}
\includegraphics[angle=0, width=0.32\textwidth]{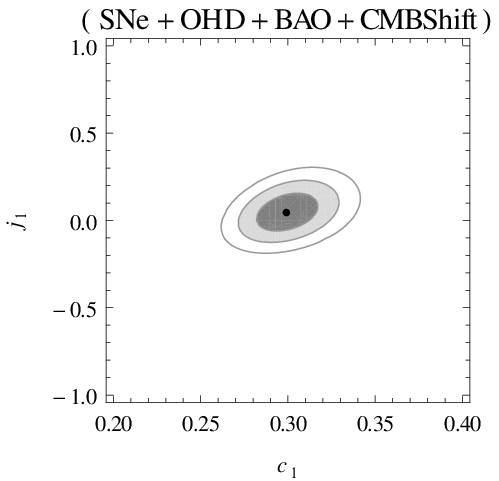}
\end{center}
\caption{{\small The confidence contours on 2D parameter space of Model II. The 1$\sigma$, 2$\sigma$ and 3$\sigma$ confidence contours are presented from inner to outer regions and the central black dots represent the corresponding best fit points. The upper left panel is obtained for (SNe+OHD), upper right panel is for (SNe+OHD+BAO), lower left panel is for (SNe+OHD+CMBShift) and lower right panel is for (SNe+OHD+BAO+CMBShift).}}
\label{jz2contoursch4}
\end{figure}

\begin{figure}[H]
\begin{center}
\includegraphics[angle=0, width=0.32\textwidth]{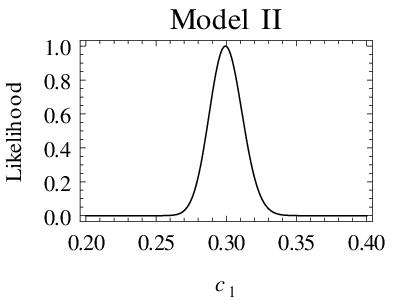}
\includegraphics[angle=0, width=0.32\textwidth]{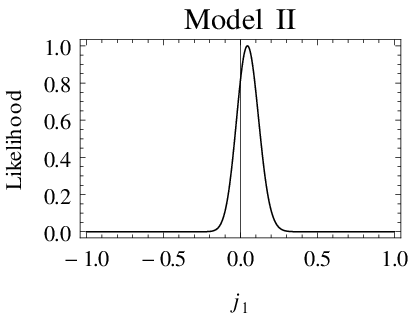}
\end{center}
\caption{{\small  The marginalised likelihood functions of Model II obtained for SNe+OHD+BAO +CMBShift.}}
\label{jz2likelihoodch4}
\end{figure}

\begin{figure}[H]
\begin{center}
\includegraphics[angle=0, width=0.32\textwidth]{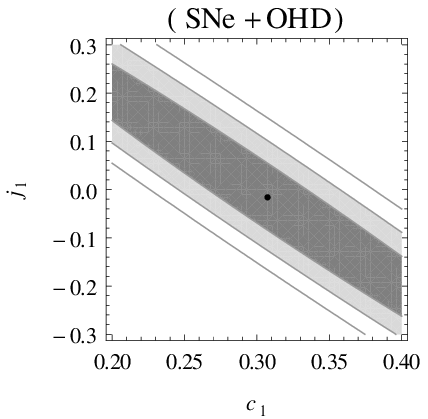}
\includegraphics[angle=0, width=0.32\textwidth]{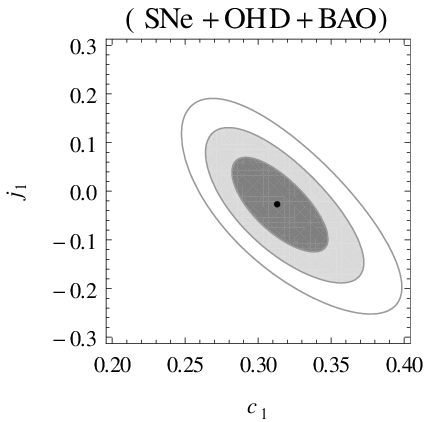}\\
\includegraphics[angle=0, width=0.32\textwidth]{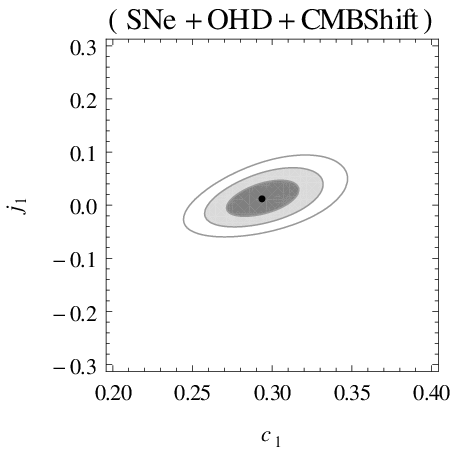}
\includegraphics[angle=0, width=0.32\textwidth]{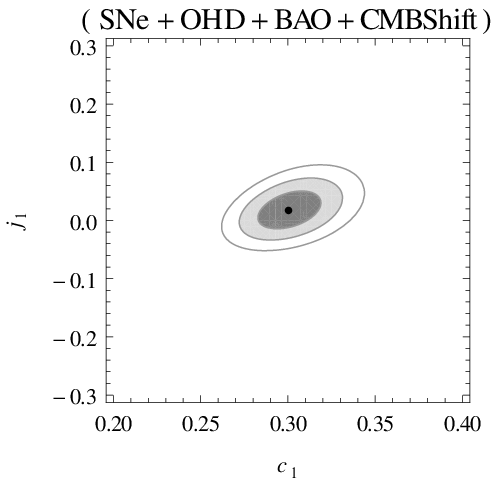}
\end{center}
\caption{{\small The confidence contours on 2D parameter space of Model III. The 1$\sigma$, 2$\sigma$ and 3$\sigma$ confidence contours are presented from inner to outer regions and the central black dots represent the corresponding best fit points. The upper left panel is obtained for (SNe+OHD), upper right panel is for (SNe+OHD+BAO), lower left panel is for (SNe+OHD+CMBShift) and lower right panel is for (SNe+OHD+BAO+CMBShift).}}
\label{jz3contoursch4}
\end{figure}

\begin{figure}[H]
\begin{center}
\includegraphics[angle=0, width=0.32\textwidth]{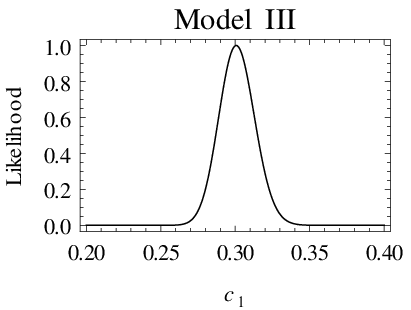}
\includegraphics[angle=0, width=0.32\textwidth]{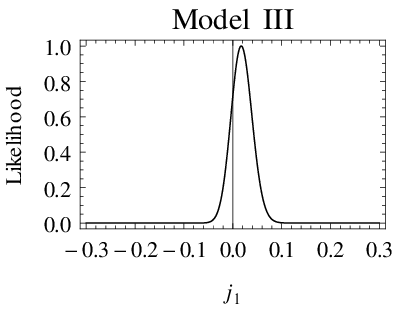}
\end{center}
\caption{{\small The marginalised likelihood functions of Model III obtained for SNe+OHD+BAO +CMBShift.}}
\label{jz3likelihoodch4}
\end{figure}

\begin{figure}[H]
\begin{center}
\includegraphics[angle=0, width=0.32\textwidth]{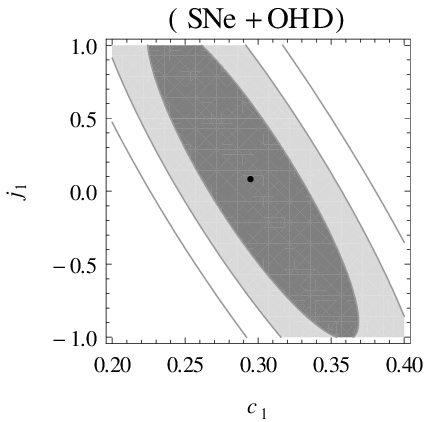}
\includegraphics[angle=0, width=0.32\textwidth]{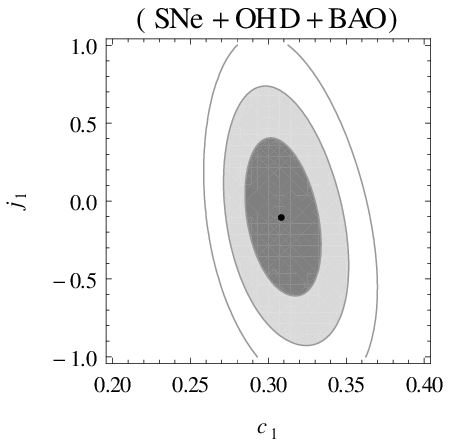}\\
\includegraphics[angle=0, width=0.32\textwidth]{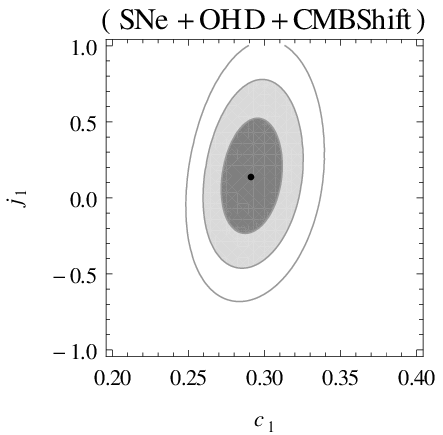}
\includegraphics[angle=0, width=0.32\textwidth]{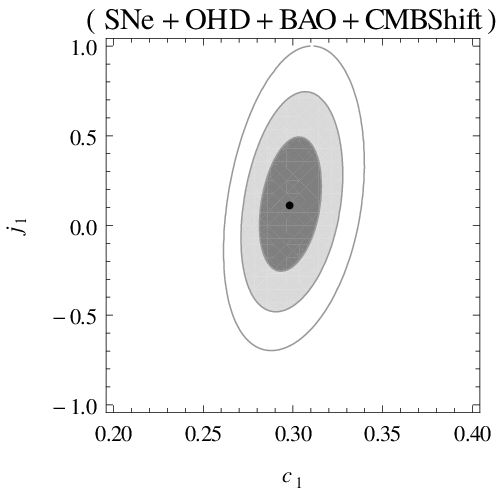}
\end{center}
\caption{{\small  The confidence contours on 2D parameter space of Model IV. The 1$\sigma$, 2$\sigma$ and 3$\sigma$ confidence contours are presented from inner to outer regions and the central black dots represent the corresponding best fit points. The upper left panel is obtained for (SNe+OHD), upper right panel is for (SNe+OHD+BAO), lower left panel is for (SNe+OHD+CMBShift) and lower right panel is for (SNe+OHD+BAO+CMBShift).}}
\label{jz4contoursch4}
\end{figure}

\begin{figure}[H]
\begin{center}
\includegraphics[angle=0, width=0.32\textwidth]{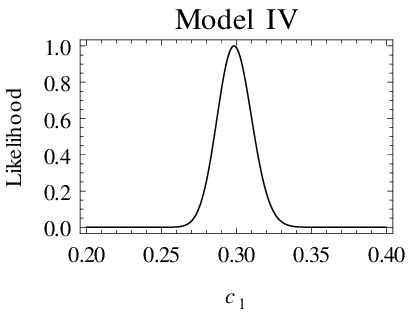}
\includegraphics[angle=0, width=0.32\textwidth]{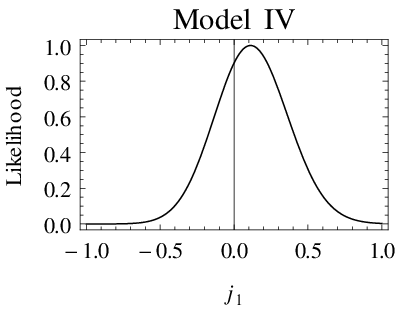}
\end{center}
\caption{{\small The marginalised likelihood functions of Model IV obtained for SNe+OHD+BAO +CMBShift.}}
\label{jz4likelihoodch4}
\end{figure}

 Similarly the figure \ref{jz2contoursch4} presents the confidence contours on the parameter space for Model II and figure \ref{jz2likelihoodch4} shows the marginalised likelihood functions. In table \ref{tablejzM2ch4} the results of the statistical analysis are presented. Figure \ref{jz3contoursch4} and \ref{jz3likelihoodch4} are of Model III and table \ref{tablejzM3ch4} presents the results of corresponding statistical analysis. And figure \ref{jz4contoursch4} and \ref{jz4likelihoodch4} and table \ref{tablejzM4ch4} correspond to Model IV.

%%%%%%%%%%%%%%%%%%%%%
%%%%%%%%%%%%%%%%%%%%%%
%%%%%%%%%%%%%%%%%%%%%
%%%%%%%%%%%%%%%%%%%%%%

%%%%%%%%%%%%%%%%%%%%%
%%%%%%%%%%%%%%%%%%%%%%
%%%%%%%%%%%%%%%%%%%%%
%%%%%%%%%%%%%%%%%%%%%%

\par It is clear from the contour plots on the parameter space that the addition of CMB shift parameter data leads to tighter constraints on the model parameters. The parameters show a negative correlation between then in the analysis with SNe, OHD and BAO data. But the addition of CMB shift parameter data changes the nature of correlation and makes it slightly positive. All the likelihood function plots are well fitted to Gaussian distribution. As the model parameter $j_1$ indicates the deviation of the models from $\Lambda$CDM (for $\Lambda$CDM $j_1=0$), it is important to note that $\Lambda$CDM remains within the $1\sigma$ confidence regions of all the models.

%%%%%%%%%%%%%%%%%%%%%%%%%
\begin{table*}[htb]
\caption{{\small Results of statistical analysis of  Model I}}
\begin{center}
\resizebox{0.75\textwidth}{!}{  
\begin{tabular}{ c |c |c c c } 
 \hline
 \hline
 Data  & $\chi^2_{min}/d.o.f.$ & $c_1 $ & $j_1$ \\ 
 \hline
 \hline
 SNe+OHD & $47.02/53$ & $0.295\pm0.052$ & $0.045\pm0.628$\\ 
 \hline
  SNe+OHD+BAO & $47.11/53$ & $0.309\pm0.012$ &  $-0.080\pm0.222$\\ 
 \hline
   SNe+OHD+CMBShift & $47.03/51$ & $0.292\pm 0.012$ &  $0.080\pm0.145$\\ 
 \hline
   SNe+OHD+BAO+CMBShift & $47.99/51$ & $0.298\pm0.010$ &  $0.078\pm0.140$\\ 
 \hline
 \hline
\end{tabular}
}
\end{center}

\label{tablejzM1ch4}
\end{table*}
%%%%%%%%%%%%%%%%%%%%%%%%%%%%%%%%%%%

%%%%%%%%%%%%%%%%%%%%%%%%%
\begin{table*}[htb]
\caption{{\small Results of statistical analysis of Model II}}
\begin{center}
\resizebox{0.75\textwidth}{!}{  
\begin{tabular}{ c |c |c c c } 
 \hline
 \hline
 Data  & $\chi^2_{min}/d.o.f.$ & $c_1 $ & $j_1$ \\ 
 \hline
 \hline
 SNe+OHD & $47.03/53$ & $0.299\pm0.050$ & $0.002\pm0.267$\\ 
 \hline
  SNe+OHD+BAO & $47.08/53$ & $0.310\pm0.012$ &  $-0.051\pm0.093$\\ 
 \hline
   SNe+OHD+CMBShift & $47.04/51$ & $0.292\pm 0.009$ &  $0.038\pm0.051$\\ 
 \hline
   SNe+OHD+BAO+CMBShift & $47.85/51$ & $0.299\pm0.008$ &  $0.045\pm0.050$\\ 
 \hline
 \hline
\end{tabular}
}
\end{center}

\label{tablejzM2ch4}
\end{table*}
%%%%%%%%%%%%%%%%%%%%%%%%%%%%%%%%%%%

\begin{figure}[tb]
\begin{center}
\includegraphics[angle=0, width=0.3\textwidth]{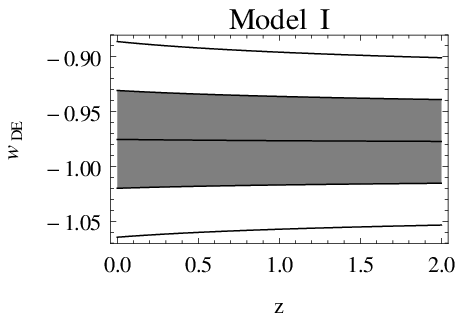}
\includegraphics[angle=0, width=0.3\textwidth]{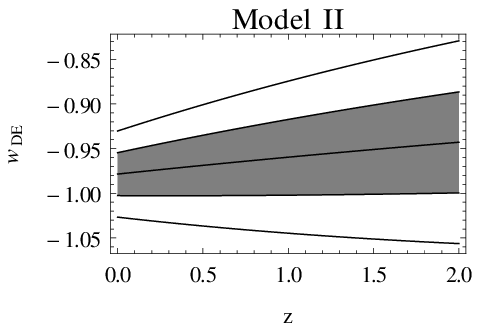}\\
\includegraphics[angle=0, width=0.3\textwidth]{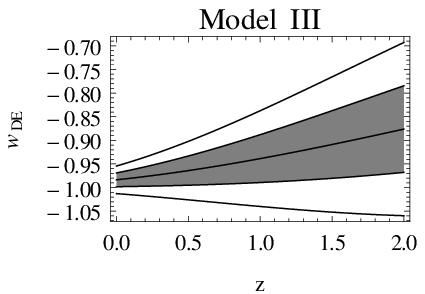}
\includegraphics[angle=0, width=0.3\textwidth]{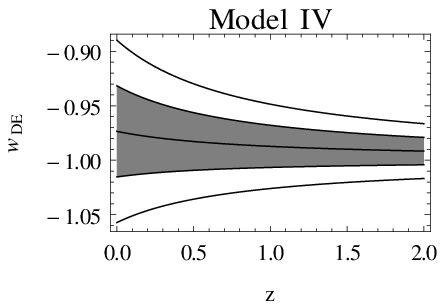}
\end{center}
\caption{{\small Plots of dark energy equation of state parameter $w_{DE}(z)$ against redshift $z$ for different parametrizations of jerk parameter $j(z)$. The 1$\sigma$ and 2$\sigma$ confidence regions, obtained from combined $\chi^2$-analysis, have been shown and the central  dark lines represent the best fit curves.}}
\label{jzwzplotch4}
\end{figure}

\begin{figure}[tb]
\begin{center}
\includegraphics[angle=0, width=0.3\textwidth]{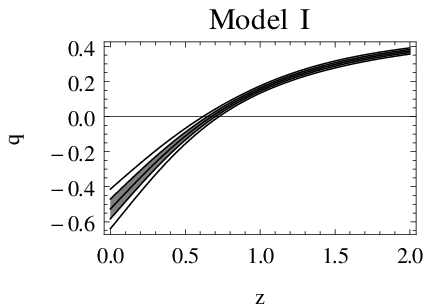}
\includegraphics[angle=0, width=0.3\textwidth]{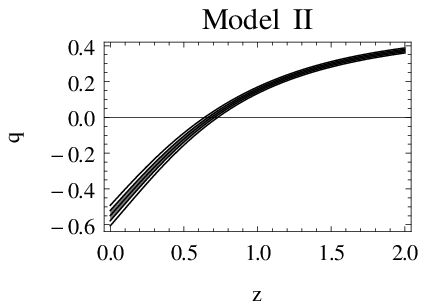}\\
\includegraphics[angle=0, width=0.3\textwidth]{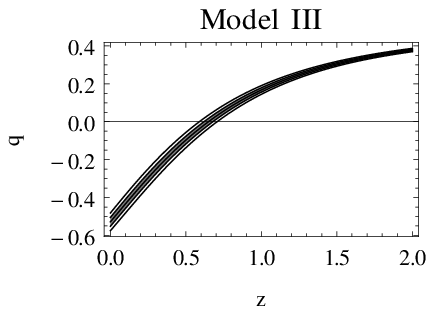}
\includegraphics[angle=0, width=0.3\textwidth]{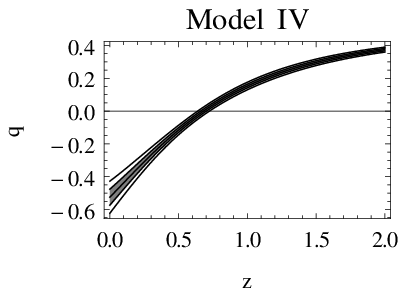}
\end{center}
\caption{{\small Plots of deceleration parameter $q(z)$ against redshift $z$ for different parametrizations of jerk parameter $j(z)$. The 1$\sigma$ and 2$\sigma$ confidence regions, obtained from combined $\chi^2$-analysis, have been shown and the central  dark lines represent the best fit curves.}}
\label{jzqzplotch4}
\end{figure}

Figure \ref{jzwzplotch4} shows the plots of dark energy equation of state parameter as a function of redshift $z$ and figure \ref{jzqzplotch4} presents the plots of deceleration parameter $q(z)$ for different models discussed in the present work. The deceleration parameter plots clearly show that the models successfully generate the late time acceleration along with the decelerated expansion in the past. The plots show the transition from decelerated to accelerated expansion phase took place in the redshift range $0.6<z<0.8$. This is consistent with the recent analysis by Farooq and Ratra \cite{farooqratra} where constraints on the transition redshift are achieved for different dark energy scenario using observational Hubble data. All the models presented in this work show the dark energy equation of state parameter to  be almost constant and the best fit values at present are slightly higher than $-1$, meaning the models prefer the non-phantom nature of dark energy. 

%%%%%%%%%%%%%%%%%%%%%
%%%%%%%%%%%%%%%%%%%%%%
%%%%%%%%%%%%%%%%%%%%%
%%%%%%%%%%%%%%%%%%%%%%

%%%%%%%%%%%%%%%%%%%%%
%%%%%%%%%%%%%%%%%%%%%%
%%%%%%%%%%%%%%%%%%%%%
%%%%%%%%%%%%%%%%%%%%%%

%%%%%%%%%%%%%%%%%%%%%%%%%
\begin{table*}[htb]
\caption{{\small Results of statistical analysis of Model III}}
\begin{center}
\resizebox{0.75\textwidth}{!}{  
\begin{tabular}{ c |c |c c c } 
 \hline
 \hline
 Data  & $\chi^2_{min}/d.o.f.$ & $c_1 $ & $j_1$ \\ 
 \hline
 \hline
 SNe+OHD & $47.02/53$ & $0.307\pm0.100$ & $-0.016\pm0.205$\\ 
 \hline
  SNe+OHD+BAO & $47.06/53$ & $0.313\pm0.015$ &  $-0.027\pm0.045$\\ 
 \hline
   SNe+OHD+CMBSfiht & $47.04/51$ & $0.294\pm 0.011$ &  $0.012\pm0.016$\\ 
 \hline
   SNe+OHD+BAO+CMBShift & $47.60/51$ & $0.300\pm0.008$ &  $0.017\pm0.015$\\ 
 \hline
 \hline
\end{tabular}
}
\end{center}

\label{tablejzM3ch4}
\end{table*}
%%%%%%%%%%%%%%%%%%%%%%%%%%%%%%%%%%%

%%%%%%%%%%%%%%%%%%%%%%%%%
\begin{table*}[htb]
\caption{{\small Results of statistical analysis of Model IV}}
\begin{center}
\resizebox{0.75\textwidth}{!}{  
\begin{tabular}{ c |c |c c c } 
 \hline
\hline
 Data  & $\chi^2_{min}/d.o.f.$ & $c_1 $ & $j_1$ \\ 
 \hline
 \hline
 SNe+OHD & $47.02/53$ & $0.295\pm0.034$ & $0.083\pm0.510$\\ 
 \hline
  SNe+OHD+BAO & $47.13/53$ & $0.308\pm0.011$ &  $-0.106\pm0.237$\\ 
 \hline
   SNe+OHD+CMBShift & $47.02/51$ & $0.291\pm0.009$ &  $0.137\pm0.176$\\ 
 \hline
   SNe+OHD+BAO+CMBShift & $48.06/51$ & $0.298\pm0.008$ &  $0.112\pm0.176$\\ 
 \hline
\hline
\end{tabular}
}
\end{center}

\label{tablejzM4ch4}
\end{table*}
%%%%%%%%%%%%%%%%%%%%%%%%%%%%%%%%%%%

%%%%%%%%%%%%%%%%%%%%%
%%%%%%%%%%%%%%%%%%%%%%

%%%%%%%%%%%%%%%%%%%%%
%%%%%%%%%%%%%%%%%%%%%%

%%%%%%%%%%%%%%%%%%%%%
\begin{figure}[tb]
\begin{center}
\includegraphics[angle=0, width=0.3\textwidth]{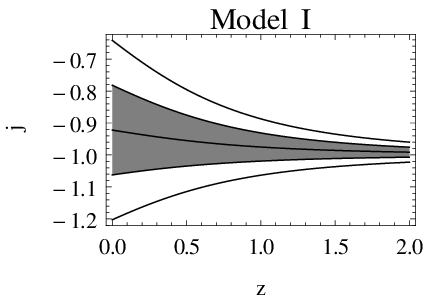}
\includegraphics[angle=0, width=0.3\textwidth]{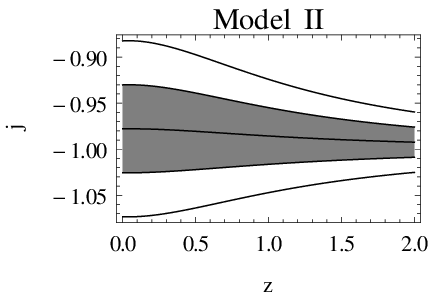}\\
\includegraphics[angle=0, width=0.3\textwidth]{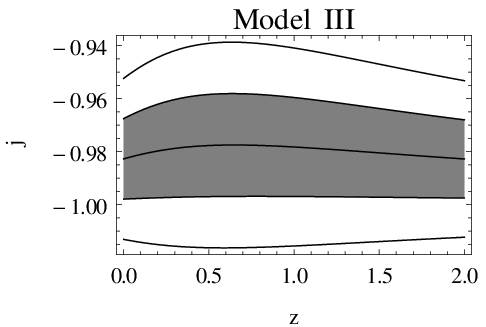}
\includegraphics[angle=0, width=0.3\textwidth]{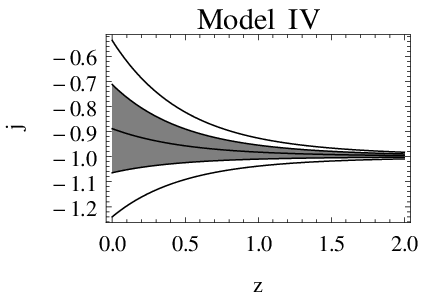}
\end{center}
\caption{{\small Plots of jerk parameter $j(z)$ against redshift $z$ for the models. The 1$\sigma$ and 2$\sigma$ confidence regions, obtained from combined $\chi^2$-analysis, have been shown and the central  dark lines represent the best fit curves.}}
\label{jzjzplotch4}
\end{figure}
%%%%%%%%%%%%%%%%%%%%%%
%%%%%%%%%%%%%%%%%%%%%
\begin{figure*}[htb]
\begin{center}
\includegraphics[angle=0, width=0.24\textwidth]{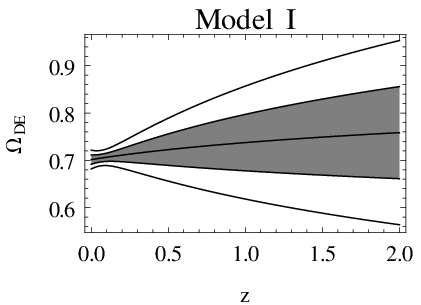}
\includegraphics[angle=0, width=0.24\textwidth]{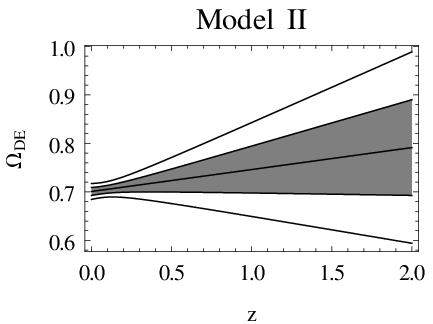}
\includegraphics[angle=0, width=0.24\textwidth]{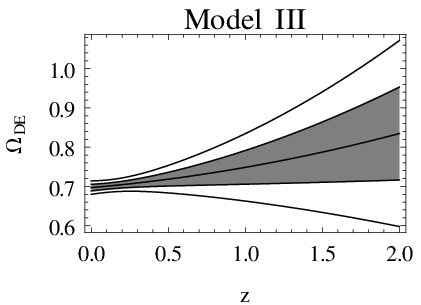}
\includegraphics[angle=0, width=0.24\textwidth]{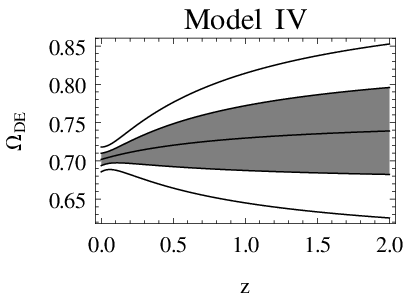}\\
\includegraphics[angle=0, width=0.24\textwidth]{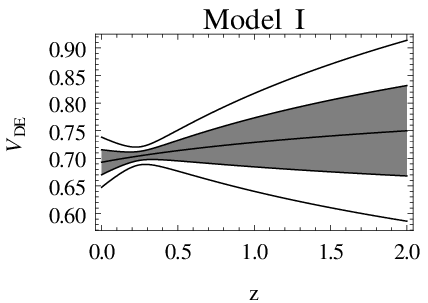}
\includegraphics[angle=0, width=0.24\textwidth]{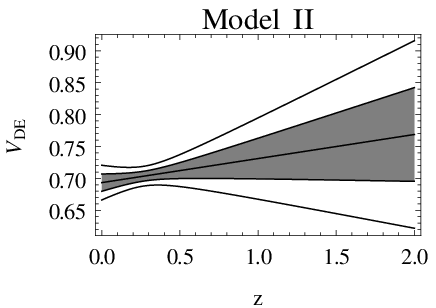}
\includegraphics[angle=0, width=0.24\textwidth]{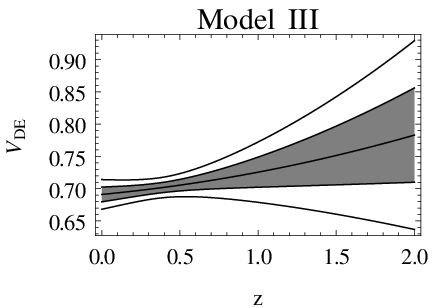}
\includegraphics[angle=0, width=0.24\textwidth]{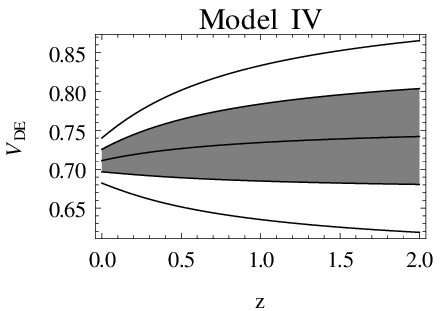}
\end{center}
\caption{{\small  Plots of dark energy density parameter ($\Omega_{DE}$) (upper panels) and the quintessence potential (lower panels) as functions of redshift $z$ for the models discussed in the present work.}}
\label{OmegaDEVzch4}
\end{figure*}
%%%%%%%%%%%%%%%%%%%%%%

\vskip 20 cm

\par Figure \ref{jzjzplotch4} shows the plots of jerk parameter $j(z)$ for the models. It is interesting to note that $j(z)$ is allowed by the models to take values different from that in the case of $\Lambda$CDM at $z=0$. All the four different models show a tendency close the $\Lambda$CDM  along with a range of possibilities for $j(z)$ at the present epoch, in the 2$\sigma$ confidence region.

\section{A Bayesian analysis}
\label{baych4}

The two commonly used criteria for model selection are  Akaike Information Criterion (AIC) \cite{aic} and  Bayesian Information Criterion (BIC)\cite{bic}. These are defined as,
\begin{equation}
AIC=-2\log{L_{max}}+2\kappa,
\label{aicch4}
\end{equation}
\begin{equation}
BIC=-2\log{L_{max}}+2\kappa\log{N},
\label{bicch4}
\end{equation}
where $L_{max}$ is the maximum likelihood, $\kappa$ is the number of free parameters in  the model and $N$ is the number of data points used for the statistical analysis of the model. We note that these two criteria  can hardly provide any information regarding the model selection amongst the four presented here because the values of $\chi^2_{min}$ do not differ significantly for the models and all the models have same number of parameters as well as same number of data points have been used for the statistical analysis of the models. 
\par
As there is hardly any model to choose based on the information criteria, it is thus useful to look for an evidence estimate for the model selection. The Bayesian evidence $E$ is defined as 
\begin{equation}
 \label{bayesianch4}
E=\int {(Likelihood \times Prior) d{\theta}_{1} d{\theta}_{2}....d{\theta}_{n}},
\end{equation}
where ${\theta}_{i}$ are model parameters. There are two parameters in the models $c_1$ and $j_1$. With a flat prior approximation, the evidences calculated for the models are \\
\begin{center}
Model I: ~~~  $ E_{1} = P \times 1.81358 \times 10^{-13}$, \\

Model II:  ~~~$ E_{2} = P \times 8.79425 \times 10^{-14}$, \\

Model III: ~~~$ E_{3} = P \times 2.65140 \times 10^{-14}$, \\

Model IV:  ~~~$ E_{4} = P \times 3.05074 \times 10^{-13}$, \\

\end{center}

where $P$ is the constant prior. These evidences show that there is hardly any model, amongst the four presented, does better than any of the other three. However, if there is any one to choose amongst these, Model IV is the one which does marginally better than the other three. On  the other hand, Model III is less preferred than the others.

\section{Discussion}
\label{disch4}

The reconstruction discussed in the present chapter deals with a parametric reconstruction of the jerk parameter $j(z)$ which is the dimensionless representation of the third order time derivative of the scale factor. As the deceleration parameter $q(z)$ is now an observational quantity and found to be evolving, jerk, amongst the kinematical quantities, appears to be the natural choice as the quantity of interest, as this determines the evolution of $q$. The philosophy is to build up the model from the evolution history of the universe. As such this is just another way of reconstruction, but it might indicate about the nature of matter distribution and the possible interaction amongst them without any {\it a priori} assumption on them. This may particularly be useful as there is not yet a clear verdict in favour of any model. \\

The formalism proposed  by  Zhai {\it et al} \cite{zhaijerk} has been utilized in the present work, with a major difference that $j$ at $z=0$ is allowed to pick up any value depending on a parameter to be fixed by the data as opposed to the work of Zhai {\it et al} where $j$ is constrained to mimic a $\Lambda$CDM at the present epoch given by $z=0$. One interesting feature of this formalism is that the matter density parameter ($\Omega_{m0}$) automatically selects itself as a model parameter.

\par The plots of the dark energy equation of state parameter ($w_{DE}(z)$) and the deceleration parameter ($q(z)$) for the proposed models (figure \ref{jzwzplotch4} and figure \ref{jzqzplotch4} respectively) clearly show that the models can successfully generate  late time acceleration along with a decelerated expansion in the past. The range of redshift of transition from decelerated to accelerated expansion as indicated in the present work is consistent with the result of a recent analysis by Farooq and Ratra \cite{farooqratra}. The model parameter $j_1$ is an indicator  of the deviation of the model from $\Lambda$CDM. For all the four models, the best fit present value of jerk parameter estimated from the observational data are slightly greater than $-1$ but $j=-1$ remains within 1$\sigma$ confidence region. Thus all these models are very close to the $\Lambda$CDM, but with an inclination towards a non-phantom nature of dark energy.

\par The values estimated for the parameter $c_1$, which is equivalent to the matter density parameter, are consistent with the results of the recent analysis of $\Lambda$CDM and $w$CDM models using the CMB temperature anisotropy and polarization data along with the other non-CMB data \cite{xiaLCDMwCDM}.

\par  A constant value of jerk is in fact allowed in all the four models within 1$\sigma$ confidence level (figure \ref{jzjzplotch4}). But the particular value estimated by Rapetti {\it et al} \cite{rapettijerk} is out of 1$\sigma$ confidence region of the present models. An evolving jerk parameter  had been discussed by Zhai {\it et al} \cite{zhaijerk} where only the supernova distance modulus data (SNe) and observational Hubble data (OHD) were used for the statistical analysis of the models.  In the present work, though the same mathematical formulation has been used as Zhai {\it et al}, tighter constraints on the parameter $j_1$ have been achieved by introducing the BAO and CMB shift parameter data along with SNe and OHD.

\par We can construct a quintessence potential associated to the models. If we consider the dark energy contribution is coming from a quintessence scalar field $\phi(t)$,  then the dark energy density ($\rho_{DE}$) and the pressure ($p_{DE}$) can be expressed as $\rho_{DE}=\frac{1}{2}\dot{\phi}^2+V(\phi)$ and $p_{DE}=\frac{1}{2}\dot{\phi}^2-V(\phi)$ where  $V(\phi)$ is the quintessence potential. As the expressions of Hubble parameter are known for the models, the quintessence potential can be expressed as a function of the model parameters and redshift using  equation (\ref{friedmann1ch4}) and (\ref{friedmann2ch4}). The upper panels of figure \ref{OmegaDEVzch4} show the evolution of $\Omega_{DE}$, where $\Omega_{DE}=8\pi G\rho_{DE}/3H_0^2$.   The lower panels of figure \ref{OmegaDEVzch4} show the evolution of the potential V, scaled by critical density ($3H_0^2/8\pi G$), as a function of $z$. The best fit curve of the potential remains almost constant. Thus it is  neither freezing nor thawing \cite{caldwelllinder,scherrersen} but rather a constant, leading to a  slow-roll scalar field.

\par The systematic uncertainties of supernova observations are considered in the statistical analysis presented here as  some of them might have their say as well on the results, such as the colour-luminosity parameter might depend on the redshift, and hence affects the magnitude in the analysis of Supernova data \cite{wangwangSNeSys}. There are some recent discussion \cite{rubinSNeSys,shaferSNeSys} on the effects of systematics which may be worthwhile in any analysis. It deserves mention that CMB data has been used to remove the dependence of the sound horizon in the case of the BAO data. The measurement of the acoustic scale $l_{A}$ and the CMB shift parameter are somewhat correlated. This correlation, calculated from the normalized covariance matrix given by Wang and Wang \cite{wangwangcmb}, is not too large and not likely to change the results significantly. So this correlation is ignored in the present work.

\par The main conclusion, therefore, is that the $\Lambda$CDM is very close to be the winner as the candidate for the favoured model with a marginal inclination towards a non-phantom behaviour of the universe. However, the present work deals with situations each of which yields the $\Lambda$CDM model as a special case ($j_{1}=0$). Anyway, a large departure from the $\Lambda$CDM has not been ruled out ab inito, the reconstructed value of $j_{1}$ shoulders the task of the determination of the departure. The statistical analysis, however shows that $j_1$ is very close to zero, thus the existence of the varying dark energy with a negative equation state is not really favoured. As revealed by equations (\ref{jmodelh3ch4}), in the model III,  $j_1 = 0$ is equivalent to the flatness constraint (curvature index $k=0$). So the small values of $j_1$ in this model also indicates that spatial flatness is favoured, although a a nonzero spatial curvature is not ruled out.

\chapter{In search of the dark matter dark energy interaction: a kinematic approach}

% **************************** Define Graphics Path **************************
\ifpdf
    \graphicspath{{Chapter5/Figs/Raster/}{Chapter5/Figs/PDF/}{Chapter5/Figs/}}
\else
    \graphicspath{{Chapter5/Figs/Vector/}{Chapter5/Figs/}}
\fi

\section{Introduction}
The present chapter contains an investigation regarding the kinematical approach to the dark energy reconstruction where the possibility of interaction between dark energy and dark matter has been included. The normal practice in cosmology is to write down Einstein equations, $G_{\mu\nu}=-8\pi G T_{\mu\nu}$ for a spatially homogeneous and isotropic model with the right hand side taking care of the matter sector. The present work has a completely different approach. We assume a spatially flat homogeneous and isotropic metric and define the usual kinematical quantities like the Hubble parameter $H$, the deceleration parameter $q$ and the jerk parameter $j$, which are respectively the first order, second order and third order time derivatives of the scale factor $a$. The derivatives are all fractional derivatives and furthermore $q$ and $j$ are dimensionless. We have observational results of the evolution of $q$, in the sense that the parameter is negative at the present epoch and had been positive in a recent past, the epoch of transition from the decelerated to the accelerated expansion is also known. The natural choice of the kinematical quantity of interest is thus the evolution of $q$, which is the jerk parameter $j$. We now assume a slowly varying jerk and find the evolution of the other kinematical quantities from the definition of jerk. The values of the model parameters are then estimated from observational data sets and the nature of evolution of different cosmological quantities are also figured out.

\par As already mentioned, the jerk parameter had hardly been used until very recently. However, its importance in building a cosmological model had been emphasized long back in terms of a {\it state-finder} parameter \cite{sahnistatefinder}. The indication of importance of jerk as a future tool for the reconstruction of cosmological models was also indicated by Alam {\it et al} \cite{alamstatefinder}. The reason for its being overlooked as the starting point of reconstruction was perhaps the unavailability of clean data. Reconstruction of dark energy model using deceleration parameter ($q$) and jerk ($j$) as model parameters was discussed by Rapetti {\it et al.} \cite{rapettijerk} where the present values of the kinematical parameters have been constrained using observational data. Parametrization of time evolving jerk parameter models  have been discussed recently by Zhai {\it et al.} \cite{zhaijerk}.

\par If the agent driving the present acceleration is the cosmological constant, then certainly the dark matter sector follows its own conservation equation. However, if the dark energy is an evolving one, there is always a possibility that the two dark sectors interact with each other, one may grow at the expense of the other. Naturally there is a lot of work in the literature where the interacting dark energy model has been discussed. An interacting dark energy model has been given by Pan, Bhattacharaya and Chakraborty \cite{PanBhattacharaya2015} where the interaction term is assumed to be proportional to the total energy density. Interacting holographic dark energy model has been discussed by Zimdahl and Pavon \cite{ZimdahlPavon2007}. For a {\it graceful entry} of the universe from a decelerated to an accelerated phase in Brans-Dicke theory, the interaction of the Brans-Dicke scalar field and the quintessence scalar field had been discussed by Das and Banerjee \cite{DasBanerjee2006}. An attempt towards a covariant Lagrangian formulation of the interaction has been made by Faraoni, Dent and Saridakis \cite{FaraoniDentSaridakis2014}.

\par Reconstruction of the interaction rate in holographic dark energy model has been discussed by Sen and Pavon \cite{SenPavon2008} where the reconstruction has been done with a prior assumption about the dark energy equation of state. Recently a non-parametric reconstruction of dark energy interaction using Gaussian process has been discussed by Yang, Guo and Cai \cite{YangGuoCai2015} where the signature of dark energy interaction has been obtained for a deviation of dark energy equation of state parameter from the value $-1$.  

\par The present work is an attempt to reconstruct the possible interaction of various matter components from the data sets via a kinematical approach. The starting point is the assumption of a slowly varying jerk parameter which is effectively a constant at low redshift regime. The result is that any deviation from the $\Lambda$CDM model indicates a possibility of an interaction amongst various matter sectors. The best fit values, however, are tantalizingly close to the $\Lambda$CDM scenario. Another important result is that the allowance of any interaction is more stringent at recent times, but slightly more relaxed in the past. 

\par It should also be mentioned at the outset that the entire work depends upon the dogma that a $\Lambda$CDM model should be included as a possibility, in an endeavour leading to the reconstruction of the dark energy, at least as a limit.

\section{Reconstruction of the model}
The fractional rate of expansion of the linear size of the universe, dubbed as the Hubble parameter, is defined as
\begin{equation}
H(t)=\frac{\dot{a}}{a},
\label{hubbleparameterch5}
\end{equation}
where the overhead dot denotes a derivative with respect to the cosmic time $t$. To understand the nature of the expansion, higher order time derivatives of the scale factor are to be invoked. The measure of cosmic acceleration is presented in a dimensionless way by the deceleration parameter $q$, defined as
\begin{equation}
q(t)=-\frac{\ddot{a}/a}{\dot{a}^2/a^2}=-1-\frac{\dot{H}}{H^2}.
\label{decelerationparameterch5}
\end{equation}
If the value of the deceleration parameter is negative, then the expansion is accelerated.
\par The cosmological {\it jerk parameter}, which is the dimensionless representation of the third order time derivative of the scale factor, is defined as
\begin{equation}
j(t)=-\frac{1}{aH^3}\Bigg(\frac{d^3a}{dt^3}\Bigg).
\label{jerkparameterch5}
\end{equation}
From the equation (\ref{jerkparameterch5}), the expression for the jerk parameter can also be expressed as,
\begin{equation}
j(z)=-1+(1+z)\frac{(h^2)'}{h^2}-\frac{1}{2}(1+z)^2\frac{(h^2)''}{h^2},
\label{jerkequationch5}
\end{equation}   
where $h(z)=\frac{H(z)}{H_0}$, ($H_0$ being the present value of the Hubble parameter) and a prime denotes the derivative with respect to redshift $z$.  In the present work, the reconstruction is done with the assumption that $j$ is a slowly varying quantity, and will be considered a constant in the subsequent discussion. The solution of the differential equation (\ref{jerkequationch5}) yields the expression of $h^2(z)$ as
\begin{equation}
h^2(z)=A(1+z)^{\frac{3+\sqrt{9-8(1+j)}}{2}}+B(1+z)^{\frac{3-\sqrt{9-8(1+j)}}{2}},
\label{h2zABjch5}
\end{equation}
where $A$ and $B$ are the constant dimensionless coefficients. Now the relation between $A$ and $B$ is obtained from the boundary condition $h(z=0)=1$ as $A+B=1$. So $h^2(z)$ is written as a function of redshift $z$ and two parameters $j$ and $A$ as
\begin{equation}
h^2(z)=A(1+z)^{\frac{3+\sqrt{9-8(1+j)}}{2}}+(1-A)(1+z)^{\frac{3-\sqrt{9-8(1+j)}}{2}}.
\label{h2zAjch5}
\end{equation}
Therefore, this is effectively a two parameter model where $j$ and $A$ are the model parameters. The value of $j$ obtained from the statistical analysis of the reconstructed model using different observational data would indicate the consistency or deviation of this model from the $\Lambda$CDM. It actually mimics the $\Lambda$CDM for $j=-1$. The deceleration parameter (defined in equation (\ref{decelerationparameterch5})) can also be expressed for the present model in terms of the model parameters and the redshift as,

\begin{equation}
\nonumber
q(z)=-1+\frac{A\Bigg(\frac{3+\sqrt{9-8(1+j)}}{2}\Bigg)(1+z)^{\frac{3+\sqrt{9-8(1+j)}}{2}}+(1-A)\Bigg(\frac{3-\sqrt{9-8(1+j)}}{2}\Bigg)(1+z)^{\frac{3-\sqrt{9-8(1+j)}}{2}}}{h^2(z)}.
\end{equation}

\par One component of the matter sector of the Universe, whether it interacts with the dark energy sector or not, is generally believed to be a cold dark matter with an equation of state $p=0$. If we stick to this presupposition, and attempt to recover a non-interacting pressureless fluid at least as a limit from equation (\ref{h2zAjch5}) for some value of $j$, we find that the second term of the right hand side of equation (\ref{h2zAjch5}) can yield a highest power of $(1+z)$ as $3/2$ and can not serve the purpose. The only possibility that remains is $j=-1$ which yield the standard $(1+z)^3$ behaviour in the first term. So we identify the first term to represent the contribution from the cold dark matter, which, in the non-interacting limit yields a $(1+z)^3$ behaviour as in the standard dust model. For the rest of the work will depend on this identification. One should note that this is not the only possibility. It may well be possible to find a corresponding pressure to each of the contribution to the matter sector so that both the components conserve by themselves. One can easily calculate the equation of state parameter $w$ (given by $w=\frac{pressure}{density}$) for both the contributions. A straightforward calculation for a constant $w$ will yield 

\begin{equation}
9w(w+1)+2(1+j)=0,
\label{weqech5}
\end{equation}

It is easy to see that $j=-1$ again gives two solutions of equation (\ref{weqech5}) $w=0$ and $w=-1$ leading to a $\Lambda$CDM behaviour where $w=0$ corresponds to the cold dark matter and $w=-1$ corresponds to the cosmological constant. However, this will not lead to any interaction. One should also note that starting from the definition of the jerk parameter (equation (\ref{jerkparameterch5})), with the assumption that $j$ is a constant, one actually recovers one of the Einstein's equations for the system, where the nature of the matter sector depends on the value of the parameter $j$.

\par Another important  point to be noted is that the parameter $A$ is equivalent to the matter density parameter $\Omega_{m0}$ because for $j=-1$, the power of $(1+z)$ in the first term on the right hand side of equation (\ref{h2zAjch5}) is 3 and the second term is a constant, equivalent to the constant vacuum energy density. If the power of $(1+z)$ in the first term is different from  3, the dark matter is not separately conserved. This invokes the possibility of interaction between the dark matter and the dark energy.

\par To investigate the nature of interaction for the present model, the total conservation equation which is a direct consequence of contracted Bianchi identity, can be divided into two parts as the following,
\begin{equation}
\dot{\rho}_m+3H\rho_m=\eta,
\label{mattconch5} 
\end{equation}
and 
\begin{equation}
\dot{\rho}_{DE}+3H(1+w_{DE})\rho_{DE}=-\eta.
\end{equation}
The over head dots represent the differentiation with respect to cosmic time, $\rho_m$ and $\rho_{DE}$ are the matter density and dark energy density respectively, $w_{DE}$ is the dark energy equation of state parameter. As the dark energy and dark matter interact with themselves, they are not conserved individually. The growth rate of one component, namely $\eta$, is the decay rate of the other. For a dimensionless representation, $\eta$ has been scaled by ($3H_0^3/8\pi G$) and written as 
\begin{equation}
Q=\frac{8\pi G}{3H_0^3}\eta.
\label{Qzch5}
\end{equation}
For the $\Lambda$CDM model, the value of $Q$ is zero. In the present work, observational constraints on the late time evolution of $Q(z)$ has been obtained. As in the expression of the Hubble parameter (equation (\ref{h2zAjch5})), the first term of the right hand side, i.e. $A(1+z)^{\frac{3+\sqrt{9-8(1+j)}}{2}}$, is considered to be the matter density scaled by the present critical density,  the interaction term $Q$ can be expressed in terms of the parameters and redshift from equation (\ref{mattconch5}), as
\begin{equation}
Q(z)=A\Bigg(\frac{3-\sqrt{9-8(1+j)}}{2}\Bigg)(1+z)^{\frac{3+\sqrt{9-8(1+j)}}{2}}h(z).
\end{equation}
It shows that the interaction term is proportion to $\rho_mH$. Interacting dark energy models with this type of interaction term are there in literature \cite{xiawa} where the form of interaction term is pre-assumed. But in the present work, this type of interaction has been obtained from a purely kinematic approach.

It deserves mention that the tacit assumption is that the dark energy interacts with only the dark matter and not the baryons. As the baryon density is much smaller than the dark matter density, the former is ignored. This is important so as not to disturb the big bang nucleosynthesis.

\par The second  term on the right hand side of equation (\ref{h2zAjch5}) is considered to be the contribution from the dark energy density as the first term is the only possibility for being identified to the dark matter. Thus the expression of the dark energy equation of state parameter ($w_{DE}=p_{DE}/\rho_{DE}$) looks like,

\begin{equation}
w_{DE}(z)=-\Bigg(\frac{3+\sqrt{9-8(1+j)}}{6}\Bigg)-\Big(\frac{A}{1-A}\Big)\\
\Bigg(\frac{3-\sqrt{9-8(1+j)}}{6}\Bigg)(1+z)^{\sqrt{9-8(1+j)}}.
\end{equation}

\section{Results of statistical analysis}

Now the remaining task is to estimate the parameter values from observational data sets. In the present work, three different data sets have been adopted. These are (i) the observational Hubble parameter data (OHD), (ii) the distance modulus data of type Ia supernova (SNe) and (iii) baryon acoustic oscillation data (BAO).

%%%%%%%%%%%%%%%%%%%%%%%%%
\begin{table}
\begin{center}
\resizebox{0.64\textwidth}{!}{  
\begin{tabular}{ c |c |c c } 
\hline
 \hline
  Data & $\chi^2_{min}/d.o.f.$  & $A$ & $j$ \\ 
 \hline
  OHD+SNe & 47.30/54 & 0.305$\pm$0.023 & -0.861$\pm$0.127\\ 

  SNe+BAO & 33.95/28 & 0.297$\pm$0.024 & -1.014$\pm$0.045\\ 

  OHD+SNe+BAO & 48.28/54 & 0.286$\pm$0.015 & -1.027$\pm$0.037\\ 
 \hline
\hline
\end{tabular}
}
\end{center}
\caption{{\small  Results of statistical analysis with different combinations of the data sets. The value of $\chi^2_{min}/d.o.f.$ and the best fit values of the parameters along with the  associated 1$\sigma$ uncertainties are presented.}}
\label{tableAjch5}
\end{table}
%%%%%%%%%%%%%%%%%%%%%%%%%%%%%%%%%%%
\par The $\chi^2$ minimization technique, which is equivalent to the maximum likelihood analysis, has been adopted to find the best fit values of the model parameters. Detail discussions regarding the observational data sets and the statistical analysis is available in  section \ref{obsdata} and \ref{statanalysis}.

\par The expression of the Hubble parameter  obtained for the present model (equation (\ref{h2zAjch5})) shows that the matter sector has two components. The first one, with constant coefficient $A$, is the dark matter density and the other one is the dark energy density. As the  energy density of relativistic particles, mainly the photon and nutrino, have an effective contribution to the dynamics of the of the universe at very high redshift, an additional energy density term, evolving as $(1+z)^4$ for radiation, has been taken into account in equation (\ref{h2zAjch5}) while using the angular diameter distance measurement in the analysis with BAO data. The present value of the energy density of relativistic particles scaled  by the present critical density is taken to be $\Omega_{r0}=9.2\times10^{-5}$ with the adopted fiducial value of current CMB temperature $T_0=2.7255$K. The adopted fiducial value of $T_0$ is based on the measurement of current CMB temperature $T_0=2.7255\pm0.0006$K \citep{Fixsen2009}.  The radiation energy density has a negligible contribution at present epoch. 

\par
\begin{figure}[H]
\begin{center}
\includegraphics[angle=0, width=0.28\textwidth]{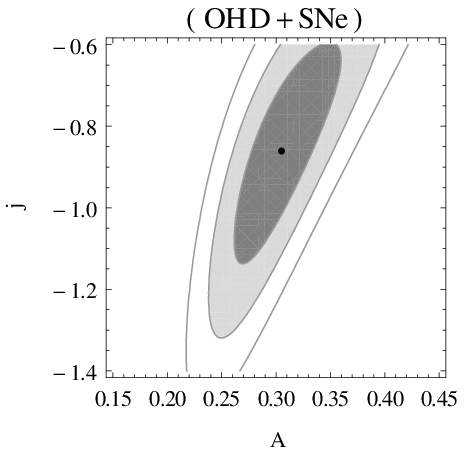}
\includegraphics[angle=0, width=0.28\textwidth]{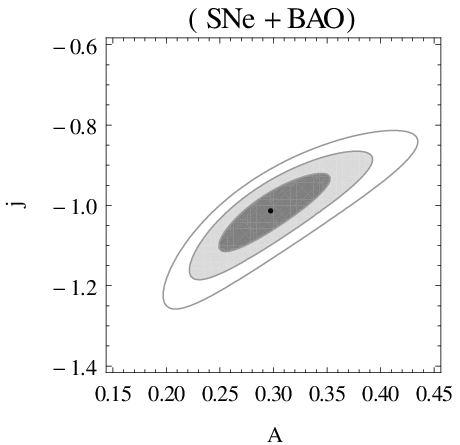}
\includegraphics[angle=0, width=0.28\textwidth]{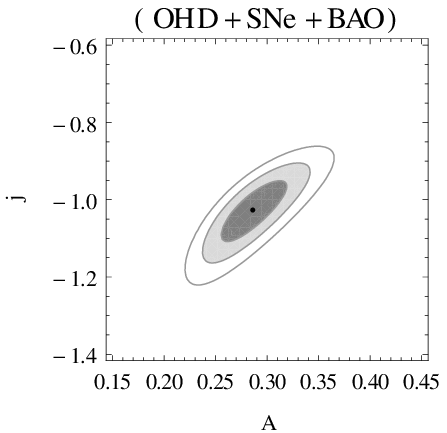}
\end{center}
\caption{{\small Confidence contours on the 2D parameter space of the reconstructed model. The 1$\sigma$, 2$\sigma$ and 3$\sigma$ confidence regions have been presented from inner to outer area and the central black dots represent the corresponding best fit point. The left panel shows the confidence contours obtained for the statistical analysis using OHD+SNe data, the middle panel is  obtained SNe+BAO and the right panel is for OHD+SNe+BAO.}}
\label{contourplot1ch5}
\end{figure}

\begin{figure}[H]
\begin{center}
\includegraphics[angle=0, width=0.35\textwidth]{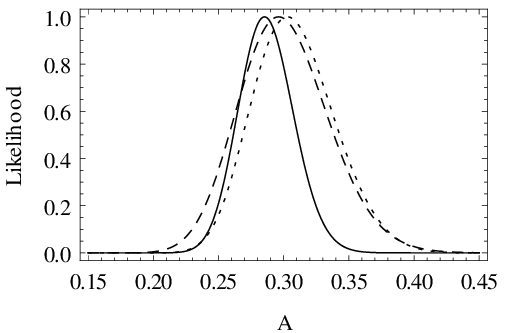}
\includegraphics[angle=0, width=0.35\textwidth]{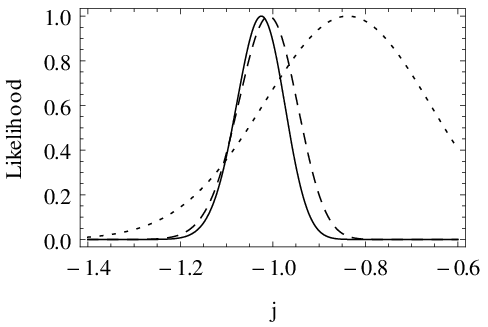}
\end{center}
\caption{{\small Plots of marginalized likelihood functions of the reconstructed model. The dotted curves represents the likelihood obtained for OHD+SNe, dashed curves represents the likelihood for SNe+BAO and the solid curves represents the likelihood for OHD+SNe +BAO.}}
\label{likelihoodplot1ch5}
\end{figure}

 Figure \ref{contourplot1ch5} shows the confidence contours on the two dimensional (2D) parameter space of the model for different combinations of the data sets and figure \ref{likelihoodplot1ch5} presents the plots of marginalized likelihoods as functions of the model parameters. The likelihods are well fitted to Gaussian distributions. Table \ref{tableAjch5} shows the results of the statistical analysis for different combinations of the data sets. The reduced $\chi^2$ (i.e. $\chi^2/d.o.f.$) values are also presented as an estimation of the goodness of fitting.

%%%%%%%%%%%%%%%%%%%%%
%%%%%%%%%%%%%%%%%%%%%%

%%%%%%%%%%%%%%%%%%%%%
%%%%%%%%%%%%%%%%%%%%%%

Both figure \ref{contourplot1ch5} and table \ref{tableAjch5} clearly show that the best fit value of $j$ is very close to $-1$, indicating clearly that the model with a constant jerk parameter is close to $\Lambda$CDM model.

\par
\begin{figure}
\begin{center}
\includegraphics[angle=0, width=0.35\textwidth]{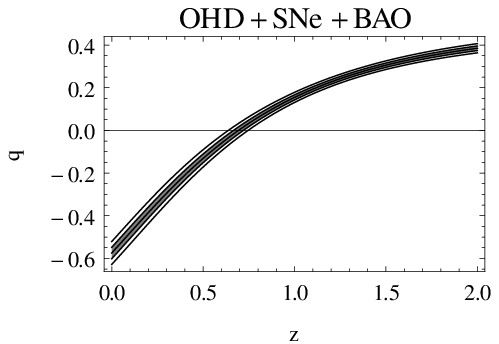}
\includegraphics[angle=0, width=0.35\textwidth]{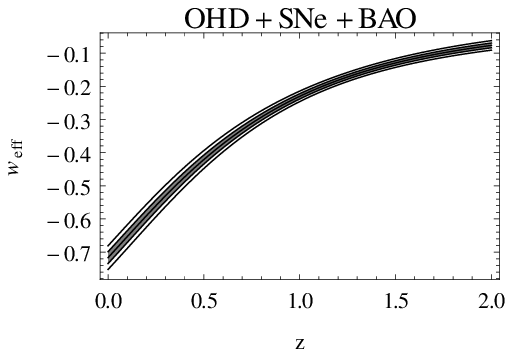}
\end{center}
\caption{{\small The plots of the deceleration parameter ($q(z)$) (left panel) and the effective equation of state parameter ($w_{eff}(z)$) (right panel) for the reconstructed model. The corresponding 1$\sigma$ and 2$\sigma$ confidence regions and the best fit curves obtained in the analysis combining OHD, SNe and BAO data sets, are presented.}}
\label{qzweffzch5}
\end{figure}

 Figure \ref{qzweffzch5} presents the plots of deceleration parameter $q(z)$ (left panel) and the effective or total equation of state $w_{eff}(z)$ (right panel) where $w_{eff}=p_{DE}/(\rho_m+\rho_{DE})$. The deceleration parameter plot clearly shows that the reconstructed model successfully generates the recent cosmic acceleration along with the decelerated expansion phase that prevailed in the past. The redshift of transition from decelerated to accelerated expansion phase obtained for the present model is 0.6 to 0.8 which is consistent with the result of recent analysis by Farooq and Ratra \cite{farooqratra}.

%%%%%%%%%%%%%%%%%%%%%
%%%%%%%%%%%%%%%%%%%%%%

%%%%%%%%%%%%%%%%%%%%%
\begin{figure}[H]
\begin{center}
\includegraphics[angle=0, width=0.3\textwidth]{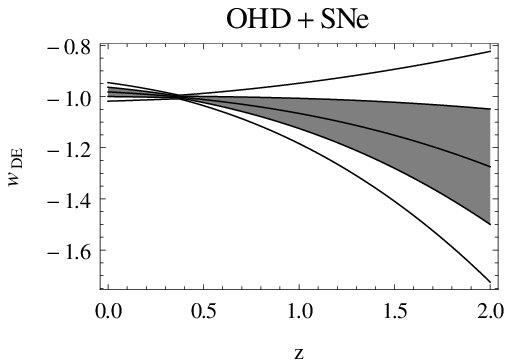}
\includegraphics[angle=0, width=0.3\textwidth]{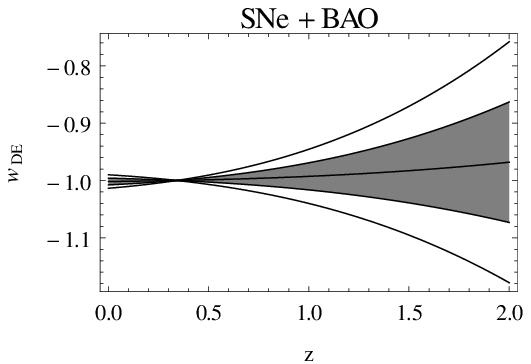}
\includegraphics[angle=0, width=0.3\textwidth]{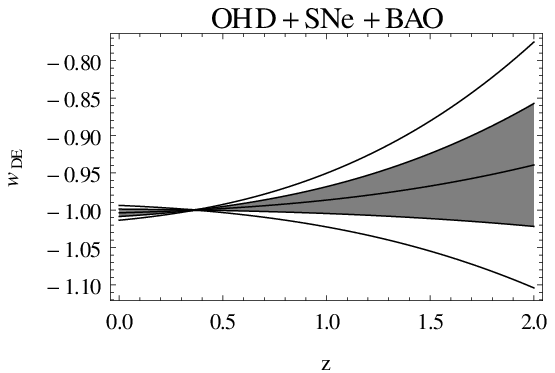}
\end{center}
\caption{{\small The plots of the dark energy equation of state parameter ($w_{DE}(z)$), obtained from the analysis with different combination of the data sets are presented. The corresponding 1$\sigma$ and 2$\sigma$ confidence regions and the best fit curves are shown.}}
\label{wDEzch5}
\end{figure}
%%%%%%%%%%%%%%%%%%%%%%

%%%%%%%%%%%%%%%%%%%%%
\begin{figure}[H]
\begin{center}
\includegraphics[angle=0, width=0.3\textwidth]{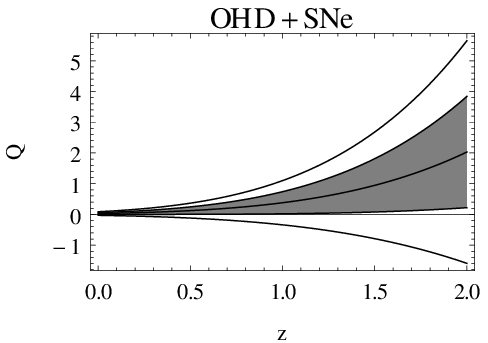}
\includegraphics[angle=0, width=0.3\textwidth]{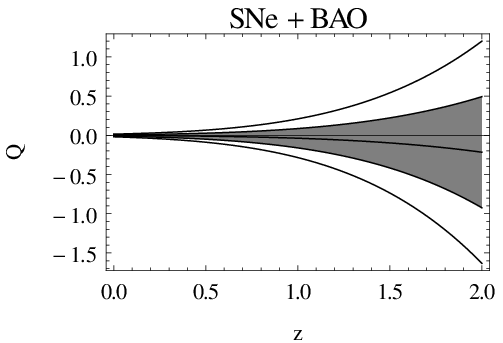}
\includegraphics[angle=0, width=0.3\textwidth]{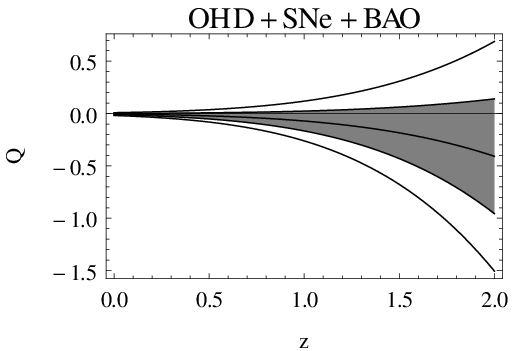}
\end{center}
\caption{{\small The plots of interaction term $Q(z)$, obtained from the analysis with different combination of the data sets are presented.  The corresponding 1$\sigma$ and 2$\sigma$ confidence regions and the best fit curves are shown. The $Q=0$ straight line represents the $\Lambda$CDM model.}}
\label{Qzch5}
\end{figure}
%%%%%%%%%%%%%%%%%%%%%%

Figure \ref{wDEzch5} presents the plots of dark energy equation of state parameter $w_{DE}$, obtained from the analysis with different combinations of the data sets. It is essential to note that the nature of the dark energy equation of state, particularly the behaviour of the best fit curve is sensitive to the choice of the data sets. The analysis combining SNe, OHD and BAO data (right panel of figure \ref{wDEzch5}) shows that the dark energy has a slight inclination toward the non-phantom nature.  The plots also show that the $w_{DE}(z)$ is constrained better at low redshift and the uncertainty increases at high redshift.

\par Figure \ref{Qzch5} shows the evolution of  the interaction term $Q(z)$, defined in equation (\ref{Qzch5}). For the present model, any deviation from $\Lambda$CDM indicates a possibility of interaction between dark energy and dark matter. For non interacting models, the interaction term $Q(z)$ is zero. The plots of $Q(z)$ of the present model obtained from the analysis with different combination of the data sets show that the evolution of  $Q(z)$ is also sensitive to the choice of data sets. However, for all combinations of the data sets taken in the present work, show the possibility of interaction between dark energy and dark matter is very low at low redshift. But the possibility of interaction is high at high redshift. The $\Lambda$CDM always remains within the 1$\sigma$ confidence region.  The result obtained from the combination of SNe, OHD  shows a higher preference towards the interaction at high redshift (left panel of figure \ref{Qzch5}). But the addition of BAO data brings best fit curve closer to $\Lambda$CDM. It is also easy to note that the present analysis allows both positive and negative value for the interaction term $Q(z)$. Table \ref{tableQ0ch5} presents the present values of the interaction term obtained from the analysis with different combination of the data sets. The result obtained from the analysis combining SNe, OHD and BAO data shows that the best fit curve of $Q(z)$ has an inclination towards negative value. But the possibility of a positive $Q(z)$ is also well within the 1$\sigma$ confidence region. The requirement of a positive $Q(z)$ in the context of thermodynamics has been discussed by Pavon and Wang \cite{PavonWang2009}.

%%%%%%%%%%%%%%%%%%%%%%%%%
\begin{table}[h!]
\begin{center}
\resizebox{0.45\textwidth}{!}{  
\begin{tabular}{ c |c  } 
\hline
 \hline
  Data  & $Q(z=0)$ \\ 
 \hline
  OHD+SNe  & $0.0292\pm0.0293$\\ 

  SNe+BAO &  $-0.0026\pm0.0087$\\ 

  OHD+SNe+BAO &  $-0.0051\pm0.0067$\\ 
 \hline
\hline
\end{tabular}
}
\end{center}
\caption{{\small The present value of the interaction term i.e. $Q(z=0)$ obtained for different combinations of the data sets. The corresponding best fit values and the associated 1$\sigma$ uncertainties are presented.}}
\label{tableQ0ch5}
\end{table}
%%%%%%%%%%%%%%%%%%%%%%%%%%%%%%%%%%%

We can defined another dimensionless quantity $\xi$, which is the interaction term normalized by $H^3(z)$, as

\begin{equation}
\xi(z)=\frac{8\pi G}{3H^3}\eta.
\end{equation}

This $\xi$ actually gives a better estimate of the possibility of the interaction. It can be written in terms of the model parameters as,

\begin{equation}
\xi(z)=A\Bigg(\frac{3-\sqrt{9-8(1+j)}}{2}\Bigg)\frac{(1+z)^{\frac{3+\sqrt{9-8(1+j)}}{2}}}{h^2(z)}.
\end{equation}

\begin{figure}[ht]
\begin{center}
\includegraphics[angle=0, width=0.3\textwidth]{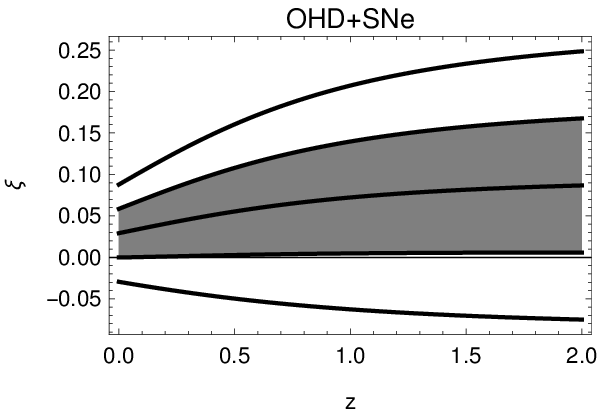}
\includegraphics[angle=0, width=0.3\textwidth]{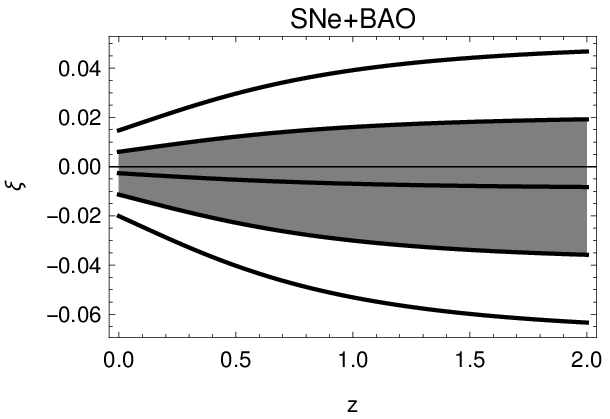}
\includegraphics[angle=0, width=0.3\textwidth]{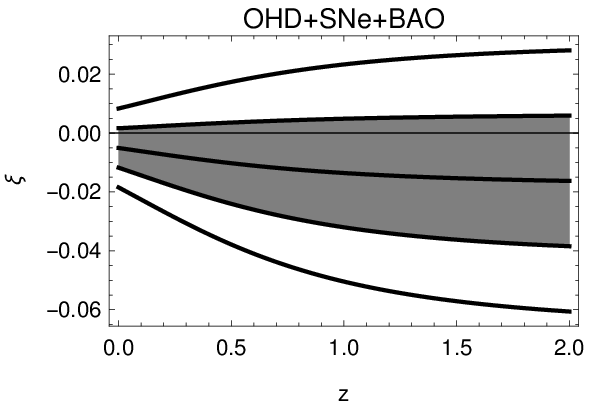}
\end{center}
\caption{{\small The plots of $\xi(z)$, obtained from the analysis with different combination of the data sets. The corresponding 1$\sigma$ and 2$\sigma$ confidence regions and the best fit curves are shown. The $\xi=0$ straight line represents the $\Lambda$CDM model.}}
\label{xich5}
\end{figure}

Figure \ref{xich5} shows the plots of the $\xi$. It is clear that the uncertainty of $\xi$ is not vary high at high redshift compared to uncertainty at present, thus it puts tighter constraints the possibility of interaction also at high redshift.
%%%%%%%%%%%%%%%%%%%%%
%%%%%%%%%%%%%%%%%%%%%%

\section{Discussion}
The present work is an attempt to search for the possibility of interaction between the dark matter and the so-called dark energy with a kinematic approach. The crucial factor is that we start from the dimensionless jerk parameter $j$ which is a third order derivative of the scale factor $a$. This choice is of a natural interest, as the evolution of $q$, the second order time derivative of $a$, is an observational quantity now. We start from the geometrical definition of jerk, and do not use the Einstein equations to start with. We reiterate that the conclusion that any deviation of the jerk parameter from $-1$ indicates an interaction between dark matter and dark energy sectors is actually based on the identification of the matter density term in equation (\ref{h2zAjch5}).  

\par The result obtained clearly shows that the best fit value of $j$ (chosen as a constant parameter) is very close to $-1$, which is consistent with a $\Lambda$CDM model. The interaction term $Q$, in a dimensionless representation, is very close to zero at the present era. This is completely consistent as $\Lambda$, being a constant, does not exchange energy with dark matter. Table \ref{tableQ0ch5} shows the best fit values of $Q$ at $z=0$ for various combinations of data sets. It is easily seen that the values are two orders of magnitude smaller than $\Omega_{m0}$ and $\Omega_{DE0}$, which are approximately $0.3$ and $0.7$ respectively. All these quantities are expressed in a dimensionless way. So this comparison is possible.

\par As already mentioned, investigations regarding a reconstruction of interaction are not too many. But the very recent work by Yang, Guo and Cai\cite{YangGuoCai2015} is a rigorous and elaborate one. The method adopted is the Gaussian processes. Although the work is model dependent, as the equation of state parameter is not specified, it can be applied to a large variety of dark energy models. The $w$CDM model is particularly emphasized. The basic result is the same as that of the present work, the interaction appears to be negligible and consistent with the $\Lambda$CDM model at $z=0$.

\par An intriguing feature in both the present work and that in ref \cite{YangGuoCai2015}, is that although the best fit value still hovers around being negligible, it is allowed to have a non-trivial value for $Q$ at higher $z$ even in the 1$\sigma$ confidence region. But the interaction term normalized by the Hubble parameter (figure \ref{xich5}) shows that the uncertainty at high redshift is not huge compaired to the present epoch. So it can only be concluded that the interaction, if any, has slightly higher possibility at high redshift. The physics of this is not yet quite understood. As already mentioned that the interaction term obtained in the present work is proportional to the matter density. In a recent analysis with this type of interaction term by Xia and Wang \cite{xiawa} shows that indication towards an interacting dark energy is not significantly strong.

\par Another interesting result in the present work is the fact that $Q$, if it has a sizeable value, it can be both positive or negative, so the pumping of energy is possible both ways. Intuitively it might appear that the dark energy should grow at the expense of dark matter ($Q<0$). However, the thermodynamic considerations demand that the flow of energy should be the other way round, from dark energy to dark matter \cite{PavonWang2009}.

\par It  has already been mentioned that the reconstructed model mimics the $\Lambda$CDM for the value of cosmological jerk parameter $j=-1$ and prevent the possibility of interaction between dark energy and dark matter. Any observational measurement which is based on the fiducial assumption of a $\Lambda$CDM cosmology, might affect the results of statistical analysis by making the parameter values highly biased towards the corresponding $\Lambda$CDM values and leading to far too optimistic error bars. Hence such kind of data, like the CMB distance prior measurement, has not been introduced directly in the likelihood analysis. The correlations of distance modulus measurement of type Ia supernova have been taken into account as it might  have its signature on the results.

%%%%%%%%%%%%%%%%%%%%%%%%%%%%%%%%%%%%%%%%%%%%%%%%%%%%%%%%%%%%%%%%%%%%%%%%%%%%%%%%%%%%%%%%%%

\chapter{Reconstruction of interaction rate in holographic dark energy}

% **************************** Define Graphics Path **************************
\ifpdf
    \graphicspath{{Chapter6/Figs/Raster/}{Chapter6/Figs/PDF/}{Chapter6/Figs/}}
\else
    \graphicspath{{Chapter6/Figs/Vector/}{Chapter6/Figs/}}
\fi

\section{Introduction}
\par In the present chapter, reconstruction of the rate of interaction between dark energy and dark matter in a {\it holographic dark energy} has been discussed. The basic idea of holographic dark energy is based on fundamental thermodynamic consideration, namely the {\it holographic principal},  introduced by 't Hooft \cite{hooft} and Susskind \cite{susskind}. To avoid the violation of the second law of thermodynamics in the context of quantum theory of gravity, Bekenstein suggested that the maximum entropy of the system should be proportional to its area instead  of its volume \cite{bekenstein}. Form this idea, t'Hooft conjectured that the phenomena within a volume can be explained by a set of degrees of freedom residing on its boundary and the degrees of freedom of a system is determined by the area of the boundary instead of the volume of the system. In quantum field theory it relates a short distance cut-off (ultraviolet (UV) cut-off) to a long distance cut off (infrared (IR) cu-toff) in the limit set by the formation of a black hole \cite{cohenhol}. The total quantum zero point energy of a system should not exceed the mass of a black hole of the same size. If $\rho_{\Lambda}$ be the quantum zero point energy density with the short distance cut-off, the total energy is $L^3\rho_{\Lambda}$, where $L$ is the size  of the system. Thus it can be written as \cite{limhol},

\begin{equation}
L^3\rho_{\Lambda}\leq LM_P^2, 
\end{equation}
where $M_P^2=(8\pi G)^{-1}$. The inequality saturates for the largest allowed value of the system size $L$, which is the long distance cut-off or the infrared cut-off. Thus the energy density $\rho_{\Lambda}$ is proportional to inverse square of the infrared cut-off. This idea have been adopted in the context of dark energy by Li \cite{limhol}. Thus the holographic dark energy density is written as,
\begin{equation}
\rho_{H}=3C^2M_P^2/L^2,
\label{rhohch6}
\end{equation}
where $C^2$ is a dimensionless constant. Different attempts are  there in literature with different selections of the IR cut-off length scale, the particle horizon \cite{parhorizon}, the future event horizon \cite{limhol,futhorizon} and the Hubble horizn \cite{hubhorizon} etc. Xu has studied holographic dark energy with the Hubble horizon  cut-off with constant as well as time varying coupling parameter ($C^2$) \cite{xuholo}. Reconstruction of interaction rate in holographic dark energy has earlier been discussed by Sen and Pavon \cite{SenPavon2008}, where the interaction rate has been reconstructed assuming a particular form of the dark energy equation of state.  A comparative study of the holographic dark energy  with different length scale cut-off has been carried out by del Campo {\it et al.} \cite{campoholo}. Recently Hu {\it et al.} \cite{huyholo} has attempted to build up the model combining cosmological constant and holographic energy density. Holographic dark energy from minimal supergravity has been discussed by Landim \cite{landimholo}. Holographic dark energy in Brans-Dicke theory has been discussed by Banerjee and Pavon \cite{banpavholo}. Stability analysis of holographic dark energy model has been discussed by Banerjee and Roy \cite{banroyholo}. 

\par In the present chapter, the Hubble horizon has been adopted as the infrared (IR) cut-off for the holographic dark energy meaning the cut-off length scale $L=(H)^{-1}$, where $H$ is the Hubble parameter. The interaction rate of holographic dark energy has been reconstructed from three different parameterizations of the deceleration parameter. The expressions of Hubble parameter obtained for these models hardly gives any indication towards the independent conservation of dark matter and dark energy. It is important to note that the holographic dark energy model with Hubble horizon as the IR cut-off can generate late time acceleration along with the matter dominated decelerated expansion phase in  the past only if there is some interaction between the dark energy and dark matter. The prime endeavour of the present work is to study the nature of interaction and the evolution of the interaction rate for these three models assuming the holographic dark energy with Hubble horizon as the IR cut-off.

\section{Reconstruction of the interaction rate}
\label{reconstch6}

From contracted Bianchi identity, the conservation equation of the total energy density can be written as,
\begin{equation}
\dot{\rho}_{total}+3H(\rho_{total}+p_{total})=0,
\label{totconservationch6}
\end{equation}
where $\rho_{total}=\rho_m+\rho_{DE}$ and $p_{total}=p_{DE}$ as the dark matter is pressureless. Now the conservation equation (equation \ref{totconservationch6}) can be decomposed into two parts,

\begin{equation}
\dot{\rho}_m+3H\rho_m=\eta,
\label{matterconservationch6}
\end{equation}
and 
\begin{equation}
\dot{\rho}_{DE}+3H(1+w_{DE})\rho_{DE}=-\eta,
\label{darkenergyconservationch6}
\end{equation}
where $w_{DE}$ is the equation of state parameter  of dark energy and the $\eta$ is the interaction term. If there is no interaction between dark energy and dark matter, then the interaction term $\eta=0$, and the matter evolves as, $\rho_m\propto\frac{1}{a^3}$.

\par The dark energy density $\rho_{DE}$ for a holographic model with the Hubble horizon as the IR cut-off (denoted as $\rho_H$) is given, according to equation (\ref{rhohch6}), as,
\begin{equation} 
\rho_H=3C^2M_P^2H^2,
\end{equation}
where $C$, the coupling parameter is assumed to be a constant in the present work and $M_P=\frac{1}{\sqrt{8\pi G}}$. Now the interaction term $\eta$ is written as, $\eta=\rho_H\Gamma$, where $\Gamma$ is the rate at which the energy exchange occurs between dark energy and dark matter. The ratio of dark matter and dark energy density, sometimes called the {\it coincidence parameter}, is written as, $r=\rho_m/\rho_H$, and its time derivative can be expressed as \cite{SenPavon2008},
\begin{equation}
\dot{r}=(1+r)\Big[3Hw_{DE}\frac{r}{1+r}+\Gamma\Big].
\label{rdotch6}
\end{equation} 
For a spatially flat geometry, it can also be shown that the ratio $r$ remains constant for a holographic dark energy with Hubble horizon as the  IR cut-off. As the ratio of dark matter and dark energy remains constant in this case, it can potentially resolve the cosmic coincidence problem. But it might be confusing as one may think that it contradicts the standard scenario of structure formation during the dark matter dominated epoch. Actually this is not the case. The matter dominated phase is automatically recovered as the interaction rate is very small at high and moderate redshift and thus the dark energy equation of state resembles the non-relativistic matter \cite{hubhorizon}. For a constant value of $r$, $\dot{r}=0$, from which the interaction rate can be expressed using equation (\ref{rdotch6}) as,
\begin{equation}
\Gamma=-3Hr\frac{w_{DE}}{1+r}.
\label{intratech6}
\end{equation}  
The effective or total equation of state parameter ($w_{eff}=\frac{p_{total}}{\rho_{total}}$), is related to the dark energy equation of state parameter as,
\begin{equation} 
w_{eff}=\frac{w_{DE}}{1+r}.
\label{weffch6}
\end{equation}
Finally the interaction rate can be written as, 

\begin{equation}
\Gamma=-3Hrw_{eff},
\end{equation} 
and it can be represented in a dimensionless way,
\begin{equation}
\frac{\Gamma}{3H_0}=-(H/H_0)rw_{eff}.
\label{intraterech6}
\end{equation}
The interaction rate has been reconstructed for three different parameterizations of the deceleration parameter. These three parameterizations of deceleration parameter have been discussed in the following. It should be mentioned that for the reconstruction of the interaction rate, it is required to fix the value of the coincident parameter $r$. The value of $r$ is taken according to the recent measurement of the dark energy density parameter $\Omega_{DE0}$ from Planck using Planck+WP+highL+BAO \cite{planck2013}. For a spatially flat universe $r$ can be written as $r=(1-\Omega_{DE0})/\Omega_{DE0}$. It is important to note that the interaction rate $\Gamma$ does not depend upon the coupling parameter ($C^2$). The effective equation of state parameter ($w_{eff}(z)$) can be obtained from the Hubble parameter using the Friedmann equations.

\par The deceleration parameter, a dimensionless representation of the second order time derivative of  the scale factor, is defined as $q=-\frac{1}{H^2}\frac{\ddot{a}}{a}$. It can also be written using redshift $z$ as the argument of differentiation as,
\begin{equation}
q(z)=-1+\frac{1}{2}(1+z)\frac{(H^2)'}{H^2}.
\end{equation} 
The parametric forms of the deceleration parameter, adopted in the present work, are given as,
\begin{equation}
Model ~~I.~~~~~~~~q(z)=q_1+\frac{q_2}{(1+z)^2},
\label{qz1ch6}
\end{equation}
\begin{equation}
Model ~~II.~~~~~~~~q(z)=\frac{1}{2}+\frac{q_1+q_2z}{(1+z)^2},
\label{qz2ch6}
\end{equation}
\begin{equation}
Model ~~III.~~~~~~~~q(z)=-1+\frac{q_1(1+z)^2}{q_2+(1+z)^2},
\label{qz3ch6}
\end{equation}
where $q_1$ and $q_2$ are the parameters for the models. However, $q_1$ and $q_2$ do not have the same physical significance in the three different models. The second model of deceleration parameter adopted in the present work has already been discussed by Gong and Wang \cite{GongWang2007} in the context of reconstruction of the late time dynamics of the Universe. The parametrization of Model III has some similarity with one of the parametrizations based on thermodynamic requirement discussed by del Campo {\it et al.} \cite{campopavon}. The expressions of Hubble parameter for the models can be obtained by integrating equation (\ref{qz1ch6}) to (\ref{qz3ch6}) as,
\begin{equation}
Model~~I.  ~~~~~~h^2(z)=\frac{H^2(z)}{H_0^2}=(1+z)^{2(1+q_1)}\exp{\Bigg[-q_2\Bigg(\frac{1}{(1+z)^2}-1\Bigg)\Bigg]},
\label{hubbleparameter1ch6}
\end{equation}
\begin{equation}
Model ~~ II. ~~~h^2(z)=\frac{H^2(z)}{H_0^2}=(1+z)^3\exp{\Bigg[\frac{q_2-q_1}{(1+z)^2}-\frac{2q_2}{(1+z)}+(q_1+q_2)\Bigg]},
\label{hubbleparameter2ch6}
\end{equation}
\begin{equation}
Model ~~ III. ~~~h^2(z)=\frac{H^2(z)}{H_0^2}=\left(\frac{q_2+(1+z)^2}{1+q_2}\right)^{q_1},
\label{hubbleparameter3ch6}
\end{equation}
and consequently the effective equation of state parameter ($w_{eff}(z)$) for the models are expressed as,
\begin{equation}
Model~~I. ~~~~ w_{eff}(z)=-1+\frac{2}{3}\left((1+q_1)+\frac{q_2}{(1+z)^2}\right),
\end{equation}
\begin{equation}
Model ~~ II. ~~~w_{eff}(z)=-1+\frac{1}{3}\left(3+\frac{2q_2}{(1+z)}-\frac{2(q_2-q_1)}{(1+z)^2}\right), 
\end{equation}
\begin{equation}
Model ~~ III. ~~~w_{eff}(z)=-1+\frac{2}{3}\left(\frac{q_1(1+z)^2}{q_2+(1+z)^2}\right). 
\end{equation}

Utilizing the expression of the effective equation of state, the interaction rate of holographic dark energy can be reconstructed using equation (\ref{intraterech6}). These expressions of Hubble parameter (equation (\ref{hubbleparameter1ch6}) to (\ref{hubbleparameter3ch6})) hardly give any indication regarding the independent conservation of dark matter and dark energy as components are not separately identified. Another point is important to note that it is clear from the expressions of Hubble parameter that $\Lambda$CDM can not be recovered as a limiting case of these models.

\section{Results of statistical analysis}
\label{resultsch6}

The values of the model parameters have been estimated by maximum likelihood analysis. The method of parameter estimation and exhaustive discussion about the observational data are presented in section \ref{obsdata} and \ref{statanalysis}.

%%%%%%%%%%%%%%%%%%%%%
\begin{figure}[H]
\begin{center}
\includegraphics[angle=0, width=0.3\textwidth]{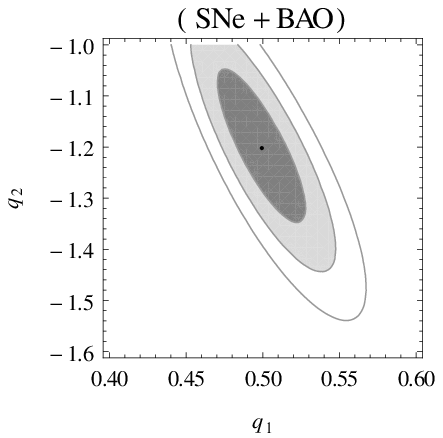}
\includegraphics[angle=0, width=0.3\textwidth]{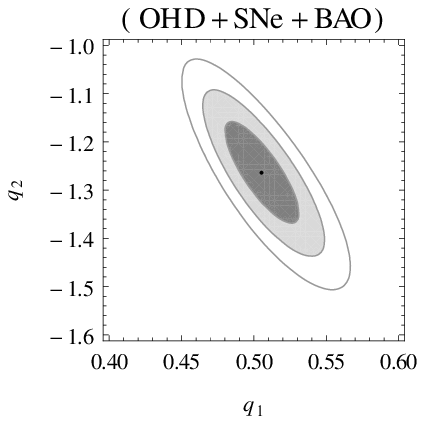}
\includegraphics[angle=0, width=0.3\textwidth]{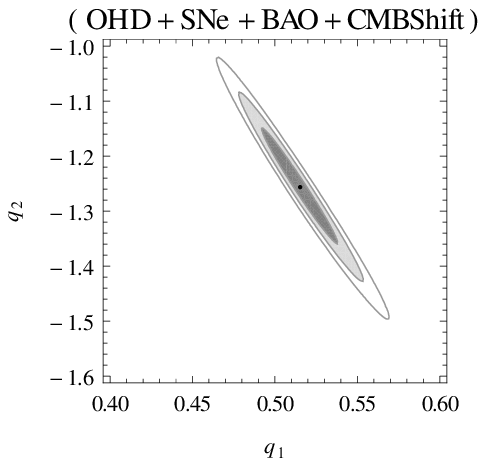}
\end{center}
\caption{{\small The confidence contours on the 2D parameter space of Model I. The 1$\sigma$, 2$\sigma$, and 3$\sigma$ confidence contours are presented from inner to outer regions, and the central black dots represent the corresponding best fit points. The left panel is obtained for SNe+BAO, the moddle panel is obtained for OHD+SNe+BAO and  the right panel is for OHD+SNe+BAO+CMBShift.}}
\label{contourplotqz1ch6}
\end{figure}
%%%%%%%%%%%%%%%%%%%%%%
%%%%%%%%%%%%%%%%%%%%%
\begin{figure}[H]
\begin{center}
\includegraphics[angle=0, width=0.32\textwidth]{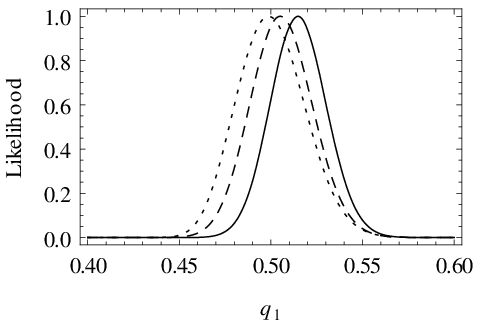}
\includegraphics[angle=0, width=0.32\textwidth]{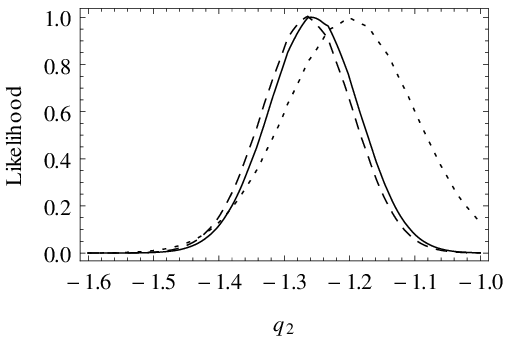}
\end{center}
\caption{{\small The marginalized likelihood as function of the model parameters $q_1$  (left panel) and $q_2$ (right panel) for Model I. The dotted curves are obtained for SNe+BAO, the dashed curves are obtained for OHD+SNe+BAO and the solid curves are obtained for OHD+ SNe+BAO+CMBShift.}}
\label{likelihoodplotqz1ch6}
\end{figure}
%%%%%%%%%%%%%%%%%%%%%%

%%%%%%%%%%%%%%%%%%%%%%%%%
\begin{table}
\caption{{\small  Results of statistical analysis of Model I with different combinations of the data sets. The value of $\chi^2_{min}/d.o.f.$ and the best fit values of the parameters along with the  associated 1$\sigma$ uncertainties are presented.}}
\begin{center}
\resizebox{0.75\textwidth}{!}{  
\begin{tabular}{ c |c |c c } 
\hline
 \hline
  Data & $\chi^2_{min}/d.o.f.$  & $q_1$ & $q_2$ \\ 
 \hline
  SNe+BAO & 35.18/28 & 0.499$\pm$0.051 & -1.202$\pm$0.367\\ 

  OHD+SNe+BAO & 50.57/54 & 0.505$\pm$0.014 & -1.264$\pm$0.064\\ 

  OHD+SNe+BAO+CMBShift & 51.97/52 & 0.515$\pm$0.013 & -1.256$\pm$0.062\\ 
 \hline
\hline
\end{tabular}
}
\end{center}

\label{tablemod1ch6}
\end{table}
%%%%%%%%%%%%%%%%%%%%%%%%%%%%%%%%%%%

%%%%%%%%%%%%%%%%%%%%%%%%%
\begin{table}[h!]
\caption{{\small  Results of statistical analysis of Model II with different combinations of the data sets. The value of $\chi^2_{min}/d.o.f.$ and the best fit values of the parameters along with the  associated 1$\sigma$ uncertainties are presented.}}
\begin{center}
\resizebox{0.75\textwidth}{!}{  
\begin{tabular}{ c |c |c c } 
\hline
 \hline
  Data & $\chi^2_{min}/d.o.f.$  & $q_1$ & $q_2$ \\ 
 \hline
  SNe+BAO & 35.18/28 & -1.189$\pm$0.067 & -0.024$\pm$0.086\\ 

  OHD+SNe+BAO & 50.64/54 & -1.242$\pm$0.050 & -0.007$\pm$0.078\\ 

  OHD+SNe+BAO+CMBShift & 51.17/52 & -1.231$\pm$0.049 & 0.022$\pm$0.073\\ 
 \hline
\hline
\end{tabular}
}
\end{center}

\label{tablemod2ch6}
\end{table}
%%%%%%%%%%%%%%%%%%%%%%%%%%%%%%%%%%%

%%%%%%%%%%%%%%%%%%%%%
\begin{figure}[H]
\begin{center}
\includegraphics[angle=0, width=0.3\textwidth]{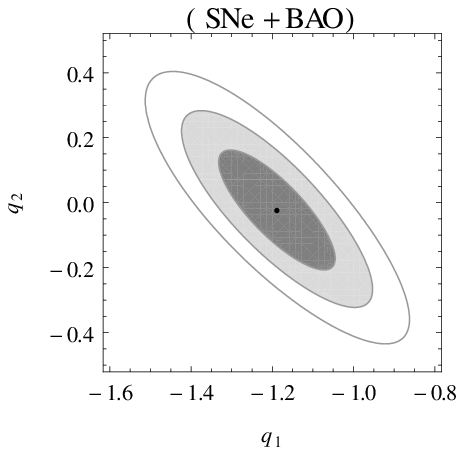}
\includegraphics[angle=0, width=0.3\textwidth]{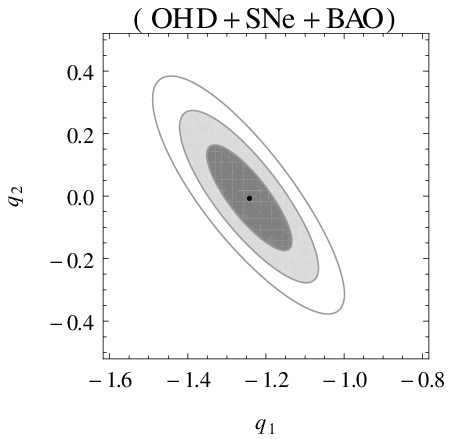}
\includegraphics[angle=0, width=0.3\textwidth]{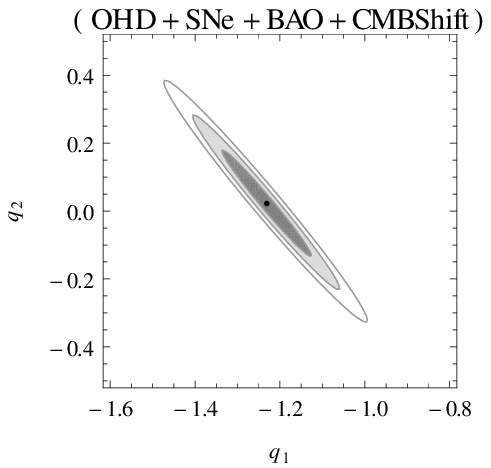}
\end{center}
\caption{{\small The confidence contours on the 2D parameter space of Model II. The 1$\sigma$, 2$\sigma$, and 3$\sigma$ confidence contours are presented from inner to outer regions, and the central black dots represent the corresponding best fit points. The left panel is obtained for SNe+BAO, the middle panel is obtained for OHD+SNe+BAO and the right panel is for OHD+SNe+BAO+CMBShift.}}
\label{contourplotqz2ch6}
\end{figure}
%%%%%%%%%%%%%%%%%%%%%%
%%%%%%%%%%%%%%%%%%%%%
\begin{figure}[H]
\begin{center}
\includegraphics[angle=0, width=0.32\textwidth]{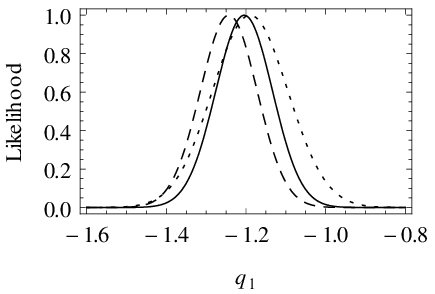}
\includegraphics[angle=0, width=0.32\textwidth]{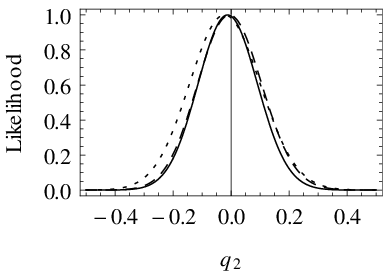}
\end{center}
\caption{{\small  The marginalized likelihood as function of the model parameters $q_1$  (left panel) and $q_2$ (right panel) for Model II. The dotted curves are obtained for SNe+BAO, the dashed curves are obtained for OHD+SNe+BAO and the solid curves are obtained for OHD+SNe+BAO+CMBShift..}}
\label{likelihoodplotqz2ch6}
\end{figure}
%%%%%%%%%%%%%%%%%%%%%%

%%%%%%%%%%%%%%%%%%%%%
\begin{figure}[tb]
\begin{center}
\includegraphics[angle=0, width=0.3\textwidth]{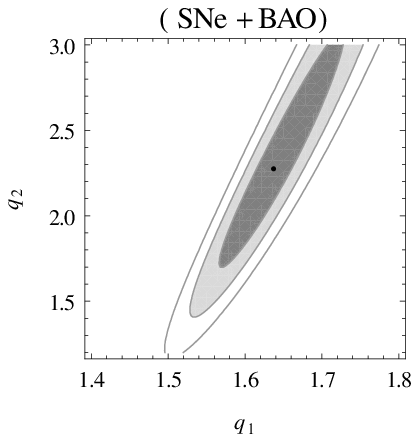}
\includegraphics[angle=0, width=0.3\textwidth]{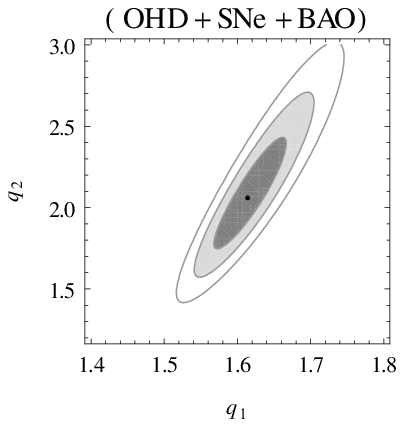}
\includegraphics[angle=0, width=0.3\textwidth]{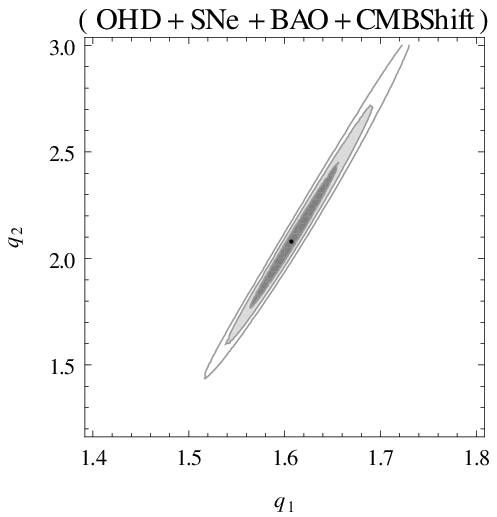}
\end{center}
\caption{{\small The confidence contours on the 2D parameter space of Model III. The 1$\sigma$, 2$\sigma$, and 3$\sigma$ confidence contours are presented from inner to outer regions, and the central black dots represent the corresponding best fit points. The left panel is obtained for SNe+BAO, the middle panel is obtained for OHD+SNe+BAO and the right panel is for OHD+SNe+BAO+CMBShift.}}
\label{contourplotqz3ch6}
\end{figure}
%%%%%%%%%%%%%%%%%%%%%%
%%%%%%%%%%%%%%%%%%%%%
\begin{figure}[tb]
\begin{center}
\includegraphics[angle=0, width=0.32\textwidth]{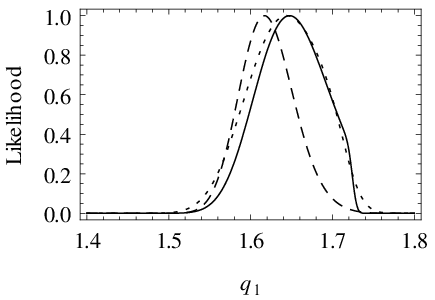}
\includegraphics[angle=0, width=0.32\textwidth]{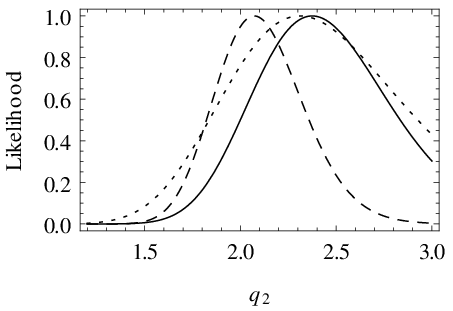}
\end{center}
\caption{{\small The marginalized likelihood as function of the model parameters $q_1$  (left panel) and $q_2$ (right panel) for Model III. The dotted curves are obtained for SNe+BAO, the dashed curves are obtained for OHD+SNe+BAO and the solid curves are obtained for OHD+SNe+BAO+CMBShift.}}
\label{likelihoodplotqz3ch6}
\end{figure}
%%%%%%%%%%%%%%%%%%%%%%

%%%%%%%%%%%%%%%%%%%%%%%%%
\begin{table}[h!]
\caption{{\small  Results of statistical analysis of Model III with different combinations of the data sets. The value of $\chi^2_{min}/d.o.f.$ and the best fit values of the parameters along with the  associated 1$\sigma$ uncertainties are presented.}}
\begin{center}
\resizebox{0.75\textwidth}{!}{  
\begin{tabular}{ c |c |c c } 
\hline
 \hline
  Data & $\chi^2_{min}/d.o.f.$  & $q_1$ & $q_2$ \\ 
 \hline
  SNe+BAO & 33.18/28 & 1.637$\pm$0.037 & 2.275$\pm$0.315\\ 

  OHD+SNe+BAO & 47.80/54 & 1.614$\pm$0.023 & 2.059$\pm$0.162\\ 

  OHD+SNe+BAO+CMBShift & 48.31/52 & 1.607$\pm$0.022 & 2.079$\pm$0.160\\ 
 \hline
\hline
\end{tabular}
}
\end{center}

\label{tablemod3ch6}
\end{table}
%%%%%%%%%%%%%%%%%%%%%%%%%%%%%%%%%%%

%%%%%%%%%%%%%%%%%%%%%
\begin{figure}[t]
\begin{center}
\includegraphics[angle=0, width=0.32\textwidth]{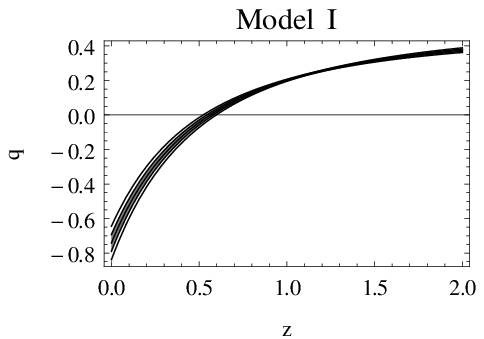}
\includegraphics[angle=0, width=0.32\textwidth]{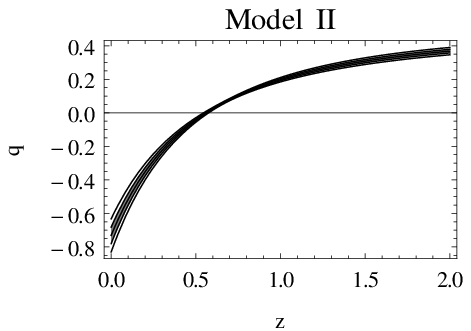}
\includegraphics[angle=0, width=0.32\textwidth]{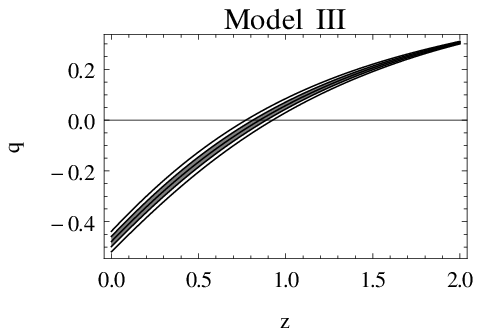}
\end{center}
\caption{{\small Plots of deceleration parameter for the models obtained from the analysis combining OHD, SNe, BAO and CMB shift parameter data. The best fit curve and the associated 1$\sigma$ and 2$\sigma$ confidence regions are presented.}}
\label{decelplotch6}
\end{figure}
%%%%%%%%%%%%%%%%%%%%%%

\begin{figure}[H]
\begin{center}
\includegraphics[angle=0, width=0.32\textwidth]{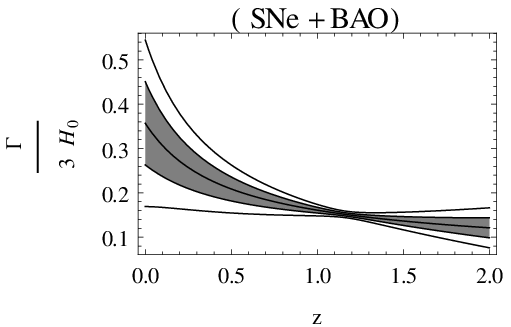}
\includegraphics[angle=0, width=0.32\textwidth]{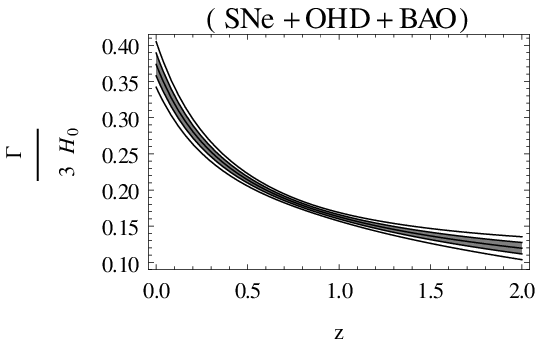}
\includegraphics[angle=0, width=0.32\textwidth]{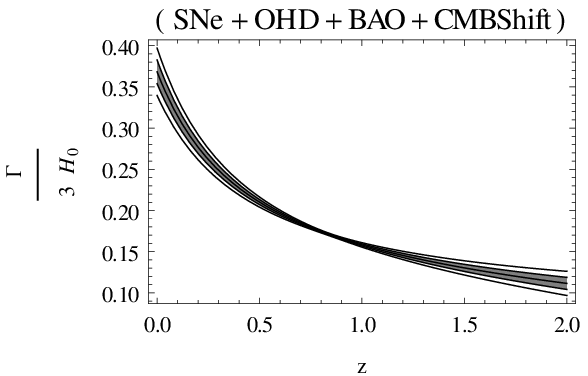}
\end{center}
\caption{{\small The plots of  interaction rate $\Gamma(z)$ scaled by $3H_0$ for Model I. Plots are obtained for three different combinations of the data sets. The left panel is obtained for SNe+BAO, the middle panel is obtained for OHD+SNe+BAO and the right panel is obtained for OHD+SNe+BAO+CMBShift. The 1$\sigma$ and 2$\sigma$ confidence regions and the corresponding best fit curves (the central dark line) are shown.}}
\label{intrateplotqz1ch6}
\end{figure}

\begin{figure}[H]
\begin{center}
\includegraphics[angle=0, width=0.32\textwidth]{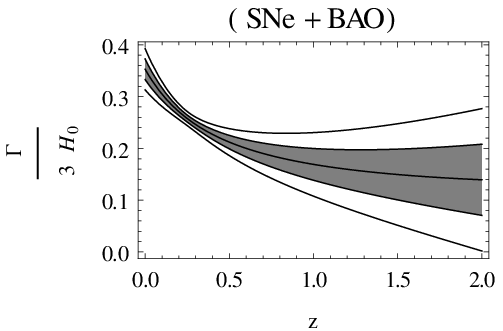}
\includegraphics[angle=0, width=0.32\textwidth]{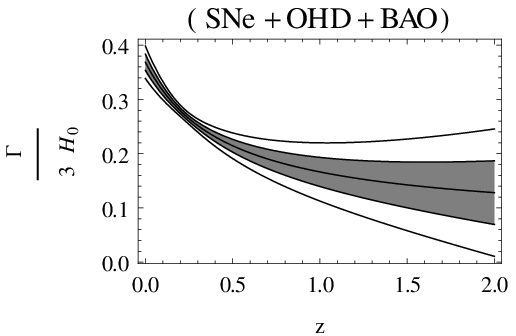}
\includegraphics[angle=0, width=0.32\textwidth]{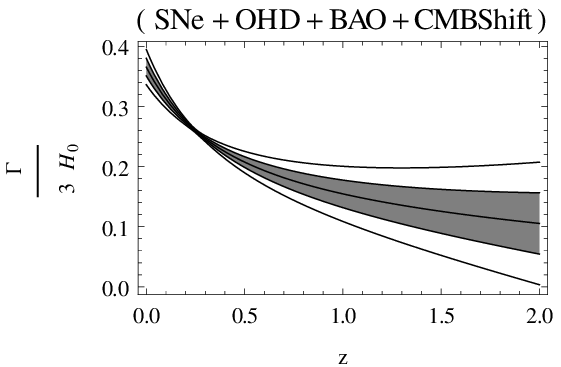}
\end{center}
\caption{{\small The plots of  interaction rate $\Gamma(z)$ scaled by $3H_0$ for Model II. Plots are obtained for three different combinations of the data sets. The left panel is obtained for SNe+BAO, the middle panel is obtained for OHD+SNe+BAO and the right panel is obtained for OHD+SNe+BAO+CMBShift. The 1$\sigma$ and 2$\sigma$ confidence regions and the corresponding best fit curves (the central dark line) are shown.}}
\label{intrateplotqz2ch6}
\end{figure}

\begin{figure}[H]
\begin{center}
\includegraphics[angle=0, width=0.32\textwidth]{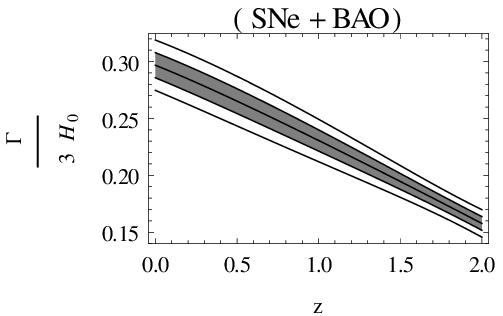}
\includegraphics[angle=0, width=0.32\textwidth]{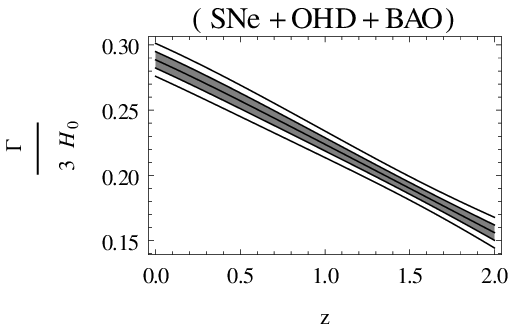}
\includegraphics[angle=0, width=0.32\textwidth]{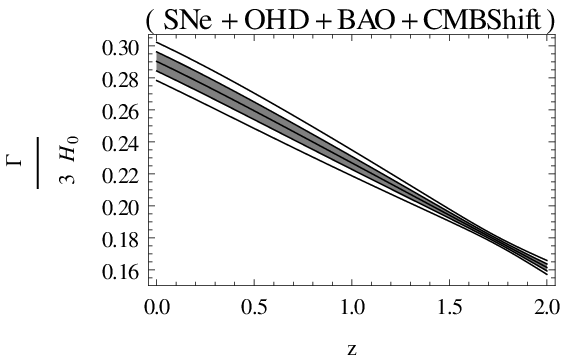}
\end{center}
\caption{{\small  The plots of  interaction rate $\Gamma(z)$ scaled by $3H_0$ for Model III. Plots are obtained for three different combinations of the data sets. The left panel is obtained for SNe+BAO, the middle panel is obtained for OHD+SNe+BAO and the right panel is obtained for OHD+SNe+BAO+CMBShift. The 1$\sigma$ and 2$\sigma$ confidence regions and the corresponding best fit curves (the central dark line) are shown.}}
\label{intrateplotqz3ch6}
\end{figure}

\begin{figure}[H]
\begin{center}
\includegraphics[angle=0, width=0.32\textwidth]{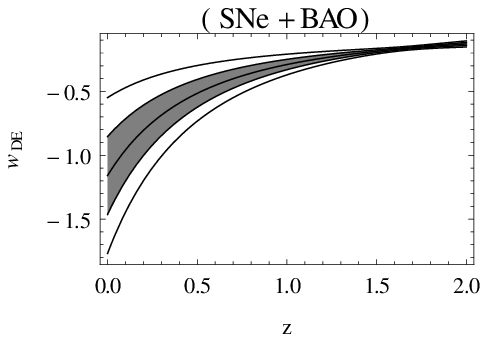}
\includegraphics[angle=0, width=0.32\textwidth]{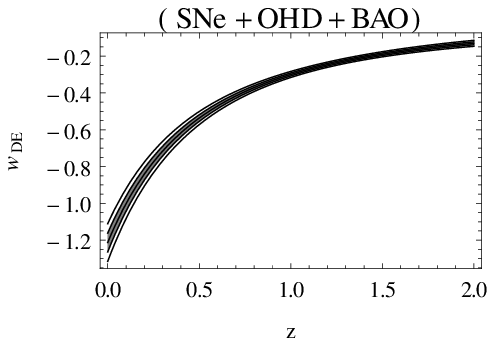}
\includegraphics[angle=0, width=0.32\textwidth]{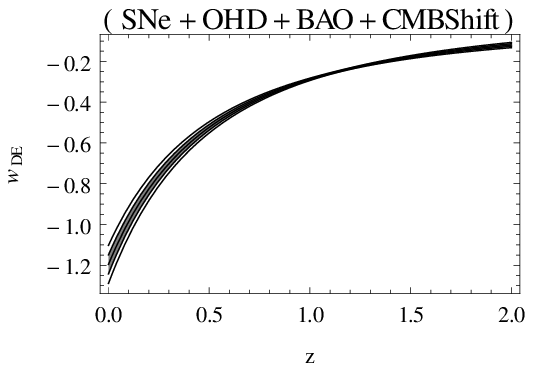}
\end{center}
\caption{{\small The plots of dark energy equation of state parameter $w_{DE}(z)$ for Model I.  The left panel is obtained for SNe+BAO, the middle panel is obtained for OHD+SNe+BAO and the right panel is obtained for OHD+SNe+BAO+CMBShift. The 1$\sigma$ and 2$\sigma$ confidence regions and the corresponding best fit curves (the central dark line) are shown.}}
\label{wDEplotqz1ch6}
\end{figure}

\begin{figure}[H]
\begin{center}
\includegraphics[angle=0, width=0.32\textwidth]{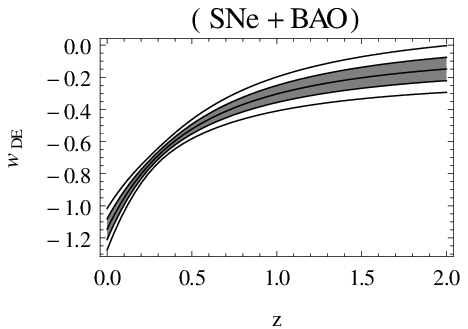}
\includegraphics[angle=0, width=0.32\textwidth]{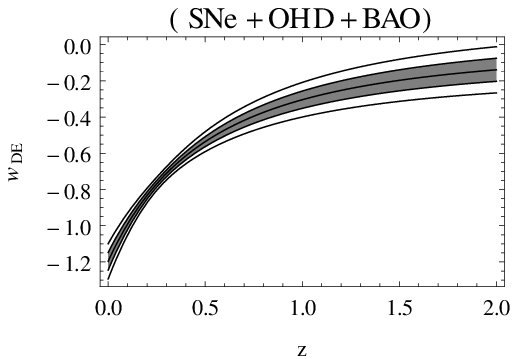}
\includegraphics[angle=0, width=0.32\textwidth]{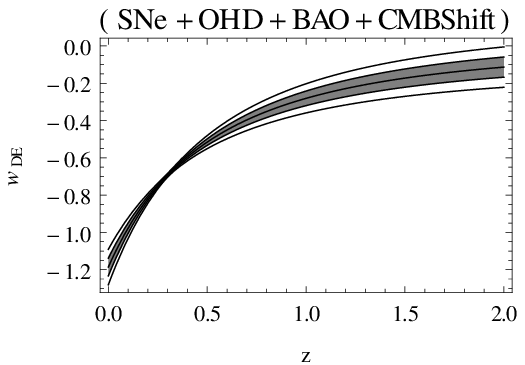}
\end{center}
\caption{{\small The plots of dark energy equation of state parameter $w_{DE}(z)$ for Model II.  The left panel is obtained for SNe+BAO, the middle panel is obtained for OHD+SNe+BAO and the right panel is obtained for OHD+SNe+BAO+CMBShift. The 1$\sigma$ and 2$\sigma$ confidence regions and the corresponding best fit curves (the central dark line) are shown.}}
\label{wDEplotqz2ch6}
\end{figure}

\begin{figure}[H]
\begin{center}
\includegraphics[angle=0, width=0.32\textwidth]{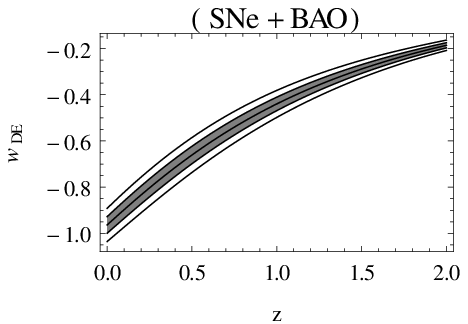}
\includegraphics[angle=0, width=0.32\textwidth]{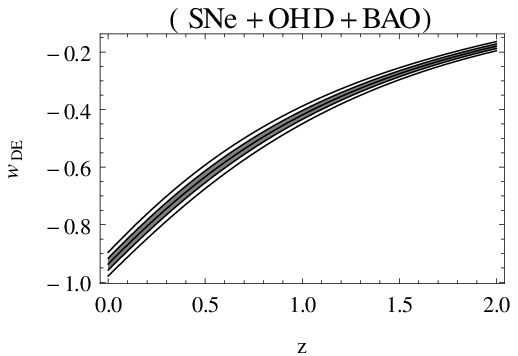}
\includegraphics[angle=0, width=0.32\textwidth]{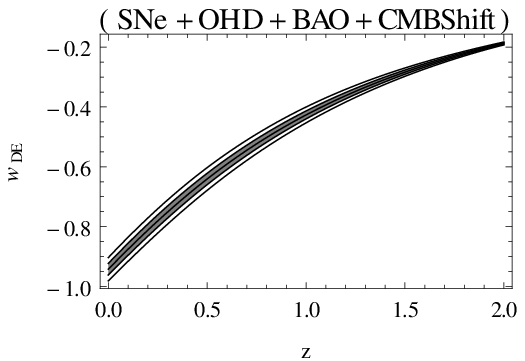}
\end{center}
\caption{{\small  The plots of dark energy equation of state parameter $w_{DE}(z)$ for Model III.  The left panel is obtained for SNe+BAO, the middle panel is obtained for OHD+SNe+BAO and the right panel is obtained for OHD+SNe+BAO+CMBShift. The 1$\sigma$ and 2$\sigma$ confidence regions and the corresponding best fit curves (the central dark line) are shown.}}
\label{wDEplotqz3ch6}
\end{figure}

Figure \ref{contourplotqz1ch6} shows the confidence contours on the 2D parameter space of Model I obtained from analysis with different combinations of the data sets. Figure \ref{likelihoodplotqz1ch6} shows the plots of the marginalized likelihood as functions of the model parameters for Model I. Similarly, figure \ref{contourplotqz2ch6} shows the confidence contours on the 2D parameter space of Model II and figure \ref{likelihoodplotqz2ch6} shows the marginalized likelihoods of Model II. Figure \ref{contourplotqz3ch6} and figure \ref{likelihoodplotqz3ch6} present the contour plots and marginalized likelihood plots for Model III. It is apparent from the contour plots and the likelihood function plots that the addition of the CMB shift parameter data does not lead to much improvement of the constraints on the model parameters. The likelihood  functions are well fitted to Gaussian distribution.  Table \ref{tablemod1ch6} presents the results of statistical analysis of Model I. The reduced $\chi^2$ i.e. $\chi^2_{min}/d.o.f.$, where the $d.o.f.$ is the degrees of freedom associated to the analysis, the best fit values of the parameters along with the associated 1$\sigma$ error bars are presented. In the similar way, table \ref{tablemod2ch6} and \ref{tablemod3ch6} present the results of the statistical analysis of Model II and Model III respectively. Figure \ref{decelplotch6} shows the plots of deceleration parameter for the models obtained in the combined analysis with OHD, SNe, BAO and CMB shift parameter data. The plots of the interaction rate ($\Gamma(z)/3H_0$) (figure \ref{intrateplotqz1ch6} to \ref{intrateplotqz3ch6}) show that the interaction was low at earlier and it increases significantly at recent time. For Model I and Model III, the nature of constraint on the interaction rate, obtained in the analysis combining OHD, SNe, BAO and CMB shift data, is similar at present time and at high redshift. But for Model II, the uncertainty increases at high redshift. The plots of the dark energy equation of state parameter $w_{DE}(z)$ also shows a very similar behaviour for the models, (figure \ref{wDEplotqz1ch6} to \ref{wDEplotqz3ch6}). It is imperative to note that for Model I and Model II, the dark energy equation of state parameter indicates a phantom nature at present as $w_{DE}(z=0)<-1$ at 2$\sigma$ confidence level  and for Model III, it is slightly inclined towards the non-phantom nature. At high redshift, the value of $w_{DE}(z)$ be close to zero and thus allows a matter dominated epoch in the past.

%%%%%%%%%%%%%%%%%%%%%
%%%%%%%%%%%%%%%%%%%%%%
%%%%%%%%%%%%%%%%%%%%%
%%%%%%%%%%%%%%%%%%%%%%
%%%%%%%%%%%%%%%%%%%%%
%%%%%%%%%%%%%%%%%%%%%%

%%%%%%%%%%%%%%%%%%%%%
%%%%%%%%%%%%%%%%%%%%%%
%%%%%%%%%%%%%%%%%%%%%
%%%%%%%%%%%%%%%%%%%%%%
%%%%%%%%%%%%%%%%%%%%%
%%%%%%%%%%%%%%%%%%%%%%

\par The interaction rate $\Gamma(z)$ remains positive throughout the evolution and increases with the expansion of  the Universe. As the interaction term $Q$ is assumed to be $Q=\rho_H\Gamma$, $Q$ is also positive. This reveals that in the interaction, the energy gets transferred from dark energy to dark matter. It is consistent with the thermodynamic requirement of a positive $Q$ \cite{PavonWang2009}. It is important to note that though the parametrization for Model III is significantly different from Model I and Model II, the basic nature of the interaction rate is same in all the case. Similar results have been obtained by Sen and Pavon \cite{SenPavon2008} where the interaction rate of holographic dark energy has been reconstructed from a parametrization of dark energy equation of state parameter. Though tighter constraints have been achieved in the present work as it is based on larger data sets, the basic nature of the interaction rate is very similar to the results obtained in the previous findings.

\section{Bayesian evidence and model selection}
\label{bayesianch6}

The Bayesian evidence is defined as,

\begin{equation}
E=\int(Prior\times Likelihood)d\theta_1d\theta_2...\theta_n,
\end{equation}
where $\theta_i$ are the parameters of the model considered. In the present analysis, a constant prior has been assumed for the parameter values for which the posterior is proportional to the likelihood. The evidence calculated for these models are,

\begin{equation}
Model~~~I:~~E_1=P_1\int Likelihood.dq_1dq_2=5.134\times10^{-14},
\end{equation}
\begin{equation}
Model~~~II:~~E_2=P_2\int Likelihood.dq_1dq_2=7.773\times10^{-14},
\end{equation}
\begin{equation}
Model~~~III:~~~E_3=P_3\int Likelihood.dq_1dq_2=21.79\times10^{-14},
\end{equation}
where $P_1$, $P_2$ and $P_3$ are the constant prior of Model I, Model II and Model III respectively. The calculation of Bayesian evidence does not give any significant information about the model selection as the value of $E_1$, $E_2$ and $E_3$ are not significantly different. It can only be concluded that the the Model III is marginally preferred than other two models.

\section{Discussion}
\label{discussionch6}
This chapter deals with an attempt to reconstruct the interaction rate for holographic dark energy. The models are based on the parameterizations of the deceleration parameter $q(z)$. The expressions for the Hubble parameter, obtained for these parametrizations of the deceleration parameter (equation (\ref{hubbleparameter1ch6})-(\ref{hubbleparameter3ch6})), give absolutely no clue to identify the dark matter and the dark energy components searately and hence indicates towards an interaction between them. The idea of the present work is to study the nature of interaction, mainly the interaction rate, for these three cases assuming the dark energy to be holographic with Hubble horizon as the IR cut-off. As mentioned earlier, the holographic dark energy with Hubble horizon as the IR cut-off requires an interaction between dark energy and dark matter to generate the late time acceleration along with the matter dominated, characterized by a decelerated expansion, phase that prevailed in the past. 

\par It has also been mentioned earlier that in a spatially flat geometry, the ratio of dark matter and dark energy density in a holographic dark energy model with Hubble horizon as the IR cut-off remains constant. Thus it could be a reasonable answer to the cosmic coincidence problem. As the dark energy equation of state parameter tends to zero at high redshift, the dark energy behaved like dust matter in the past. Thus it produces the matter dominated phase in the past which is consistent with the standard models of structure formation. The interaction rate ($\Gamma$) and consequently the interaction term $\eta$, where $\eta=\rho_H\Gamma$, remain positive through the evolution for the reconstructed models. It indicates that in the interaction, the energy gets transferred from dark energy to dark matter which is consistent with the second law of thermodynamics \cite{PavonWang2009}. Though the parametrizations are different, the basic natute of interaction rate remains same in all the cases. Similar results have also been found by Sen and Pavon \cite{SenPavon2008} where the interaction rate has been reconstructed from parametrization of dark energy equation of state. The dark energy equation of state parameter shows a highly phantom nature at present for the Model I and Model II. For Model III, however,  it is inclined towards a non-phanton nature. The plots of deceleration parameter for this models (figure \ref{decelplotch6}) show that the early decelerated expansion phase phase is successfully recovered by these models. The value of deceleration parameter is close to 0.5 at high redshisty which is equivalent to the dust matter dominate phase and the transition form decelerated to accelerated phase occurred in the redshift range 0.6 to 0.8.

\par The plots of interaction rate  for these models (figure \ref{intrateplotqz1ch6} to \ref{intrateplotqz3ch6}) show that the best fit curves for Model I and Model II behave in a very similar way and for Model III, it is slightly different. The nature of the associated uncertainty is different for these three models. For Model II, the uncertainty increases at high redshift. Similar behaviour can also be found in the dark energy equation of state parameter ($w_{DE}(z)$) plots of the models (figure \ref{wDEplotqz1ch6} to \ref{wDEplotqz3ch6}).

\par Three different combinations of  the data sets have been used in the analysis. The first one is  the combination of SNe and BAO, the second combination is of OHD, SNe and BAO. The CMB shift parameter data has been added to it in the third combination.  It is apparent that the addition of CMB shift parameter data does not lead to much improvement to the constraints on the model parameters. In case of the supernova data, the systematics have also been taken into account in the statistical analysis as the systematics might have its signature on the results. 

\par For a comparison of models, the Bayesian evidence calculation have been invoked. The Bayesian evidences for the models are of the same order of magnitude. It can only be concluded by looking at the ratio of the Bayesian evidences of these three models, that Model III is slightly preferred than Model I and Model II, but they are comparable to each other in terms of model selection.

\chapter{Conclusion}

The present thesis contains the investigations on the reconstruction of cosmological models. The reconstructions are mainly based on the parametric approach where the prime endeavour is to estimate the values of the model parameters and then to figure out the evolution of different cosmological quantities like the Hubble parameter, deceleration parameter, dark energy equation of state parameter etc.

\par The indispensable part of a reconstruction is the statistical analysis to estimate the parameter values and the associated uncertainties. Maximum likelihood analysis method has been adopted here to estimate the values of the model parameters using different observational data sets. Mainly the distance modulus data of type Ia supernova (SNe), observational measurements of Hubble parameter (OHD), baryon acoustic oscillation data (BAO) and CMB distance prior, namely the CMB shift parameter (CMBShift) have been utilized. These are the most relevant data sets for the reconstruction of late time cosmology. The models, discussed in the present work, are well constrained in combined analysis with these four data sets.

\par The reconstruction of the models, which have been discussed in the chapters of the thesis, are based on the parametrization of various cosmological quantities. These models can be classified into two sections, based on the nature of conservation of the dark energy and the dark matter. One type of models allow the dark matter and dark energy to have independent conservation without any possibility of interaction between them. The other type of models allows the possibility of interaction between the dark matter and the dark energy.  

\par The results obtained for most of the models show that the models are in close proximity of $\Lambda$CDM (in most of the cases the $\Lambda$CDM remains within 1$\sigma$ confidence region). Some of the models have inclination towards the phantom nature of dark energy (i.e. $w_{DE}<-1$) and in some cases, non-phantom nature is preferred. Though the results are model dependent, it can be concluded that the dark energy equation of state parameter hovers around the limit $w_{DE}=-1$. At this point, it is important to mention that though the $\Lambda$CDM cosmology is well consistent with most of the observational data, mainly with the observation at very low redshift, but it might not be true for the observations at slightly higher redshift. The recent measurement of BAO from the flux-correlation of the Lyman-alpha (Ly$\alpha$) forest of BOSS quasars \cite{delubacohd} and cross-correlation function of Ly$\alpha$ forest with quasars \cite{afRibera} show some discrepancy with Planck $\Lambda$CDM cosmology. The estimation, from  Ly$\alpha$ forest BAO, of the quantity $c/H(2.34)r_s(z_d)=9.14\pm0.20$ \cite{eAubourg} is at $2.7\sigma$ level of discrepancy with the Planck $\Lambda$CDM prediction of $8.586\pm0.021$ \cite{planck2015}. Though it is not clear at the moment whether this discrepancy is due to the systematics of  Ly$\alpha$ forest BAO, which is more complicated than the galaxy BAO data, or it indicates towards a new physics.

\par Another important aspect, which has been emphasised in the present work, is the kinematic approach to the reconstruction. The cosmological quantities which are constructed from the scale factor and its time derivatives, are the kinematical quantities, for example the Hubble parameter, the deceleration parameter, the cosmological jerk parameter etc. The kinematic approach to the reconstruction of cosmological models is independent of any prior assumption about the gravity theory and the nature of dark energy. Thus it is different from the standard dynamical approach to the reconstruction of dark energy where the model is reconstructed from the prior assumption about the  dark energy equation of state, the scalar field energy density or potential. The reconstruction of dark energy equation of state and the reconstruction of effective or total equation of state which have been discussed in chapter 2 and chapter 3 respectively, belong to the dynamical approach. On the other hand, kinematical quantities, mainly the cosmological jerk parameter has been specially focused in some cases (in chapter 4 and chapter 5). In chapter 4, time evolving jerk parameter models have reconstructed. In chapter 5, possibility of interaction between dark energy and dark matter has been investigated through the assumption of a slowly varying jerk parameter. In both the cases, the models behave like $\Lambda$CDM for certain values of the model parameters. 

\par Holographic dark energy model has also been studied (chapter 6). The rate of interaction between dark energy and dark matter has been reconstructed for a holographic model with Hubble horizon as the infrared cut-off. In this context, it is important to  note that the interaction is essential for a holographic dark energy model with Hubble horizon cut-off to generate a late time cosmic acceleration. The dark energy equation of state parameter in this case shows a huge deviation from the $\Lambda$CDM scenario.    
 
\par It is important to note that the models which allows the interaction between the dark energy and dark matter, have non-zero contribution of dark energy component in the energy budget at high redshidt. If the dark energy component is not negligible in the pre-recombination era, then the sound horizon ($r_s$) is reduced be  a factor $(1-\Omega_{de}^e)^{\frac{1}{2}}$, where $\Omega_{de}^e$ is the dark energy density at pre-recombination scaled by the present critical density \cite{doramsam}. Again the Hubble parameter is higher by a factor of $(1-\Omega_{de}^e)^{-\frac{1}{2}}$ due to the existence of early dark energy \cite{eAubourg}. The BAO scale, namely the {\it dilation scale} (equation \ref{dVscale}), has been scaled by the sound horizon at photon drag epoch ($r_s(z_d)$). Thus the factor  $(1-\Omega_{de}^e)^{\frac{1}{2}}$ get cancelled and the BAO scale remains insensitive to the existence of any dark energy contribution at pre-recombination era.

\par The present thesis the based on {\it reconstruction} of dark energy models. The basic idea of the reconstruction of lies in the assumption that the dark energy equation of state parameter, energy density or the kinematical quantities (like the deceleration parameter, cosmological jerk parameter) must be a smooth function of time or redshfit. In case of a parametric reconstruction, these functions are represented with some simple functional form in terms of redshift and other model parameters. The functional from reduces to $\Lambda$CDM for some limit of the model parameters and in most of the cases, the $\Lambda$CDM value of the model parameters remains within 1$\sigma$ error bar of the parameter values estimated in the statistical analysis.  But it is not possible to judge whether a model parameter has different values at different redshift regime, that means we can not figure out whether the parameters itself has some evolution. If another functional form is introduced to figure out the evolution of the parameter, then it will introduce new parameters and the model will be penalized due to the increase in the number of parameters. One possible way to overcome this problem is to estimate the parameters at different redshift regime separately. But there is no proper way to find the correlation between different redshift bins. A non-parametric reconstruction of cosmological quantities like deceleration parameter or equation of state parameter \cite{nonparreconst} can potentially resolve the problem, but it suffers from the lack of data points at high redshift. As there is very less number of data points at high redshift, the uncertainty associated to the reconstructed parameter becomes very high at high redshift. Besides, non-parametric reconstruction of the parameters which involves higher order differentiation of the data (like the jerk parameter) makes the confidence region very large and anything can hardly be concluded about the model. So it may be possible that the $\Lambda$CDM is favoured due to our lack of knowledge about the time evolution.

%%%%%%%%%%%%%%%%%%%%%%%%%%

% ********************************** Back Matter *******************************
% Backmatter should be commented out, if you are using appendices after References
%\backmatter

% ********************************** Bibliography ******************************
\begin{spacing}{0.9}

%\bibliographystyle{myapsrev}
%\bibliography{dissertation}
%\addcontentsline{toc}{chapter}{Bibliography}
% To use the conventional natbib style referencing
% Bibliography style previews: http://nodonn.tipido.net/bibstyle.php
% Reference styles: http://sites.stat.psu.edu/~surajit/present/bib.htm

\bibliographystyle{apalike}
\bibliographystyle{plainnat} % use this to have URLs listed in References
\cleardoublepage
\bibliography{References/thesis} % Path to your References.bib file

% If you would like to use BibLaTeX for your references, pass `custombib' as
% an option in the document class. The location of 'reference.bib' should be
% specified in the preamble.tex file in the custombib section.
% Comment out the lines related to natbib above and uncomment the following line.

%\printbibliography[heading=bibintoc, title={References}]

\end{spacing}

% ********************************** Appendices ********************************

%\begin{appendices} % Using appendices environment for more functunality

%\include{Appendix1/appendix1}
%\include{Appendix2/appendix2} 

%\end{appendices}

% *************************************** Index ********************************
\printthesisindex % If index is present

\end{document}